\documentclass[mnsc,nonblindrev]{informs3} % current default for manuscript submission

\usepackage{tcolorbox}

\newtcolorbox[auto counter]{bigdef}[1][]{
  fonttitle=\scshape,
  title={Mechanism \thetcbcounter: The square root agreement rule (SRA). Assumes $|\cM_i|\geq 2$ for all $i\in\cN$.},
  #1
}

\newtcolorbox[auto counter]{bigdef2}[1][]{
  fonttitle=\scshape,
  title={Mechanism 2: The Peer-Truth-Serum (PTS) mechanism \citep{radanovic2016incentives}. Assumes $|\cM_i|\geq 2$ for all $i\in\cN$. },
  #1
}
\usepackage{enumitem}
\usepackage{booktabs}
    \usepackage{epsfig}
    \usepackage{amsmath}
    \usepackage{amsfonts}
    \usepackage{amssymb}
\usepackage{graphicx}
\usepackage{algpseudocode}
\usepackage{algorithm}
\usepackage{bm}
\usepackage{float}
    \usepackage{eucal}
\usepackage{subcaption}
\usepackage{capt-of}
\usepackage{xcolor}
\usepackage{caption}
\usepackage{pifont}
\usepackage{bbm}
\usepackage[T1]{fontenc}

\newcommand{\cS}{\mathcal{S}}
\newcommand{\cX}{\mathcal{X}}
\newcommand{\cH}{\mathcal{H}}
\newcommand{\cY}{\mathcal{Y}}
\newcommand{\cW}{\mathcal{W}}
\newcommand{\cN}{\mathcal{N}}
\newcommand{\cM}{\mathcal{M}}

\newcommand{\bu}{\mathbf{u}}
\newcommand{\bt}{\mathbf{t}}
\newcommand{\bp}{\mathbf{p}}
\newcommand{\bq}{\mathbf{q}}
\newcommand{\bv}{\mathbf{v}}

\newcommand\numberthis{\addtocounter{equation}{1}\tag{\theequation}}
\newtheorem{prop}{Proposition}[section]

\usepackage{natbib}
 \bibpunct[, ]{(}{)}{,}{a}{}{,}%
 \def\bibfont{\small}%
\DeclareMathOperator{\sgn}{sgn}
\usepackage{url}

\TheoremsNumberedThrough     % Preferred (Theorem 1, Lemma 1, Theorem 2)
\EquationsNumberedThrough    % Default: (1), (2), ...
\newcommand{\cmark}{\ding{51}}%
\newcommand{\xmark}{\ding{55}}%
\begin{document}
%%%%%%%%%%%%%%%%

%\setcounter{page}{1}
%\DOI{10.1287/xxxx.0000.0000}%

 \RUNAUTHOR{Kamble et. al.}

\RUNTITLE{The Square Root Agreement Rule for Incentivizing Truthful Feedback on Online Platforms}

% \TITLE{Bundling Information Goods of Decreasing Value}
\TITLE{The Square Root Agreement Rule for Incentivizing Truthful Feedback on Online Platforms}
\ARTICLEAUTHORS{
\AUTHOR{Vijay Kamble\\ The University of Illinois at Chicago\\ \textit{kamble@uic.edu}}
\AUTHOR{Nihar Shah\\ Carnegie Mellon University\\ \textit{nihars@cs.cmu.edu}}
\AUTHOR{David Marn\\ University of California, Berkeley\\ \textit{marn@berkeley.edu}}
\AUTHOR{Abhay Parekh\\ University of California, Berkeley\\ \textit{yahbaa@gmail.com}}
\AUTHOR{Kannan Ramchandran\\ University of California, Berkeley\\ \textit{kannanr@eecs.berkeley.edu}}
}
%\AFF{Department of Information and Decision Sciences,\\ University of Illinois at Chicago\\ \EMAIL{kamble@uic.edu}}
%\AUTHOR{Patrick Loiseau}
%\AFF{Univ. Grenoble Alpes, CNRS, INRIA, Grenoble INP, LIG, France\\ Max-Planck Institute for Software Systems (MPI-SWS), Germany\\
%\EMAIL{patrick.loiseau@univ-grenoble-alpes.fr}}
%\AUTHOR{Jean Walrand}
%\AFF{Dept. of Electrical Engineering and Computer Science,\\ University of California, Berkeley\\ \EMAIL{walrand@berkeley.edu}}
%} % end of the block

%Tiergartenstr. 17, 69121 Heidelberg, Germany\\
%\url{http://www.springer.com/lncs}}
%
%
%\toctitle{Lecture Notes in Computer Science}
%\tocauthor{Authors' Instructions}

\ABSTRACT{
A major challenge in obtaining evaluations of products or services on e-commerce platforms is eliciting informative responses in the absence of verifiability. This paper proposes the {\it Square Root Agreement Rule} (SRA): a simple reward mechanism that incentivizes truthful responses to objective evaluations on such platforms. In this mechanism, an agent gets a reward for an evaluation only if her answer matches that of her peer, where this reward is inversely proportional to a {\it popularity index} of the answer.
This index is defined to be the square root of the empirical frequency at which any two agents performing the same evaluation agree on the particular answer across evaluations of similar entities operating on the platform. Rarely agreed-upon answers thus earn a higher reward than answers for which agreements are relatively more common. 

We show that in the many tasks regime, the truthful equilibrium under SRA is strictly payoff-dominant across large classes of natural equilibria that could arise in these settings, thus increasing the likelihood of its adoption. While there exist other mechanisms achieving such guarantees, they either impose additional assumptions on the response distribution that are not generally satisfied for objective evaluations or they incentivize truthful behavior only if each agent performs a prohibitively large number of evaluations and commits to using the same strategy for each evaluation. SRA is the first known incentive mechanism satisfying such guarantees without imposing any such requirements. Moreover, our empirical findings demonstrate the robustness of the incentive properties of SRA in the presence of mild subjectivity or observational biases in the responses. These properties make SRA uniquely attractive for administering reward-based incentive schemes (e.g., rebates, discounts, reputation scores, etc.) on online platforms. %Importantly, these incentive properties are the strongest known and SRA is the unique mechanism that achieves them while requiring only one evaluation per agent.
}

\maketitle

\section{Introduction}\label{sec:intro}
Reputation systems, in which people provide feedback for products or services based on their personal experiences, are a critical component of online platforms and marketplaces~\citep{resnick2000reputation,josang2007survey,tadelis2016reputation,luca2017designing}. These systems improve the overall quality of transactions, increase trust, and thus play a key role in determining the success of these platforms in the long run. A major practical challenge in these systems is that of eliciting truthful and high-quality responses from the agents. In the absence of appropriate incentives, agents could shirk investing effort, provide uninformative feedback, or even exploit these systems for selfish motives, thus undermining their utility. For instance, significant empirical evidence of bias in user ratings has been found on many online platforms \citep{hu2017self,filippas2018reputation,nosko2015limits}. This work describes a simple and intuitive reward mechanism that attempts to address this concern.

We consider a setting where an online platform is interested in obtaining responses for a large number of evaluations pertaining to the products or the services being offered on the platform from a pool of customers, whom we refer to as agents. We focus on \emph{objective but unverifiable evaluations}, i.e., evaluations in which the answers can, in principle, be objectively verified, but such verification is infeasible for the platform. This is the case for evaluations comprising of questions like:
\begin{enumerate}
\item What was your waiting time to get a table in the restaurant? (Less than 15 mins/Between 15-30 mins/More than 30 mins)
\item Did the plumber show up within 5 mins of your appointed time? (Yes/No)
\item How long did the moving company take to respond with a quote? (1 day/2 days/3 days/more than three days)
\item Did the dimensions of the received product exceed the dimensions given by the seller? (Yes/No)
\item How long did it take for the product to arrive after the purchase was made from the seller? (1 week/2 weeks/3 or more weeks)
\end{enumerate}

In each of these questions, the evaluating agent is being asked to truthfully report an {\it observation} about the entity being evaluated. The main property of such evaluations is that each evaluating agent's observation is an independent sample from an unknown distribution of behaviors specific to the entity being evaluated. In other words, the true responses of agents for a fixed evaluation task are \emph{conditionally independent and identically distributed} (conditional on the unknown distribution of responses). For example, in the first situation, we can assume that each customer experiences an independently sampled waiting time from a common unknown distribution specific to that restaurant. In the second situation, the customer's experience is sampled from the distribution of whether or not the plumber is punctual. Similarly, in the remaining questions, the customer's true experience is a sample of the moving company's or the seller's business practices. We will refer to this property as the responses being {\it homogeneous} for the rest of the paper, informally referring to the fact that the true responses of any set of agents to a fixed evaluation task are statistically exchangeable. 
%In all three cases, the true responses are independent of the individual characteristics of the customers.

In such scenarios, we hope to achieve the following informal goals through the design of an effective incentive mechanism: (a) incentivize agents to participate in the provision of feedback in online platforms, i.e., improve response rates, and (b) conditional on participation, incentivize agents to report true observations while overcoming any observation or reporting bias. The second goal is arguably more critical and challenging since an easy way of achieving the first goal is to give everyone a fixed reward for participation. As one can imagine, such a na\"ive reward scheme may not necessarily lead to a high quality of responses.

If the platform could verify the responses to the evaluations, it can simply reward the agents based on whether or not they reported their true observations. But such verification is infeasible for questions such as the ones mentioned above since these evaluations are based on interactions that take place outside the platform. In these cases, inducing truthful behavior is a challenging problem. A common approach to this problem, first described in the pioneering work of \cite{miller2005eliciting}, is to reward the agents' responses based on comparisons with the responses of other agents who have performed the same evaluation task. Such mechanisms have come to be referred to as {\it peer-prediction mechanisms} in subsequent literature (after the original mechanism called the {\it peer-prediction method} described in \cite{miller2005eliciting}).
Informally, such reward mechanisms leverage the property that the true response of any agent is correlated with the response of some other agent for the same question.\footnote{With homogeneous responses, the structure of the correlation between an agent's true response and the true response of a typical agent in the population is identical across agents. This feature contrasts with the case when the responses are \emph{heterogeneous}, i.e., when the agents' true responses strongly depend on their characteristics that vary widely across the population. 
In these cases, designing mechanisms without obtaining requisite fine-grained information about agent heterogeneity or without making any regularity assumptions on the agent responses, e.g., `self-predicting responses' \citep{radanovic2016incentives} or `categorical responses' \citep{dasgupta2013crowdsourced, shnayder2016informed} (these assumptions are discussed in Section~\ref{sec:related}), is known to be impossible \citep{radanovic2015incentives}.} 
The situation is then inherently strategic, in which one hopes to sustain truthful reporting as an equilibrium of the game that the reward mechanism induces. It is additionally desirable that such an equilibrium is preferable to the agents over other non-truthful equilibria that may arise in the game. 

{\bf Our contribution.} In this paper, we propose the {\it Square Root Agreement Rule} (SRA): a new peer-prediction mechanism for online platforms that truthfully elicits objective but unverifiable responses at equilibrium. In the setting of our interest, i.e., elicitation on online platforms, we show that the truthful equilibrium under SRA satisfies a key dominance property; namely, it yields the agents the highest payoff amongst all symmetric equilibrium payoffs in the system limit where there are a large number of evaluation tasks. In addition, the truthful equilibrium payoff is {\it strictly} higher than that under any symmetric equilibrium strategy profile that incurs any degree of information loss in the reports. In keeping with the existing terminology in the literature, we refer to this property as SRA being asymptotically {\it strongly truthful across symmetric equilibria}. Such a dominance property is crucial in these settings since it hinders the emergence of low-effort equilibria with poorly informative reports -- such as everyone reporting the same answer irrespective of their true evaluation -- known to plague many other incentive mechanisms, e.g., the peer-prediction method of \cite{miller2005eliciting}.

Moreover, under certain additional assumptions satisfied in applications such as crowdsourcing,\footnote{Crowdsourcing on labor platforms such as Amazon Mechanical Turk is an important means for sourcing the large-scale execution of information-oriented micro-tasks, such as obtaining labeled data for training machine learning algorithms. Incentivizing truthful, high-quality responses from participants is a key concern in these applications. } we show that the truthful equilibrium under SRA gives the highest payoff amongst {\it all} equilibrium payoffs (and not just symmetric equilibrium payoffs) in the large system limit, i.e., SRA is asymptotically {\it strongly truthful} under these assumptions.

While such strong truthfulness guarantees are satisfied by existing mechanisms, they either impose conditions on the response distributions that are not satisfied in general for objective evaluations \citep{dasgupta2013crowdsourced, radanovic2016incentives} or they require a prohibitively large number of evaluations from each agent and assume that the agent uses the same reporting strategy for each evaluation \citep{kong2016framework, kong2020dominantly}. SRA is the first known mechanism that achieves this guarantee for objective evaluations without imposing any such constraints; in particular, SRA is the first mechanism satisfying strong truthfulness guarantees for objective evaluations that incentivizes truthful behavior even among agents who perform a {\it single} evaluation. 

This result is arguably non-trivial. The dominant existing framework for designing strongly truthful peer-prediction mechanisms is due to \cite{kong2016framework}, which incentivizes truthful behavior only if the agents perform multiple evaluations (at least twice the number of possible responses to any evaluation; see \cite{kong2020dominantly}) and additionally commit to using the same reporting strategy for each evaluation. In Section~\ref{sec:related} and Section D.4 in the Appendix, we show that even if one leverages the homogeneous responses property satisfied by objective evaluations, it is not possible to generically adapt the approach of \cite{kong2016framework} to incentivize truthfulness in a single evaluation. By designing SRA, we nevertheless demonstrate that this is indeed possible. In showing this result, we make novel information-theoretic contributions that are of interest beyond this work.

The fact that SRA strongly incentivizes even a single response is vital since requiring multiple evaluations from any agent to incentivize truthfulness is impractical in platforms where customers interact with the marketplace relatively rarely, e.g., in vacation rental platforms such as Airbnb. In these settings, requiring a single additional evaluation from each user may prohibitively increase the evaluation period's duration. Such delay and resulting changes in market characteristics over time undermine the platform's ability to procure the most up-to-date feedback information reliably. While there are mechanisms that incentivize truthfulness in a single evaluation under the homogeneous responses assumption (see Section~\ref{sec:related} for a discussion), none of these mechanisms satisfy the crucial {\it strong} truthfulness guarantees that SRA satisfies. 

%Finally, our numerical evaluations on synthetic as well as real data show that SRA generates strong incentives for truthful behavior with finite data despite the presence of subjectivity or observational biases in the agent population.

{\bf Background.} To appropriately position our contribution, we first present a brief discussion of the main existing approaches in incentive design for the elicitation of unverifiable responses. 

In a pioneering work in this domain, \cite{miller2005eliciting} considered the case of a single evaluation task and homogeneous responses and described the so-called \emph{peer-prediction method} that incentivizes truthful answers. The main requirement is that there is a commonly known prior on the unknown distribution from which the agents' true observations are sampled, and this prior is known to the principal (who, in our case, is the platform). Truthfulness is achieved by rewarding/scoring an agent's posterior probability distribution of her peer's answer conditioned on her own answer, using a \emph{Proper Scoring Rule} (PSR). PSRs are a well-known class of payment/scoring rules that incentivize truthful elicitation of probabilities of events that can be observed at a future date \citep{brier1950verification,gneiting2007strictly,savage1971elicitation}. This approach is infeasible in platforms since a prior on the distribution of evaluations is typically not a priori available, and even if it is, it may not be common knowledge across all agents and the platform. Moreover, this mechanism is known to induce uninformative equilibria that yield the agents a higher payoff than the truthful equilibrium payoff \citep{jurca2005enforcing}.
 
Another influential design in this domain, the Bayesian Truth Serum (BTS) \citep{prelec2004bayesian}, and its subsequent refinements and generalizations \citep{witkowski2012robust,radanovic2013robust,grant2020two}, preserved the common prior assumption but relaxed the requirement that the principal needs to know this prior. These mechanisms instead require the agents to make extraneous reports about their beliefs in addition to their answers. In particular, they are asked to report a prediction of the empirical distribution of answers reported by other agents. Again, requiring customers on platforms to provide such extraneous information about their beliefs is a tall order given the already low response rates seen for simpler forms of feedback, e.g., ratings. Unfortunately, such extraneous reports of beliefs, although undesirable, are indispensable for incentivizing a single evaluation; it has been shown that it is impossible to design mechanisms that incentivize truthfulness without obtaining some information about the prior distribution \citep{jurca2011incentives} (which is obtained via agents' reports of their beliefs in BTS). 

Mechanisms that do not assume that the principal knows the prior on the distribution of answers have been referred to as \emph{detail-free} in the literature and those that do not require any extraneous reporting from the agents apart from the evaluations are called \emph{minimal}. Ideally, for reputation systems, we need incentive mechanisms that are {\it both} detail-free and minimal, and that do not rely on the assumption of the existence of a commonly known prior across agents. In light of the impossibility result mentioned above, this seems like a challenging task, if not entirely impossible. The earliest known minimal and detail-free incentive mechanism for a single evaluation task was designed by \cite{jurca2008incentives}, in which respondents arrive sequentially and the distributional knowledge of responses is obtained and leveraged by the mechanism in an online fashion; however, this mechanism critically relies on the assumption of binary evaluations. %and it was unclear if extensions to general evaluations are possible.

%We mention here that in many settings where a large population of agents responds to a single evaluation task (e.g., polls, surveys, etc.) it suffices to ensure ``by and large'' truthfulness (as opposed to guaranteeing truthful reports from all agents) to obtain an accurate estimate of the average evaluation across the population for an evaluation task; in such cases, one can potentially design mechanisms that are both detail-free and minimal that obtain such estimates; this was first shown for binary evaluations by \cite{jurca2008incentives}.

This is where a key feature of online platforms can be exploited: they typically host a {\it large number of similar products or services}. For instance, there are thousands of similar restaurants listed on review platforms like Yelp that users rate. Online marketplaces like Amazon or eBay would like to obtain reviews for many existing sellers on these platforms. Online labor platforms like Thumbtack and Handy would like to get performance metrics for thousands of workers and service providers that operate on these platforms.

The presence of multiple similar evaluation tasks hints at an approach for designing detail-free mechanisms that are also minimal: the missing information about the prior can be obtained from consistent statistical estimates of the distribution of agent responses derived from the response data across multiple tasks. \cite{witkowski2013learning} first explored such a possibility in the context of crowdsourcing, for eliciting binary (e.g., yes/no) responses. A potential concern with this approach is that it assumes that the response data is truthfully generated. But it turns out that in these situations, with careful design, truthfulness can become a \emph{self-fulfilling prophecy} -- truthful behavior is an equilibrium in the induced game when the mechanism assumes that these reports are truthful. This is the basic principle underlying the design of \cite{witkowski2013learning} that forms the foundation of our design, resulting in the fact that SRA is both minimal and detail-free. Mechanisms exploiting this principle are commonly referred to as {\it multi-task} peer prediction mechanisms in the literature.

{\bf Structure of SRA.} %SRA is a {\it multi-task} peer-prediction mechanism: the reward to an agent for her answer not only depends on the answer given by her peer, but also on certain statistics of the answers reported by other agents on other similar evaluation tasks \citep{witkowski2013learning, dasgupta2013crowdsourced}. 
SRA  is a {\it multi-task} peer-prediction mechanism that builds upon the structure of \emph{output agreement mechanisms} \citep{von2008designing,von2004labeling}, which are simple and intuitive mechanisms that have been quite popular in crowdsourcing practice, except they suffer from a critical drawback of being susceptible to strategic manipulations. In an output agreement mechanism, two agents answer the same question, and they are both rewarded if their answers match. There are two critical drawbacks of this scheme: (a) truthful behavior may not necessarily be an equilibrium (see Section~\ref{sec:sramain} for an example) and (b) there is always an undesirable equilibrium in the game it induces, in which every person reports the same answer irrespective of their true evaluation. This equilibrium guarantees each person the highest possible payoff rewarded by the mechanism. Our mechanism overcomes these drawbacks by giving proportionately lower rewards for answers that turn out to be more popular on other similar evaluation tasks. This is achieved by inversely scaling the rewards for agreement by a \emph{popularity index} for each answer, thus discouraging blind convergence on a single answer. This is not a new idea: such a biased output agreement scheme, called the Peer Truth Serum (PTS), was first introduced by \cite{jurca2011incentives}, and was further refined by \cite{radanovic2016incentives} and \cite{faltings2017peer}.

The key innovation in our design is in the way these popularity indices are defined. 
All the strong incentive properties of our mechanism trace their origin to this novel definition. 
In our mechanism, these indices are certain second-order population statistics that capture how frequently two people performing the same task agree on a particular answer on average across all tasks. Formally, the popularity index of an answer is the \emph{square root of the estimate of the probability of agreement on that answer} obtained from response data. Thus rare agreements receive higher rewards than agreements that are relatively common. As the number of tasks increases, the accuracy of these indices improves, and truthfulness is obtained as a Bayes-Nash equilibrium when the number of tasks is large enough. A common prior is not necessary for this result; it should just be common knowledge amongst agents that the prior satisfies a certain non-degeneracy property.% (this property is related to the ``stochastic relevance'' condition that frequently appears in this domain; see \cite{miller2005eliciting}).

{\bf Strong truthfulness.} A crucial concern in any reward mechanism is that the induced game may possess multiple equilibria. In such cases, there needs to be an adequate rationale for the truthful equilibrium to be selected. Indeed, the theory of equilibrium selection, i.e., justifying certain equilibria as more likely to arise than others, occupies an important position in game theory; see \cite{harsanyi1988general} and \cite{van2002strategic}, and references therein. It is known that elicitation mechanisms for a single evaluation task with no extraneous reporting (which includes \cite{miller2005eliciting}) possess uninformative equilibria that give a higher expected payoff to each agent than in the truthful equilibrium \citep{jurca2005enforcing}. Moreover, these equilibria involve simple strategies such as every agent reporting the same answer, due to which these mechanisms are particularly vulnerable to uninformative feedback. 
%The ideal property that mechanism designers strive for in this regard is {\it strong truthfulness}: a mechanism is strongly truthful if the truthful equilibrium gives the highest expected payoff to any agent across all equilibria, and in particular, this payoff is strictly higher than that under any equilibrium strategy profile that is not fully informative. \cite{kong2016framework} describe a general information-theoretic framework to achieve strong truthfulness by connecting the loss in the agents' expected payoff relative to the truthful equilibrium to some form of mutual-information loss or correlation loss in the population due to deviation from truthfulness. 
The Bayesian Truth Serum demonstrated that this issue could be overcome by requiring extraneous reports of beliefs; the truthful equilibrium under BTS gives the highest expected payoff to an agent across all equilibria, and in particular, this payoff is strictly higher than that under any equilibrium strategy profile that is not fully informative. Mechanisms that satisfy this property are called {\it strongly truthful} mechanisms in the literature. 
\cite{dasgupta2013crowdsourced} first showed that such strong truthfulness properties could be obtained in the multi-task setting without requiring extraneous reports from the agents. \cite{kong2016framework} describe a general information-theoretic analysis of incentive mechanisms in this space, and show that most mechanisms achieve such properties by (implicitly or explicitly) connecting the loss in the agents' expected payoff relative to the truthful equilibrium to some form of mutual-information loss or correlation loss in the population due to deviation from truthfulness. 

We show that SRA achieves a vanishing uniform upper bound (in the number of tasks) on the difference between the expected payoff obtained by the truthful equilibrium and that obtained under any other symmetric equilibrium (equilibrium in which all players choose the same reporting strategy). As the number of tasks grows, asymptotically, the expected payoff in the limit under a truthful strategy profile is higher than that under any other symmetric strategy profile. Moreover, this payoff is strictly higher than that under any symmetric equilibrium strategy profile that is not {\it fully informative}. A fully informative strategy profile is one where each agent applies a common permutation map to her observation to generate her report (essentially amounting to relabeling the set of responses). In other words, SRA is asymptotically {\it strongly truthful across all symmetric equilibria}. As a dual to this property, under a mild assumption on the strategy spaces, we also show that any symmetric equilibrium that gives the highest expected reward to an agent across all symmetric equilibria must be close, in a well-defined sense, to being fully informative when the number of tasks is large. Such properties hinder the rise of potential ``obviously attractive'' symmetric equilibria such as all agents reporting the same answer for every evaluation. 

The restriction to symmetric equilibrium payoffs in the equilibrium dominance property of SRA may seem undesirable. %, especially in comparison to dominance properties satisfied by existing mechanisms. 
However, this restriction stems from a crucial difference in our setting compared to the settings considered by other mechanisms that are mainly motivated by crowdsourcing applications. In our setting, identifying information for the different entities to be evaluated is available to the agents (captured by the task number in our formal setup). Moreover, allocations of evaluation tasks to the agents are exogenously specified. 
Thus, in our setting, the agents are free to choose reporting strategies that depend on the identities of the tasks they perform to coordinate their reports with other agents, in addition to the (desirable) coordination that can be achieved by reporting their observations truthfully. Due to the possibility of such extraneously achieved coordination, it is well known that it is impossible to elicit truthful evaluations under a payoff-dominant truthful equilibrium in general \citep{gao2016incentivizing}.\footnote{The argument for this impossibility is the following: suppose that there is a mechanism that ensures that the payoffs that agents obtain by truthfully signaling an extraneous feature of an entity being evaluated via their reports are always lower than those obtained by truthfully reporting the feature that they are supposed to report. Then one obtains a contradiction by exchanging the role of the extraneous feature and the feature to be reported. } Other mechanisms get around this difficulty by making certain assumptions that eliminate the possibility of the agents choosing reporting strategies that depend on task identity. Although such assumptions may be justifiable in applications such as crowdsourcing, we do not rely on such assumptions since they are inappropriate in the context of eliciting feedback on platforms; see Section~\ref{sec:symm} for a discussion. In Section~\ref{sec:symm}, we additionally show that if such assumptions are made, then SRA is indeed asymptotically strongly truthful.

In a similar spirit as the framework of \cite{kong2016framework}, these strong truthfulness guarantees are obtained by showing that the expected payoff of an agent under a particular strategy profile under SRA is proportional to a novel notion of a \emph{square root agreement measure} (SRAM) between two independent responses, which we show to be monotonically decreasing in unilateral information loss in the responses. Both the measure and this monotonicity property are new and of independent interest. Moreover, in Section~\ref{sec:related}, we show that a generic adaptation of the framework of \cite{kong2016framework} to the homogeneous response setting along the lines of SRA does not yield truthfulness under any arbitrary mutual information measure.
%Crucially, the SRAM is the first known mutual information measure that induces the strong truthfulness property for objective evaluations (homogeneous responses) while only requiring one evaluation per agent, in stark contrast to other known mutual information measures that require a prohibitively large number of evaluations per agent to induce strong truthfulness, even for objective evaluations. 
SRAM is thus, arguably, the ``right" notion of an agreement measure for objective evaluations. We discuss this aspect in more detail in Section~\ref{sec:related}.

{\bf Robustness.} We finally perform numerical experiments on synthetic as well as real data to test SRA's robustness in incentivizing truthful behavior in finite-data settings featuring deviations from the homogeneous responses assumption. This is practically important since, despite making a faithful effort to obtain and report true observations, agents may have residual biases in their observations and reports. Such biases could also capture mild subjectivity in responding to objective evaluations. Hence, it is desirable that SRA incentivizes each agent to be truthful even when the agent accounts for such biases in other agents. We find that SRA exhibits a high degree of robustness to these practical concerns and generates strong incentives for truthful behavior.

{\bf Reward mechanisms in practice.} Non-monetary rewards for incentivizing informative feedback, e.g., coupons, badges, or some form of a reputation score, are commonly seen in crowdsourced review forums like Yelp, Tripadvisor, etc. A prominent example of monetary incentives is the ``Rebate for Feedback (RFF)'' program that was launched by Taobao.com (one of the world's largest e-commerce websites), on March 1, 2012.\footnote{See \url{https://bit.ly/2GVntzC}, and also \cite{li2016buying}.} In this program, sellers can set a rebate amount in the form of cashback or a store coupon for any items they sell, as a reward for a buyer's feedback after purchasing that item. If a seller opts for RFF, then Taobao ensures that the rebate is transferred from the seller's account to a buyer who leaves high-quality feedback. The feedback quality is determined by a machine learning algorithm depending on attributes like the length of the feedback, whether or not certain key features of the item (e.g., longevity, whether or not it is true to size, etc.) are mentioned, etc. The main contention of the present work is that strategic considerations are paramount in incentivizing {\it informative} feedback. For example, it is easy to give untruthful feedback that appears to be of high quality to a machine learning algorithm; this is especially a concern for objective evaluations with a fixed, finite set of answers. SRA can thus be an effective approach to administer such rebate schemes in a manner that is robust to strategic behavior.

{\bf Organization of the paper.} The remainder of the paper is organized as follows. In Section~\ref{sec:related}, we discuss related mechanisms and their comparisons with SRA. Section~\ref{sec:model} presents a formal description of the model considered in the paper. Section \ref{sec:results} presents the SRA mechanism and its main incentive property. In Section~\ref{sec:select} we address the issue of equilibrium selection. In Section~\ref{sec:numerics}, we perform numerical experiments in a practically motivated experimental setup to test the robustness of SRA in incentivizing truthful behavior to deviations from our main assumptions.
%We discuss some comparisons and relations with existing mechanisms in greater detail in Section~\ref{sec:compare}.  
We finally summarize our contributions and conclude in Section~\ref{sec:conclusion}. 
The proofs of all of our results can be found in the Appendix.

\section{Related literature}\label{sec:related} 
%To appropriately position our contribution, we first present a brief discussion of the main existing approaches in incentive design for the elicitation of unverifiable responses. We then discuss our approach in relation to key comparables.

%\subsection{Background} 
As a minimal, detail-free, multi-task mechanism, the key feature of SRA is that it {\it strongly} incentivizes truthful responses in homogeneous response settings, even among agents who have performed a single evaluation. We now discuss this property in relation to the properties satisfied by other existing multi-task mechanisms. Table~\ref{table:prop} summarizes the differences between SRA and these mechanisms at a high-level.
% Correspondingly, we discuss related mechanisms on three fronts: (1) mechanisms that do not necessarily incentivize truthfulness in the homogeneous responses setting, (2) mechanisms that require more than one evaluation per agent in the homogeneous responses setting, and (3) mechanisms that have a strictly weaker dominance property for their truthful equilibrium compared to SRA in our setting. To the best of our knowledge, these three categories cover the space of all known minimal and detail-free mechanisms for truthful elicitation (with some mechanisms falling in multiple categories), thus establishing the unique position of SRA in the literature. Table~\ref{table:prop} summarizes the differences between SRA and these mechanisms at a high-level.
\begin{table}    \resizebox{\textwidth}{!}{ \begin{tabular}{llll}
    \hline
    Mechanism                      & Incentivizes truthful homogeneous responses  & Incentivizes single evaluations &   Dominance property for truthful equilibrium                       \\\hline
        {\bf SRA}                  & \cmark            & \cmark      & Strongly truthful across symmetric equilibria        \\ 
        Vanilla output agreement & \xmark     & \cmark &None\\
       \cite{witkowski2013learning} &\xmark ~~(needs binary responses) & \cmark & None \\
     \cite{dasgupta2013crowdsourced}       & \xmark~~(needs categorical responses)     &\cmark       & Strongly truthful across symmetric equilibria\\
      Peer Truth Serum for Crowdsourcing (PTSC)  & \xmark~~(needs self-predicting responses)     &\cmark         & Strongly truthful across symmetric equilibria                     \\
    \cite{kong2016framework} & \cmark             & \xmark      &Strongly truthful* (see Section~\ref{sec:kong})\                                      \\
     Correlated Agreement (CA) & \cmark             & \xmark~~($\geq$ 2)      &Informed truthful across symmetric equilibria                                       \\
CA-HR (Appendix Section~D.1  ) & \cmark             & \cmark      &Informed truthful across symmetric equilibria                                    \\
      	Multi-task Peer Prediction Method                   & \cmark              & \cmark      & None                    \\
        \cite{radanovic2015incentives} &\cmark &\cmark &None\\ \hline
    \end{tabular}}
    \caption{Properties of different multi-task peer-prediction mechanisms in the many tasks regime in our setting.} \label{table:prop}
   \end{table}
   
\subsection{Existing strongly truthful multi-task mechanisms} \label{sec:mech}
We first discuss mechanisms that achieve strong truthfulness guarantees, focusing on distinctions from SRA.
\begin{enumerate}[wide, labelwidth=!, labelindent=0pt]

\item {\bf \cite{dasgupta2013crowdsourced}.}\label{sec:dg}  \cite{dasgupta2013crowdsourced} proposed the first known detail-free and minimal multi-task peer-prediction mechanism that is also strongly truthful, assuming that agents do not choose reporting strategies contingent on task identities; in the absence of this assumption, it is strongly truthful across symmetric equilibria. The original paper restricted the setting to binary evaluations; however, it was later shown by \cite{shnayder2016informed} that the mechanism is strongly truthful for non-binary responses under the condition that the responses are ``categorical.'' This condition says that conditional on an agent's answer, the posterior probability of all other agents' answers must reduce relative to the prior. That is, if $Pr(y)$ is the prior probability of an answer $y$ and $Pr(y|y')$ is the conditional probability that some other agent has answer $y$ given that one agent has answer $y'$, then $Pr(y'|y)\leq Pr(y')$ for all $y'\neq y$. Except for the case of binary evaluations, this condition is not satisfied in general under homogeneous responses (see Remark~\ref{notsat} in Section~\ref{sec:numerics}).\\ 

\item {\bf Peer Truth Serum for crowdsourcing (PTSC) \citep{radanovic2016incentives}.}\label{sec:selfpred} PTSC is the multi-task extension of the PTS mechanism, originally defined by \cite{jurca2011incentives} for the case where the prior distribution of responses is known to the principal. Both these mechanisms operate in the homogeneous responses setting. PTS has a biased output agreement structure requiring only one evaluation per agent, where the popularity index of each answer is the prior probability of an agent reporting that answer. In PTSC, this prior probability is replaced by its estimate computed from the response data obtained from a large number of tasks. In order to obtain truthfulness, PTS/PTSC requires that the agent responses satisfy a ``self-prediction'' condition, which says that $Pr(y|y)/Pr(y)\geq Pr(y'|y)/Pr(y')$ for any $y'\neq y$. This is equivalent to saying that $Pr(y|y)\geq Pr(y|y')$ for any $y'\neq y$ We can show that this condition is weaker than the categorical responses condition required by the mechanism of  \cite{dasgupta2013crowdsourced} for incentivizing truthfulness; see Proposition~E.1 in the Appendix. However, except for the case of binary evaluations, this condition is also not satisfied in general under homogeneous responses (see Remark~\ref{notsat}  in our numerical evaluations). If this condition is satisfied, PTSC has been shown to be asymptotically strongly truthful while restricting to symmetric strategy profiles, or in other words, it is strongly truthful across symmetric equilibria.\\

\item {\bf \cite{kong2016framework}.}\label{sec:kong} The underlying principle leading to the strong truthfulness property of SRA is closely related to the mechanism design paradigm of \cite{kong2016framework} (KS), who provide an information-theoretic framework for designing strongly truthful mechanisms for general settings with non-homogeneous responses. Their mechanism operates on a pair of agents, and the payment to each agent is defined to be a scaling of an unbiased estimate of some mutual information measure between the two agents' response distributions, constructed using their reported responses to a common set of evaluation tasks. A variety of mechanisms can be obtained by varying the information measure. Strong truthfulness follows from the fact that the mutual information measure is monotonically decreasing with respect to loss of informativeness in the agents' responses. We note that strong truthfulness here rests on the assumption that agents use the same reporting strategy for all tasks. However, in our setting, since agents are allowed to choose reporting strategies contingent on task identities, this mechanism is strongly truthful only across equilibria in which agents choose a common reporting strategy for all tasks that they perform (which includes symmetric equilibria). 

Until recently, all known mutual information measures required an unboundedly large number of responses per agent (a large fraction of which must be commonly performed by the two agents) to construct unbiased estimates.\footnote{\cite{schoenebeck2020learning} recently proposed a sample-efficient approach to directly learn an appropriate scoring-rule for scoring the agents' reports that implements the mutual-information mechanism of \cite{kong2016framework}, rather than learning the joint distribution and then estimating the mutual information.}  \cite{kong2020dominantly} recently proposed a mutual information measure, of which an unbiased estimator can be constructed using a finite number of per-agent responses. Nevertheless, this construction still requires a number of per-agent responses at least twice the number of possible answers (e.g., $4$ responses for binary evaluations). In Section~D.3 in the Appendix, we show that there can be no mutual information measure satisfying information monotonicity whose unbiased estimate can be constructed from two agents' responses to a single evaluation task, even in the homogeneous, binary responses setting. Thus, the KS mechanism design framework fails to incentivize truthful behavior in a single evaluation even in the homogeneous responses setting.

Via the design of SRA, we effectively show that if one leverages distributional information obtained from multiple tasks in the homogeneous responses setting, then there is a mutual information measure (the square root agreement measure) and a deviation from the KS mechanism design framework that utilizes the learned distributional information along with agent responses to compute payments, such that the mechanism strongly incentivizes truthful behavior in even a single evaluation.

Based on SRA's design, one may wonder if it is possible to obtain a generic adaptation of the KS mechanism to the multi-task, homogeneous responses setting, which utilizes distributional information (obtained from many tasks) to incentivize truthful single responses under {\it any} mutual information measure. In Section~D.4 in the Appendix, we show that this is not true by considering the Shannon mutual information \citep{cover2006elements}: we show that a mechanism along the lines of SRA that leverages the Shannon mutual information instead of the square root agreement measure is not truthful in general for homogenous responses. This underscores the importance of the discovery of SRAM and shows that it is arguably the ``right'' agreement measure for the purpose of strongly incentivizing objective evaluations.
\end{enumerate}

\subsection{Other prominent multi-task mechanisms}
We now discuss multi-task mechanisms that are not known to be strongly truthful in general. 
\begin{enumerate}[wide, labelwidth=!, labelindent=0pt]
\item {\bf \cite{witkowski2013learning}.} The mechanism proposed by \cite{witkowski2013learning} is minimal and requires each agent to perform only one task; however, it requires that the responses are binary, and hence it is not uniformly applicable to the homogeneous responses setting. Additionally, no equilibrium dominance properties are known for this mechanism. \\
\item {\bf Correlated Agreement mechanism (CA) \citep{shnayder2016informed}.} CA is a multi-task mechanism that incentivizes truthful behavior in the general heterogeneous responses setting. CA operates on a pair of agents, and it requires at least two evaluations per agent. As originally described, it assumes that certain information about the joint distribution of the agents' responses is known to the principal. However, this knowledge assumption can be relaxed in the multi-task setting with homogeneous responses, since this distribution can be estimated from the response data obtained from a large number of tasks (appealing to the self-fulfilling prophecy of truthful behavior). Additionally, the requirement of two evaluations per agent can also be relaxed: in Section~D.1 in the Appendix, we describe an adaptation of CA to our setting that only requires one evaluation per agent. We call this mechanism CA for homogenous responses (CA-HR).

In the setting in which CA is originally defined, it is assumed that there is no extraneous identifying information for the evaluation tasks and the task allocation is randomized across agents, in effect eliminating the need to consider task-contingent reporting strategies of the kind allowed in our setting. In this setting, CA satisfies the dominance property of {\it informed truthfulness}, which is weaker than strong truthfulness. Under informed truthfulness, the truthful equilibrium yields the highest equilibrium payoff to each agent, which is higher than an agent's payoff in any equilibrium where her reporting strategy is fully {\it uninformative}, i.e., her reports are independent of her observation. However, in contrast to strong truthfulness, there could be other equilibrium strategy profiles that are not fully informative, which yield the same payoff as the truthful equilibrium. In particular, although the informed truthfulness property precludes fully uninformative equilibria where all observations map to a single response, the CA mechanism remains vulnerable to equilibria where agents map smaller sets of responses to a single response (e.g. if the responses are $\{1,2,3,4\}$ then $\{1,2\}$ map to $1$ and $\{3,4\}$ map to $4$), which a strongly truthful mechanism precludes (i.e., strictly payoff-dominates). Such types of equilibria are payoff-equivalent to the truthful equilibrium under CA if the joint distribution of observations of a pair of agents for a fixed evaluation task is ``clustered,'' as defined in Definition~10 (from \cite{shnayder2016informed}) in the Appendix. In our practically motivated experimental setup in Section~\ref{sec:numerics}, we show that instances with clustered observations are encountered with a very high frequency; see Remark~\ref{rem:clust}.

In our setting, since reporting strategies contingent on task identities are allowed, CA and CA-HR are not informed truthful; we present an example Section~D.2 in the Appendix illustrating this fact. But they can be shown to be (asymptotically) informed truthful {\it across symmetric equilibria}. On the other hand, in Section~\ref{sec:symm}, we show that if strategies contingent on task identities are disallowed in our setting and tasks are randomly allocated across agents, the SRA is (asymptotically) {\it strongly truthful} across all equilibria. Thus, SRA satisfies a stronger equilibrium dominance property compared to CA or CA-HR for homogeneous responses.\\

\item {\bf Multi-task extension of the peer-prediction method \citep{miller2005eliciting}.} The peer-prediction method is a minimal mechanism and incentivizes truthful responses with only one evaluation per agent in the homogeneous responses setting. As originally described, it assumes that the joint distribution of the agents' responses is commonly known to the principal and the agents. However, this assumption can be relaxed in the multi-task setting with homogeneous responses since this distribution can be estimated from the response data obtained from a large number of tasks. As we discussed in Section~\ref{sec:intro}, this mechanism achieves truthfulness by rewarding an agent's predicted probability distribution of her peer's answer, as implied by her own answer, using a Proper Scoring Rule (PSR). 

This mechanism, however, does not satisfy any equilibrium dominance properties: in particular, there exist uninformative equilibria that yield the highest possible payoff to each agent, irrespective of the PSR utilized.\footnote{This observation has already been made for the original non-detail-free version of the peer-prediction method by \citep{jurca2005enforcing}.} For example, for any PSR, the highest possible utility under this mechanism (in the many tasks limit) is achieved when all agents simply report the same answer irrespective of their observation. To see this, note that under such a symmetric strategy, in the many tasks limit, the joint distribution of responses estimated by the principal puts a unit probability mass on the event of the two peer agents reporting the fixed answer. Hence, the conditional distribution on the peer's response implied by an agent's response perfectly predicts the peer's response and thus, achieves the highest score. In contrast, we show that such equilibrium leads to a strictly lower payoff than the truthful equilibrium under SRA when the number of tasks is large enough (see Remark~\ref{rem:dom}).\\

\item {\bf \cite{radanovic2015incentives}.} \cite{radanovic2015incentives} describe a mechanism for the homogeneous responses setting that only utilizes one evaluation per agent. This mechanism is a multi-task extension of the peer-prediction method that relaxes the common prior assumption: they show how an unbiased estimate of the payoff under the peer-prediction method utilizing the quadratic scoring rule (a PSR) can be constructed in the homogeneous responses setting with response data from a random but almost surely finite number of tasks. However, the mechanism inherits the chief concern regarding the peer-prediction method that we discuss above, in that the strategy profile where everyone reports a fixed answer irrespective of their observations results in the highest possible score and thus a higher payoff than the truthful equilibrium.
\end{enumerate}

\section{Model}\label{sec:model}
We now describe the details of our model.

{\bf Operational details.} We consider a setting with $N$ evaluation tasks denoted by the set $\cN$ and labeled as $i = 1,\cdots, N$. Let $\cM$ denote the population of $M$ agents, labeled as $j = 1,\dots, M$. Let $\cM_i\subseteq \cM$ denote the subset of agents that perform task $i$, and let $\cN_j$ be the set of tasks that an agent $j$ performs. We assume that the sets $\cM_i$ and $\cN_j$ are exogenously specified. The set of possible observations and the set of possible answers in each evaluation task is assumed to be the same finite set, denoted as $\cY$. A generic element of $\cY$ will be denoted as $y$.

{\bf Statistical assumptions.} The distribution of observations of agents performing task $i$ is specified by an unobservable type of the entity being evaluated in task $i$, denoted as the random variable $X_i$, which takes values in the finite set $\cX$. A generic element of $\cX$ will be denoted as $x$. This set of possible types $\cX$ is common across all tasks. The distribution of the observations of the agents for any task, as a function of the type $x\in\cX$ of the entity being evaluated in that task, is denoted as $\bp(x) = (p_y(x);y\in\cY)$. The observation of an agent $j$ in $\cM_i$ is denoted as the random variable $Y^i_j$, which is independently drawn from $\bp(X_i)$ for each such agent. This implies that the observations of different agents for a single task $i$ are independent conditioned on $X_i$, but may be dependent otherwise.\footnote{Note that $\{Y^i_j; \,j\in \cM_i\}$ is a set of exchangeable random variables by the virtue of the fact that they are i.i.d. conditioned on the unobservable entity type $X_i$. Instead of explicitly assuming the existence of such a type, we can assume that the set of observations made by any and potentially an infinite number of agents for an entity $i$ are exchangeable random variables, which a property that is expected to hold in practice for objective evaluations. De Finetti's theorem \citep{aldous1985exchangeability} would then imply the existence of a type for each entity such that the observations of the agents are conditionally i.i.d.} Further, we assume that the types of entities being evaluated in the different tasks are independently sampled from a common distribution, $P_X$. We refer to this property as the tasks being {\it statistically similar}.

Finally, we assume that from the perspective of any agent $j$, there are no other observable features of the evaluation task $i$ except the observation $Y^i_j$. %\footnote{The presence of extraneous features in the evaluation tasks can introduce equilibria of the form ``If the seller is based in the United States then report `good,' otherwise, report `bad.'" These type of equilibria trouble most payment mechanisms in this domain; it has been shown that it is impossible to elicit every feature under a payoff-dominant truthful equilibrium \citep{gao2016incentivizing}.} 
The probability distribution over types, $P_X$, and the function $\bp$ together form a \emph{probability generating model} (henceforth referred to as the generating model) of the agent observations, denoted as the pair $(P_X,\bp)$. In particular, this pair fully specifies a joint distribution on the underlying types of the different entities being evaluated and the different agents' observations across tasks. The following example illustrates this model.

\begin{example}\label{ex1}
{\it Consider a situation where a labor platform wants to obtain feedback on punctuality of plumbers that operate on the platform. Suppose that each plumber could be of 2 possible unobservable types $\cX = \{\textrm{Punctual, Not Punctual}\}$, with $P_X(\textrm{Punctual}) = P_X(\textrm{Not Punctual})=0.5$. Each plumber's type is independently sampled from the distribution $P_X$. The question is ``Did the plumber show up within 5 minutes of his/her appointed time?''. The two possible observations/answers are $\cY=\{\textrm{Yes, No}\}$. And the distributions of true answers as a function of type are $\bp(\textrm{Punctual}) = (0.95,0.05)$ and $\bp(\textrm{Not Punctual}) = (0.5,0.5)$.}
\end{example} 

We want to remark that the above example is merely illustrative: the type of plumber need not have any semantic interpretation. It may lie in some abstract space. More importantly, we emphasize that this type is unobservable, both, to the agents and the principal. 

{\bf The payment mechanism.} Our goal is to design a payment mechanism that elicits observations from the agents. For any $j\in\cM_i$, let $r^i_j$ denote agent $j$'s reported answer for task $i$. We define a payment mechanism as follows. 
\begin{definition}
A payment (or reward/scoring) mechanism is a set of functions $\{\tau_j: j\in\mathcal{M}\}$, one for each person in the population, that map the reports $\{r^i_j: j\in \mathcal{M},\,i\in\cN_j\}$ to a real valued payment (or score). 
\end{definition}

{\bf Agents' strategies. }An agent $j$'s strategy is a set of mappings $\{\bq^{ij}: i \in \cN_j\}$ where $\bq^{ij}(y) = (q^{ij}_{y'}(y);\, y'\in\cY)$ is the probability distribution over answers for evaluation task $i\in\cN_j$ conditional on the observation being $y$. We emphasize that agents are allowed to choose different reporting strategies for the different tasks, i.e., their reporting strategy can be task-contingent. We however restrict ourselves to considering task-contingent reporting strategies in which the reported answer of an evaluation only depends on the observation for {\it that} evaluation instead of potentially depending on the observations for all the other evaluations that the agent performs. This restriction is simply for the ease of exposition. All the incentive properties continue to hold for our proposed mechanism even if such strategies are allowed. This is simply because the payment in our mechanism to any agent is additive over the tasks that she performs. Hence, by the expectation operator's additivity, only the marginal distributions of the responses for the different tasks matter in determining the total expected payoff to an agent. Thus, choosing a reporting strategy for a given task that depends on the observations for other tasks is equivalent in terms of expected payoff to choosing the reporting strategy based on the output of some randomization device that produces values that are identically distributed to the observations for these other tasks. Such a reporting strategy is already included in the space of strategies for each agent.

{\bf Equilibrium notion. }We define the following notion of a Bayes-Nash equilibrium \citep{myerson2013game} in the game induced by a payment mechanism.
\begin{definition}[Bayes-Nash Equilibrium] \label{def:bne} Given a generating model $(P_X,\bp)$ that is common knowledge amongst the agents, we say that a strategy profile $\{\bq^{ij}: j\in\cM,\, i \in \cN_j\}$ comprises a Bayes-Nash equilibrium in the game induced by the payment mechanism if for each $j\in \mathcal{M}$,
\begin{align}
 &\textup{E}\bigg[\tau_j \big(\{\bq^{ij'}(Y^i_{j'}): j'\in\cM,\, i \in \cN_{j'}\}\big)\bigg]\nonumber\\
 &~~\geq \textup{E}\bigg[\tau_j\big(\{\bar{\bq}^{ij}(Y^i_j):i\in \mathcal{N}_j\}\cup\{ \bq^{ij'}(Y^i_{j'}): j'\in\cM,\,j'\neq j,\, i \in \cN_{j'}\}\big) \bigg],\label{equilibrium}
\end{align}
for each $\{\bar{\bq}^{ij}:i\in \cN_j\}\neq \{\bq^{ij}:i\in \cN_j\} $, where the expectation is with respect to the joint distribution on the responses of the population specified by the generating model $(P_X, \bp)$. We say that the strategy profile is a strict Bayes-Nash equilibrium if the above inequality is strict.
\end{definition}

In words, this says that assuming all the other agents adhere to the reporting strategy profile $\{\bq^{ij}: j\in\cM,\, i \in \cN_j\}$, each agent maximizes her expected reward by also adhering to the prescriptions of the strategy profile.
Next, we define Bayes-Nash incentive compatibility, which is the property that truthful reporting is a Bayes-Nash equilibrium of the game induced by the reward mechanism. 

\begin{definition}[Bayes-Nash Incentive compatibility]\label{def:bic} We say that a payment mechanism is Bayes-Nash incentive-compatible with respect to the generating model $(P_X,\bp)$ if the truthful reporting strategy profile, i.e., where $q_{y'}^{ij}(y) =\mathbf{1}_{\{y'=y\}}$ for all $j\in\cM$ and $i\in\cN_j$, is a Bayes-Nash equilibrium in the game induced by the mechanism. If this equilibrium is strict, we say that the mechanism is strictly Bayes-Nash incentive compatible. 
\end{definition}

{\bf Informational assumptions.} We make the following informational assumptions.
\begin{enumerate}
\item The principal is {\it not} assumed to know $P_X$ or $\bp$. Hence, $(P_X,\bp)$ is not an input to the payment mechanism.
\item We assume that the structure of the underlying generating model, i.e., the existence of some prior distribution $P_X$ from which the type of any evaluated entity is drawn, and the function $\bp$ that captures the conditional distribution of observations given the type, that is common across entities being evaluated, is common-knowledge across all agents. In particular, this means that all the agents commonly know that all the tasks are statistically similar, and the observations of agents performing each evaluation are statistically homogeneous. Additionally, we will also assume that it is commonly known to all the agents that the generating model $(P_X,\bp)$ satisfies a certain separation property, which we define in Section~\ref{sec:strict} (Definition~\ref{def:well}), where we also discuss simple interpretations of this property. Finally, we assume that the payment mechanism, once fixed by the principal, is publicly announced and is commonly known to all agents.
\end{enumerate}

Note that the definitions of Bayes-Nash equilibrium and Bayes-Nash incentive compatibility (Definitions~\ref{def:bne} and \ref{def:bic}) assume that the generating model is commonly known to the agents. However, we show that SRA is Bayes-Nash incentive-compatible with respect to any commonly known generating model that satisfies the separation property (discussed in Sections~\ref{sec:strict}; Definition~\ref{def:well}). Consequently, it is only necessary for the agents to commonly know that this separation property is satisfied by the generating model to obtain truthful behavior. Similarly, we show that the other properties we discuss concerning the Bayes-Nash equilibria in the game induced by SRA hold for {\it any} commonly known generating model that satisfies the separation property. Hence, to obtain these properties, we only need to assume that it is common knowledge amongst the agents that the generating model satisfies this property.

\section{The square root agreement rule (SRA)}\label{sec:results}
Our main proposal, the square root agreement rule (SRA), is formally defined in Mechanism \ref{tcbsra}. Informally, the mechanism can be described as follows. Consider an agent $j$ who has performed evaluation task $i$. Suppose that she submits an answer $y\in\cY$. Then, she receives payment for this answer only if another agent $j'$, who has performed the same task $i$, and who is chosen to be her peer, also reports the same answer $y$. This payment denoted as $e_j(y)$ is inversely proportional to the square root of the empirical frequency at which arbitrarily chosen agents agree on answer $y$ across all tasks that $j$ has not performed; see Equation \ref{mech}. This empirical frequency is denoted by $\bar{f}_j(y)$ and is computed in Equation \ref{popind}. To ensure that $\bar{f}_j(y)>0$ and the inverse is well defined, we use a smoothed version of empirical frequency, i.e., we add $1$ to the total number of agreements on each answer before dividing by the number of tasks $j$ has not performed. The following example illustrates the mechanism.

\begin{bigdef}[float, label=tcbsra]
The responses of agents for the different evaluation tasks are solicited. Let these be denoted by $\{r^i_j:j\in\cM, i\in\cN_j\}$. An agent $j$'s payment is computed as follows:\\
\begin{itemize}
\item For each population $\cM_{i}$ such that $i\notin \cN_j$, choose any two agents $j_1(i),\,j_2(i)\in\cM_{i}$, and for each possible evaluation $y\in\cY$, compute the quantity 
\begin{align}
\bar{f}_j(y)=\frac{1}{N-|\cN_j|}(1+\sum_{i\in\cN\setminus\cN_j}\mathbf{1}_{\{r^{i}_{j_1(i)}=y\}}\mathbf{1}_{\{r^{i}_{j_2(i)}=y\}}).\label{popind}
\end{align}

\item For each answer $y$, fix a payment $e_j(y)$ defined as

\begin{align}
e_j(y)=\frac{K}{\sqrt{\bar{f}_j(y)}}. \label{mech}
\end{align}
where $K>0$ is any positive constant. $\sqrt{\bar{f}_j(y)}$ is the popularity index of answer $y$.\\

\item For computing agent $j$'s payment for evaluation task $i\in\cN_j$, choose another agent $j'\in\cM_i$, who will be called $j$'s peer for task $i$. If their responses match, i.e., if $r^i_j=r^i_{j'}=y$, then $j$ gets a reward of $e_j(y)$. If the responses do not match, then $j$ gets 0 payment for that task. 
\end{itemize}
\end{bigdef}

\begin{example}[SRA in action] {\it Consider the labor platform presented in Example~\ref{ex1}. During an evaluation period, the platform solicits answers to the question ``Did the plumber arrive within 5 minutes of his/her appointed time?'' from all the customers who have hired a plumber from a set of a priori similar plumbers operating on the platform (e.g., new plumbers who have recently joined the platform) during this period. Suppose a customer, Susan, reports an answer ``Yes,'' meaning that she reports that the plumber that she hired, Tim, did arrive within 5 minutes of his appointed time. Then, Susan gets a reward only if a randomly chosen customer who has also hired Tim in the same period also reports the answer ``Yes.'' The reward is inversely proportional to a popularity index of the answer ``Yes'' across the customer population computed from the response data. SRA defines this popularity index as the square root of the (smoothed) empirical frequency at which two customers who hire the same plumber both report the answer ``Yes'' across all the plumbers that Susan hasn't evaluated. The payment procedure is similar if Susan reports ``No'' instead.}
\end{example}

\begin{remark} 
Note that in SRA, a separate set of popularity indices for the different answers is computed for each agent based on population responses for tasks that this agent hasn't performed. In practice, however, these indices are expected to be almost identical across agents when the total number of tasks is large relative to the number of tasks each agent performs; hence one can potentially calculate a single set of popularity indices of the answers and use them for all agents with negligible impact on incentives. Our theoretical results, however, pertain to SRA as it is formally defined.
\end{remark}

\subsection{The main ideas behind SRA}\label{sec:sramain}
SRA is a {\it biased} output agreement mechanism, i.e., an agent gets paid for her evaluation only if her answer matches the answer reported by her peer agent who has made the same evaluation, where the payment depends on the answer. The simplest description of the core idea of the mechanism is obtained in the {\it hypothetical} scenario where the generating model $(P_X, \bp)$ is known to the principal and is commonly known to the agents (this is the setting considered by \cite{miller2005eliciting}). In this setting, a biased output agreement scheme is defined as follows. 
\begin{enumerate}
\item Each agent $j$ is paired with another randomly chosen peer agent $j'$, and their responses are compared. 
\item If their responses don't match, then $j$ gets no reward. 
\item If their responses match and this common response is $y\in\cY$, then $j$ gets a positive reward $e(y)$. 
\end{enumerate}
The agreement rewards $e(y)$ for $y\in \cY$ are defined as a function of the generating model of responses. The main innovation in SRA is how these agreement rewards are defined. 
%The simplest description of the core idea behind these definitions is obtained in the case of a single evaluation task, in the {\it hypothetical} scenario where the generating model $(P_X, \bp)$ is known to the principal and is commonly known to the agents (this is the setting considered by \cite{miller2005eliciting}). In this scenario, 
To motivate their design, we first discuss why the na\"ive output agreement mechanism fails to incentivize truthful behavior.

{\bf Failure of na\"ive output agreement.} In the na\"ive output agreement mechanism, $e(y)$ is defined to be a constant $K>0$ for all $y\in\cY$. This mechanism tries to exploit an intuitive property that one may na\"ively expect to hold in many scenarios, which is that the peer agent $j'$ has the highest conditional likelihood of observing the {\it same} answer as that observed by an agent $j$. However, this property is not always true. For instance, it is violated if, irrespective of the observation made by an agent, one ``popular'' answer has an overwhelming conditional likelihood of being observed by the peer agent.  This feature can be observed in Example~\ref{ex1}.

%We provide an example to show that this mechanism does not yield a truthful equilibrium in general for homogeneous responses. %We note that this observation is not new; we include it for completeness. We also note that this fact also follows from Proposition~\ref{prop:KS} below.
\begin{example}\label{ex:oa}
{\it Consider the setting in Example~\ref{ex1}. The question is ``Did the plumber show up within 5 minutes of his/her appointed time?'' with the possible observations/answers being $\cY=\{\textrm{Yes, No}\}$. Suppose an agent observes that the plumber did not arrive within 5 minutes of her appointed time, i.e., her observation was ``No.'' Then the conditional probability of the peer agent, assumed to be truthful, replying ``Yes'' can be computed to be $0.54\overline{09}$, which is higher than the conditional probability of her replying ``No,'' which can be computed to be $0.459\overline{09}$. Hence, replying ``Yes,'' i.e., lying, results in a higher expected payoff than being truthful and replying ``No.'' }
\end{example}

In the example above, plumbers are a priori overwhelmingly likely to turn up on time, to the extent that even if an agent observes that a plumber was late, she will still find it more likely that the same plumber will be observed to be on time by her peer agent. Thus, assuming that the peer agent is truthful, it is better to lie and say that the plumber was on time. Summarily, truthful behavior is not an equilibrium in this case.

To overcome this shortcoming of the na\"ive output agreement scheme, the key obstacle that one must tackle is this tendency of regressing to the conditionally most popular answer. Formally, if the observation of an agent $j$ is $Y_j = y$ for some $y\in \cY$, her tendency is to report $\arg\max_{y'\in \cY} P(Y_{j'} = y'\mid Y_j = y)$ so as to maximize the probability of agreement. As we saw in the example above, this optimal answer need not necessarily be $y$, i.e., the inequality 
\begin{align}
P(Y_{j'} = y\mid Y_j = y) \geq P(Y_{j'} = y'\mid Y_j = y), \label{eq:oa}
\end{align}
doesn't necessarily hold for all $y,\,y'\in\cY$, even in the binary responses setting where $|\cY| = 2$. Biased output agreement schemes can tackle this issue by scaling the rewards for agreement depending on the answer, essentially lowering rewards for answers that are expected to be more (conditionally) likely and increasing rewards for answers that are less (conditionally) likely. The intuition is that if an agent observes an answer that is less likely to also be observed by her peer, she is still incentivized to report that answer since the matching reward on that answer is higher. Conversely, if she observes an answer that is more likely to also be observed by her peer, she prefers to report that answer despite the low reward for agreement relative to other answers since the probability of agreement is higher. The challenge is to design the reward scalings for different answers so that these incentives for truthful reporting for an agent are satisfied for {\it each} answer (assuming every other agent is truthful). 

A straightforward way of making an agent indifferent between different reports (and hence weakly incentivize truthful behavior) is by defining the agreement reward for answer $y'\in\cY$ to be proportional to $1/P(Y_{j'} = y'\mid Y_j = y)$, where $y$ is the answer observed by the agent. This approach is rendered infeasible by observing that $y$ is not known to the mechanism --  indeed, the entire exercise is meant for the purpose of eliciting $y$. With this observation in perspective, we discuss two approaches of defining the rewards before we present our approach in SRA. 

{\bf Approach 1: Peer Truth Serum.} The PTS mechanism  \citep{jurca2011incentives, faltings2017peer} scales the agreement reward for answer $y'\in\cY$ by $1/P(Y_{j'} = y')$, i.e., reward for agreement on an answer is inversely proportional to the probability that an evaluating agent makes that observation. %In effect, it uses the proxy $1/P(Y_{j'} = y')$ for the $1/P(Y_{j'} = y'\mid Y_j = y)$ scaling that depends on the agent's true answer $y$. 
We thus obtain truthful behavior from agent $j$ under PTS if for all $y\in\cY$ and $y' \neq y$,
\begin{align}
\frac{P(Y_{j'}=y\mid Y_j = y)}{P(Y_{j'} = y)}&\geq \frac{P(Y_{j'}=y'\mid Y_j = y)}{P(Y_{j'} = y')}, \textrm{ i.e., if,}\\\nonumber\\
P(Y_{j'}=y\mid Y_j = y)&\geq P(Y_{j'}=y\mid Y_j = y').\label{eq:selfp}
\end{align}
In the literature, this is referred to as the `self-predicting responses' condition (that we discussed earlier in Section~\ref{sec:related}), which is not satisfied in general for homogeneous responses, except when $|\cY| = 2$. Intuitively, defining the popularity index of an answer to be equal to the {\it marginal} probability of a single agent making that observation does not capture the fact that the agent evaluates the {\it conditional} probabilities of agreement for the different answers, which depend on the {\it joint probabilities} of a pair of agents making various observations.
%:  recall that in the hypothetical scaling that weakly incentivizes truthfulness, we have
%$$\frac{1}{P(Y_{j'} = y'\mid Y_j = y)} = \frac{P(Y_{j'} = y)}{P(Y_{j'} = y',\,Y_j = y)},$$ 
%which explicitly incorporates the joint observation probabilities in the denominator. 

{\bf Approach 2: A ``conditional'' peer truth serum (CPTS).} With the goal of incorporating the conditional distribution of the observations in the definition of the agreement rewards, another proxy for the ideal scaling can be defined to be $1/P(Y_{j'} = y'\mid Y_j = y')$, i.e., $e(y')$ is defined to be proportional to $1/P(Y_{j'} = y'\mid Y_j = y')$. In this case, we obtain truthful behavior from agent $j$ if for all $y\in\cY$ and $y' \neq y$,
\begin{align}
\frac{P(Y_{j'}=y\mid Y_j = y)}{P(Y_{j'}=y\mid Y_j = y)} = 1 &\geq \frac{P(Y_{j'}=y'\mid Y_j = y)}{P(Y_{j'}=y'\mid Y_j = y')}, \textrm{ i.e., if,}\\\nonumber\\
P(Y_{j'}=y'\mid Y_j = y')&\geq P(Y_{j'}=y'\mid Y_j = y).
\end{align}
This condition is exactly the self-prediction condition from Equation~\ref{eq:selfp}. Thus, the conditions that are necessary for the PTS and CPTS mechanisms to induce truthful behavior are the same. This shows that a na\"ive incorporation of conditional probabilities in the reward scalings need not provide any advantage over PTS, at least in terms of the conditions required for truthfulness.

{\bf SRA's approach.} SRA defines the rewards for agreement as $e(y') = K/\sqrt{P(Y_j = Y_{j'} =y')}$ for each $y'\in \cY$, for some $K>0$. To put it simply, \emph{the reward for an agreement is inversely proportional to the square root of the probability of that agreement}. Thus a more probable agreement earns a lower reward than an agreement that is relatively less probable. To see the relation to PTS and CPTS, observe that SRA simply replaces the scaling $1/P(Y_{j'} = y')$ of PTS and the scaling $1/P(Y_{j'} = y'\mid Y_j = y')$ of CPTS by the product of the square root of the two scalings $1/\sqrt{P(Y_{j'} = y')}\times 1/\sqrt{P(Y_{j'} = y'\mid Y_j = y')} = 1/\sqrt{P(Y_j = Y_{j'} =y')}$. By making this change, SRA explicitly incorporates the joint probabilities of agreements in the rewards. 
%The relation to the CPTS scheme above is similar. Note that the reward scaling under CPTS can be written as $e(y') = P(Y_j = y')/P(Y_j = Y_{j'} =y')$. SRA replaces the numerator $P(Y_j = y')$ in this definition by the quantity $\sqrt{P(Y_j = Y_{j'} =y')}$. 

It turns out that the scaling under SRA successfully overcomes the shortcoming of the na\"ive output agreement scheme in homogeneous responses settings. In particular, truthful behavior is a Bayes-Nash equilibrium in this mechanism. To see this, consider an agent $j$, and suppose that all other agents are truthful. Then if $j$'s true response is $y$, her expected reward for a truthful report is,
\begin{equation}
K\frac{P(Y_{j'} = y\mid Y_j = y)}{\sqrt{P(Y_j = Y_{j'} =y)}} = K\frac{\sqrt{P(Y_{j'} =Y_j = y)}}{P(Y_j =y)}.\label{eq:truthful}
\end{equation}
Similarly, her reward for any other report $y'$ is, 
\begin{equation}
K\frac{P(Y_{j'} = y'\mid Y_j = y)}{\sqrt{P(Y_j = Y_{j'} =y')}} = K\frac{P(Y_{j'} =y',\,Y_j = y)}{P(Y_j =y)\sqrt{P(Y_j = Y_{j'} =y')}}.\label{eq:notruthful}
\end{equation}
Thus being truthful gives a higher reward if the quantity in Equation~\ref{eq:truthful} is larger than the quantity in Equation~\ref{eq:notruthful}, which is, if,
\begin{align}\sqrt{P(Y_{j'} =Y_j = y)}\sqrt{P(Y_j = Y_{j'} =y')} \geq P(Y_{j'} =y',\,Y_j = y).\label{eq:csprop}\end{align}
This inequality resembles the well-known Cauchy-Schwarz inequality that relates second-order moments of two random variables $X$ and $Y$ as, $\textup{E}[XY]\leq \sqrt{\textup{E}(X^2)\textup{E}(Y^2)}$. Hence, if the joint distribution of responses of the two agents satisfies this inequality for each $y,\,y'\in \cY$, we say that the distribution satisfies the ``Cauchy-Schwarz'' (CS) property. And in this case, truthful equilibrium is a Bayes-Nash equilibrium under SRA.

Now the key interesting fact is that this property is {\it always} satisfied for homogeneous responses, i.e., for objective evaluations. To see this, the inequality in Equation~\ref{eq:csprop} can be expressed as follows in the homogeneous responses setting.
\begin{equation}
\sqrt{\sum_{x\in\cX}P_X(x)p_y(x)^2}\sqrt{\sum_{x\in\cX}P_X(x)p_{y'}(x)^2} \geq \sum_{x\in\cX}P_X(x)p_y(x)p_{y'}(x).\label{eq:CS}
\end{equation}
But this is precisely the Cauchy-Schwarz inequality that always holds. Hence, truthful behavior is incentivized as a Bayes-Nash equilibrium under SRA. For truthful behavior to be a {\it strict} Bayes-Nash equilibrium, we need the above inequality to be strict. We show that this requirement is satisfied if the generating model is `separated': a condition we discuss in Section~\ref{sec:strict}.

Now implementing the mechanism above in our setting is infeasible since the principal does not know the generating model $(P_X, \bp)$. Moreover, the generating model is not assumed to be commonly known to the agents. SRA addresses these issues by replacing the required agreement probabilities with consistent statistical estimates computed from reports obtained across multiple tasks. Observe that if everyone except agent $j$ is truthful, then in the definition of SRA, $\textup{E}[\bar{f}_j(y)] = P(Y^i_{j_1(i)} = Y^i_{j_2(i)}=y)+ 1/(N-|\cN_j|)$. In fact, as $N$ grows large, assuming $|\cN_j|$ remains bounded, $\bar{f}_j(y)$ almost surely converges to $P(Y^i_{j_1(i)} = Y^i_{j_2(i)}=y)$ by the strong law of large numbers, i.e., $\bar{f}_j(y)$ is an asymptotically consistent estimate of $P(Y^i_{j_1(i)} = Y^i_{j_2(i)}=y)$. For a large enough $N$, we can show that the estimate's quality is sufficiently high to ensure that truthfulness is recovered as a strict Bayes-Nash equilibrium for any `separated' generating model (see our main result in Section~\ref{sec:mainresult}). Thus, the agents only need to commonly know the generating model's structure and that it is separated to obtain truthful behavior. 

\subsection{Obtaining strictness}\label{sec:strict}
An important goal for any reward mechanism is to \emph{strictly} incentivize truthfulness, i.e., in the truthful equilibrium, each agent gets a strictly higher reward by being truthful than by adopting any other strategy. Without this property, trivial mechanisms like the one that gives a fixed payment to each agent regardless of her report, in principle, weakly incentivize truthfulness. For truthfulness to be a strict equilibrium under SRA, we need the Cauchy-Schwarz inequality in Equation \ref{eq:CS} to be strict for every pair $y,\, y'\in\cY$. It will be useful to define the following notion of the ``inequality gap.''

\begin{definition}[Cauchy-Schwarz inequality gap]\label{def:hom} For a generating model $(P_X,\bp)$ defined on $\cX$ and $\cY$, define 
$$\delta(P_X,\bp) = \min_{y,\,y'\in\cY,\, y \neq y'} \sqrt{(\sum_{x\in \cX}P_X(x)p_y(x)^2)(\sum_{x\in\cX}P_X(x)p_{y'}(x)^2)} - \sum_{x\in\cX}P_X(x)p_y(x)p_{y'}(x).$$
\end{definition}
\label{eq:myst}

By the Cauchy-Schwarz inequality, $\delta(P_X, \bp)\geq 0$. If $\delta(P_X, \bp)> 0$ for some generating model $(P_X,\bp)$, then truthfulness is a strict Nash equilibrium in the game induced by the mechanism we described earlier for the case where the principal knows this generating model. Consider the following definition, which will be useful for our forthcoming discussion.
\begin{definition}[Separation]\label{def:well}
We say that a generating model $(P_X,\bp)$ is separated if $\delta(P_X,\bp)>0$. We say that it is $\alpha$-separated for any $\alpha> 0$ if $\delta(P_X,\bp)\geq \alpha$.
\end{definition}

To understand whether separation is a reasonable assumption on the generating model, a little demystification of this condition is in order. 
For any answer $y\in \cY$, define the vector
\begin{equation}
\bv(y)\triangleq \big(\sqrt{P_X(x)}p_y(x);\,x\in\cX\big).
\end{equation}
Then the Cauchy-Schwarz inequality says that for any two answers $y$ and $y'$, the magnitude (in the Euclidean norm) of the projection of the vector $\bv(y)$ on the unit vector in the direction $\bv(y')$ is less than the magnitude of the vector $\bv(y)$ itself (one can reverse the roles of $y$ and $y'$), i.e., 
$$\frac{|\bv(y).\bv(y')|}{\|\bv(y)\|}\leq \|\bv(y')\|,$$
or 
\begin{equation}
|\bv(y).\bv(y')|\leq \|\bv(y)\|\|\bv(y')\|.\label{eq:vecCS}
\end{equation}
Let $\theta(\bu,\bv)$ denote the angle in radians between two vectors $\bu$ and $\bv$, defined as 
\begin{equation}
\theta(\bu,\bv)\triangleq\arccos\frac{\bu.\bv}{\|\bu\|\|\bv\|},
\end{equation}
when both $\bu$ and $\bv$ are non-zero and as $0$ when either of them is a zero vector.
We can then show that the inequality in Equation \ref{eq:vecCS} is strict if and only if the angle between the vectors $\bv(y)$ and $\bv(y')$ is positive. The following proposition gives a precise statement.
%In fact, under the condition that $\|\bv(y)\|\leq 1$ for all $y\in\cY$, which holds in our case, we can show that the gap in Equation \ref{eq:vecCS} is bounded away from $0$ if and only if the angle between the vectors $\bv(y)$ and $\bv(y')$ is bounded away from $0$. The following proposition gives a precise statement.
\begin{prop}\label{prop:equi} For a generating model $(P_X,\bp)$ defined on $\cX$ and $\cY$, the following two conditions are equivalent.
\begin{enumerate}
\item There is some $\alpha>0$ such that $(P_X,\bp)$ is $\alpha$-separated.
\item There is some $\gamma>0$ such that $\theta(\bv(y),\bv(y'))\geq \gamma$ for all $y,\,y'\in \cY$ such that $y\neq y'$. 
\end{enumerate}
\end{prop}
%Condition $2(a)$ says that an agent's probability of forming any response $y\in\cY$ is bounded away from zero. This condition is naturally satisfied in practice. If this doesn't hold for some answer, one can eliminate that answer from the admissible set.
Thus, separation is equivalent to assuming that the angle between $\bv(y)$ and $\bv(y')$ is positive for any $y\neq y'$. If this is not true for some $y$ and $y'$, then there is a $C\in\mathbb{R}$ such that $p_y(x)=Cp_{y'}(x)$ for each $x\in \cX$ such that $P_X(x)>0$. But in this case, the responses $y$ and $y'$ need not be distinguished at all, since they contain the same information about $X$, and hence about the rest of the random quantities. In particular, $P(X_i=x|Y^i_j=y)=P(X_i=x|Y^i_j=y')$ for each $x\in \cX$. Hence, the principal can simply ask the agents to map both these responses to a single response. 

In the context of our model, separation is also equivalent to the stochastic relevance condition that is imposed to obtain strictness in several works in this domain, starting from \cite{miller2005eliciting}.
An agent's answer to a question is a stochastically relevant random variable if no two answers induce the same conditional distribution on the answers of some other agent who has answered the same question. Clearly, if $\theta(\bv(y),\bv(y'))=0$, then stochastic relevance is violated, and thus stochastic relevance implies that $\theta(\bv(y),\bv(y'))>0$, which is equivalent to separation by Proposition~\ref{prop:equi}. Showing that separation implies stochastic relevance is less straightforward and we show it in Proposition~E.2 in the Appendix.

\subsection{Main result}\label{sec:mainresult}
The following result presents the main incentive property satisfied by SRA.
\begin{theorem}\label{thm:main}
Consider an $\alpha$-separated generating model $(P_X,\bp)$ that is commonly known to the agents. Further, suppose that $\cN_j\leq n$ for all $j\in\cM$ and $|\cM_i|\geq 2$ for all $i\in\cN$. Then for any $\omega \in (0,\alpha K (|\cY|-1))$, there exists a positive integer $N_0$ that depends only on $\omega$, $\alpha$, $|\cY|$, $n$, and $K$ such that if the number of evaluation tasks $N>N_0$, then
\begin{itemize}
\item SRA is strictly Bayes-Nash incentive compatible with respect to $(P_X,\bp)$, and,
\item At the truthful Bayes-Nash equilibrium, the expected payoff to an agent under the truthful strategy is at least $\omega$ higher than the expected payoff under any reporting strategy where the agent's reported response is independent of her true response.
\end{itemize}
\end{theorem} 
Note that the bounds we derive in the proof of this result can be used to explicitly calculate an $N_0$ as a function of $\omega$, $\alpha$, $|\cY|$, $n$, and $K$.
Also note that, although the theorem assumes that the generating model is commonly known to the agents, the mentioned properties hold for any $\alpha$-separated generating model. In particular, the dependence of $N_0$ on the generating model is only through $\alpha$. Hence, this result implies that even if only the fact that $(P_X,\bp)$ is $\alpha$-separated is common knowledge amongst agents, then irrespective of their individual beliefs about the specifics of the generating model, truthful reporting is strictly incentivized in the game induced by the mechanism for a large enough $N$. 

The second claim in the theorem is crucial too: it says that in the truthful equilibrium, the difference in the payoffs to an agent under the truthful strategy and under any strategy in which an agent's reported response is independent of her true response, is bounded away from zero. Such strict incentives allow the principal to account for any costs that the agents may incur for their evaluation effort by appropriately scaling the mechanism's rewards. Note that it is not possible to ensure that the difference between the expected payoff under truthful behavior and that under {\it any} other strategy is bounded away from zero, since one can choose a randomized strategy that chooses a non-truthful response with an arbitrarily small probability. However, our goal here is to deter agents who report an arbitrary answer without investing effort into making an observation. Our result ensures that any perceived cost for such effort can be absorbed in the difference in the payoff under truthful reporting and under any reporting strategy that ignores the observation, by appropriately scaling the rewards.\footnote{\cite{liu2016learning} show how to learn such a scaling when there is heterogeneity in the cost for effort in classical output agreement mechanisms. We believe a similar approach can be adopted in our case. }

Finally, we note that although the common knowledge assumption above is necessary to obtain the theoretical properties of our mechanism, our numerical experiments in Section~\ref{sec:numerics} test SRA under more practical considerations.

\section{Equilibrium selection}\label{sec:select}
In this section, we address the issue of multiplicity of equilibria in the game induced by SRA. First, observe that if truthful behavior is an equilibrium, then so is any \emph{symmetric fully informative strategy profile} in which all agents apply a common permutation map to the responses they receive. And all such equilibria are payoff-equivalent. But the significantly higher degree of coordination needed for the agents to play a fully informative equilibrium other than truthful behavior makes it unlikely that such equilibria will emerge in practice. Thus full informativeness shall be our benchmark as we focus on other equilibria that may emerge. 

The equilibria that give high expected payoffs are arguably the most attractive for the agents and thus can be assumed to have an increased likelihood of being chosen. In what follows, we show that for a large $N$, the truthful equilibrium is approximately payoff-optimal across all symmetric equilibria, with an approximation error that vanishes in $N$. In the limit, any symmetric fully informative strategy profile gives a strictly higher expected payoff to any agent than \emph{any} other symmetric strategy profile. We also show a weak dual to this result: under a certain assumption on the strategy spaces, any symmetric equilibrium that results in the highest expected payoff to an agent across all symmetric equilibria cannot be too ``uninformative'' when $N$ is large, where ``uninformativeness'' is a precise notion that we define.

\subsection{Truthfulness vs. symmetric equilibria as $N\rightarrow\infty$}
Before we discuss the result for a large but finite $N$, let us first discuss the result in the limiting case as $N\rightarrow\infty$, which is easier to obtain, and sheds light on the core idea. Consider a symmetric strategy profile in which every agent adopts a reporting strategy $\bq$, where $\bq(y) = (q_{y'}(y);\, y'\in\cY)$ is the distribution over the reported response conditional on the true response. Let us denote the reported responses under this strategy by the random variables $\{Z^i_j;\, i= 1,\cdots, N,\, j\in \cM_i\}$. Under the truthful strategy profile (or equivalently, any symmetric fully informative strategy profile), in the limit as $N\rightarrow\infty$, the expected reward of each agent performing task $i$ converges to (see Equation~\ref{eq:truthful}),
\begin{equation}
\sum_{y\in\cY}P(Y^i_j =y) K\frac{\sqrt{P(Y^i_{j'} =Y^i_j = y)}}{P(Y^i_j =y)}= K\sum_{y\in\cY}\sqrt{P(Y^i_{j'} =Y^i_j = y)} = K \sum_{y\in \cY}\sqrt{\sum_{x\in\cX}P_X(x)p_y(x)^2}. \label{eq:truthexp}
\end{equation}
Whereas, under any other symmetric strategy profile, the expected reward of each agent converges to,
\begin{equation}
K\sum_{y\in\cY}\sqrt{P(Z^i_{j'} =Z^i_j = y)} = K \sum_{y\in \cY}\sqrt{\sum_{x\in\cX}P_X(x)(\sum_{y'\in\cY}p_{y'}(x)q_y(y'))^2}. \label{eq:symexp}
\end{equation}
It turns out that the quantity in Equation \ref{eq:symexp} is, in general, lower than the quantity in Equation \ref{eq:truthexp}. How much lower depends on the `uninformativeness' of the strategy $\bq$: the more uninformative the strategy $\bq$, the higher is the difference. We will describe this phenomenon more generally since we believe it has applications beyond this work (see Appendix~B.1 for a discussion). We first define the following notion of a {\it square root agreement measure.}

\begin{definition}[Square root agreement measure (SRAM)]\label{def:agree}
Consider a generating model $(P_X,\bp)$ defined over $\cX$ and $\cY$, and consider two random responses $Y_1$ and $Y_2$ drawn from this model. Then the square root agreement measure between $Y_1$ and $Y_2$ is defined as  
\begin{equation*}
\Gamma(Y_1,Y_2) = \sum_{y\in \cY}\sqrt{P(Y_1=Y_2=y)} = \sum_{y\in \cY} \sqrt{\sum_{x\in \cX}P_X(x)p_y(x)^2}
\end{equation*}
\end{definition}

From the definition of SRAM, it is clear that under any symmetric strategy profile in which every agent adopts a reporting strategy $\bq$, the expected payoff to each agent in the limit as $N\rightarrow\infty$ is $K$ times the SRAM between the reported responses (see Equation~\ref{eq:symexp}). Some important properties of the SRAM are presented in Appendix~B.

Next, we define a new notion of {\it uninformativeness} of a reporting strategy. Informally, a reporting strategy is more uninformative if it frequently maps multiple true responses to a single reported response, the extreme case being when a report is chosen independently of the true response. The following definition formalizes this notion. 
\begin{definition}[An uninformativeness measure]
The uninformativeness of a reporting strategy $\bq$ is defined as
\begin{equation}
\Omega(\bq) = \frac{1}{|\cY|(|\cY|-1)}\sum_{y\in \cY}\sum_{y'\in \cY,\,y''\in\cY;\,y'\neq y''}\sqrt{q_y(y')q_y(y'')}.
\end{equation} 
We say that a strategy $\bq$ is $\omega-$uninformative if $\Omega(\bq)\geq \omega$.
\end{definition}
Certain important properties of the uninformativeness measure are presented in Appendix~C. In particular, $\Omega(\bq)=0$ if and only if $(\bq(y);\, y\in\cY)$ have disjoint supports across all $y\in\cY$, i.e., if and only if $\bq$ is fully informative, and $\Omega(\bq)$ attains its highest value of $1$, if and only if $\bq(y) = \bq(y')$ for any $y\neq y'$, i.e., if the report is chosen independently of the true answer. 

We finally present the following information monotonicity property, which is key to our results.
\begin{prop}[A monotonicity property]\label{prop:monotone}
Consider a generating model $(P_X,\bp)$ defined over $\cX$ and $\cY$, and consider two random responses $Y_1$ and $Y_2$ drawn from this model. Also, consider two random responses $Z_1$ and $Z_2$ obtained by applying a reporting strategy $\bq$ independently to $Y_1$ and $Y_2$ respectively. Then, 
\begin{equation}
\Gamma(Z_1,Z_2)\leq \Gamma(Y_1,Y_2) - \frac{\delta(P_X,\bp)\Omega(\bq)^2(|\cY|-1)}{2\sqrt{|\cY|}}. \label{eq:monotone}
\end{equation}
\end{prop}
To see how this property helps us, recall from Equations~\ref{eq:truthexp} and \ref{eq:symexp} that the expected payoff under any symmetric fully informative strategy profile is $K\Gamma(Y_1,Y_2)$, and that under any other symmetric strategy profile $\bq$ is $K\Gamma(Z_1,Z_2)$, where $Z_1$ and $Z_2$ is obtained by applying a reporting strategy $\bq$ independently to $Y_1$ and $Y_2$. The proposition implies that if $\delta(P_X,\bp)>0$, then $\Gamma(Y_1,Y_2) = \Gamma(Z_1,Z_2)$ \emph{only if} $\Omega(\bq)=0$, i.e., only if $\bq$ is fully informative. Thus we conclude that if $\delta(P_X,\bp)>0$, then in the limit as $N\rightarrow\infty$, any fully informative strategy profile gives a strictly higher payoff than any other symmetric strategy profile that is not fully informative. In other words, SRA is asymptotically strongly truthful across symmetric equilibria. In the next section, we use Proposition~\ref{prop:monotone} to address the case where $N$ is large but finite.

\subsection{Equilibrium selection in the finite $N$ regime}
Now we turn to the finite $N$ setting. In this case, the expected payoffs under the fully informative strategy and any other symmetric strategy will not have converged to $\Gamma(Y_1,Y_2)$ and $\Gamma(Z_1,Z_2)$ respectively. But for any fixed $N$, one can obtain concentration bounds on how far the expected payoffs will be from these target values. This, in turn, gives us vanishing bounds on how much lower the payoff under the truthful equilibrium could be compared to any other symmetric equilibrium. 

\begin{theorem}\label{thm:sym}
Consider an $\alpha$-separated generating model $(P_X, \bp)$ that is commonly known to the agents. Further, suppose that $\cN_j\leq n$ for all $j\in\cM$ and $|\cM_i|\geq 2$ for all $i\in\cN$. Then for any $\omega>0$, there is a positive integer $N_0$ that depends on $\alpha$, $\omega$, $n$, $K$, and $|\cY|$, such that for any $N>N_0$, under SRA,
\begin{enumerate}
\item Any symmetric fully informative strategy profile is a strict Bayes-Nash equilibrium, and,
\item Any other symmetric Bayes-Nash equilibrium strategy profile gives an expected payoff at most $\omega$ higher than any symmetric fully informative strategy profile.
\end{enumerate} 
\end{theorem}
Note once again that the bounds we derive in the proof of this result can be used to explicitly calculate an $N_0$ as a function of $\alpha$, $\omega$, $n$, $K$, and $|\cY|$. Also, note that similar to Theorem~\ref{thm:main}, although Theorem \ref{thm:sym} assumes that the generating model is commonly known to the agents, the result holds for any $\alpha$-separated generating model (since the dependence of $N_0$ on $(P_X,\bp)$ is only through $\alpha$). Hence, we conclude that these properties presented in Theorem \ref{thm:sym} hold even in the game where it is only commonly known to the agents that the generating model is $\alpha$-separated.

Finally, note that in the statement of Theorem \ref{thm:sym}, $N>N_0$ suffices to ensure that the truthful equilibrium gives a payoff at most $\omega$ lower than that under \emph{any} symmetric equilibrium strategy profile. This is significant since, for a large but fixed $N$, it is not possible to obtain uniformly vanishing concentration bounds on $\textup{E}(e_j(y))$ (which involves an inverse) across all symmetric reporting strategies. This is because there could be a symmetric strategy profile for which the probability of agreement for an answer $y\in\cY$ gets arbitrarily close to $0$. To overcome this issue, we utilize the fact that under a mixed equilibrium, since the problem of computing the best response is a linear optimization problem, a fixed agent is indifferent between multiple deterministic reporting strategies. This allows us to choose a best-response strategy for a single agent in a way that ensures that the probability of agreement on any answer $y$ is bounded away from zero while ensuring that the expected payoff is same as that under the given symmetric equilibrium. 

Can we say anything about the informativeness of the symmetric equilibrium profile that gives the highest expected payoff across all symmetric equilibria? Intuitively, bounds on $\textup{E}(e_j(y))$ for a large $N$, coupled with the ``inequality gap'' characterized in Proposition~\ref{prop:monotone} should result in an upper bound on the uninformativeness of any symmetric reporting strategy that gives a higher expected payoff than a fully informative strategy. It turns out that in doing so, the same difficulty that we described earlier arises in obtaining the requisite concentration bounds, where the symmetric strategy profile could lead to probabilities of agreement arbitrarily close to $0$. In this case, we cannot use the trick we used earlier, and instead, we show the following result. 

\begin{theorem}\label{thm:sym2}
Consider an $\alpha$-separated generating model $(P_X, \bp)$. Further, suppose that $\cN_j\leq n$ for all $j\in\cM$ and $|\cM_i|\geq 2$ for all $i\in\cN$. Then for any $\omega>0$ and $\eta>0$, there is a positive integer $N_0$ that depends on $\omega$, $\alpha$, $\eta$, $n$, $K$, and $|\cY|$, such that for any $N>N_0$, under SRA, any symmetric strategy profile in which the probability of reporting any answer $y\in \cY$ is either $0$ or at least $\eta$, and that gives a higher expected payoff to an agent than the truthful strategy profile, is at most $\omega-$uninformative. 
\end{theorem}

\begin{remark} \label{rem:dom}Theorem~\ref{thm:sym2} implies that for a large enough $N$, the truthful equilibrium yields a strictly higher payoff to each agent than any equilibrium where all agents report a fixed answer irrespective of the observation. To see this, note that the latter strategy profile is 1-uninformative (see properties of the uninformativeness measure presented in Appendix~C); hence, we can choose $\omega \in (0,1)$. Moreover, $\eta$ can be chosen to be any number in $(0,1)$ since mapping all observations to a single response implies that the probability of any response is either $0$ or $1$. Choosing such $\eta$ and $\omega$, we see that the above property will hold for any $N$ larger than the corresponding $N_0$. This property is in contrast to the multi-task extension of the peer-prediction method \citep{miller2005eliciting} and the mechanism of \cite{radanovic2015incentives}, in which each agent reporting a fixed answer is an equilibrium that yields the highest possible payoff. A similar argument shows that for a large enough $N$, the truthful equilibrium yields a strictly higher payoff to each agent than any equilibrium where all agents map smaller sets of responses to a single response, e.g., if the responses are $\{a,b,c,d\}$ then $\{a,b\}$ map to $b$ and $\{c,d\}$ map to $d$. This contrasts with the informed truthfulness property of CA, under which the truthful equilibrium doesn't necessarily strictly payoff-dominate equilibria of this type (particularly in scenarios where the observations are clustered; see Definition~10 in the Appendix).
\end{remark}
\subsection{Strong truthfulness vs. Strong truthfulness across symmetric equilibria}\label{sec:symm}
In this section, we show that the property of strong truthfulness across symmetric equilibria of SRA can be strengthened to strong truthfulness (across all equilibria) under certain assumptions commonly made in the peer-prediction literature. To do so, we first discuss what kind of {\it asymmetric} equilibria may arise under SRA that yield a higher reward to the agents than the truthful equilibrium. 

\begin{example}
{\it Consider the labor platform setting considered in Example~\ref{ex1}. In this example, the observation that is expected to be least frequently agreed upon is ``No'' (i.e., the plumber did {\it not} arrive within 5 mins of his/her appointed time). Let John be a plumber operating on this platform, and consider the strategy profile where the customers who evaluate John always report ``No'' for him, while all customers report truthfully for every other plumber. This is an asymmetric strategy profile: the agents who evaluate John follow a different reporting strategy from those who don't. In the limit where the number of plumbers on the platform is large, under the assumptions of Theorem~\ref{thm:main}, we claim that this strategy profile constitutes an equilibrium. The key idea is that the popularity indices are not expected to be impacted by John's reports in this limit. Hence, assuming everyone else adheres to this strategy profile, being truthful is optimal on all tasks other than evaluating John. Moreover, for John's evaluation, reporting ``No'' is optimal since it is the only way to obtain a positive payment. Thus this strategy profile constitutes an equilibrium. Moreover, this equilibrium gives a strictly higher reward to agents who have evaluated John than the truthful equilibrium, since the answer ``No'' is expected to have the lowest popularity index and hence the highest reward for agreement.}
\end{example}

The ability to construct such equilibria relies on agents coordinating their behavior on a task or a small subset of tasks. Our model assumes that task-identifying information is available and the task allocation is exogenously specified, so we cannot preclude this possibility. However, when tasks are randomly allocated to the agents and agents do not make their reporting strategy contingent on the task identity, such equilibria are not expected to arise. Formally, consider the following assumptions.

\begin{assumption}[Randomized Task Allocation]\label{asp:random}
Suppose that there are $N$ tasks and $M$ agents. Each task is to be performed by $m$ agents, where we assume that $m\geq 2$. Also, suppose that no agent should perform more than $n$ tasks on average. To ensure that this is feasible, we consider a system regime in which $N$ and $M$ simultaneously grow such that $M>mN/n$. For each task, $m$ distinct agents are uniformly sampled and assigned to that task. Note that each agent gets picked for a task with probability $m/M$ and thus performs $mN/M$ tasks on average, which is less than $n$.
\end{assumption}
Note that the average number of tasks performed by an agent, $n$, can be chosen to be as small as required (at the cost of requiring a large $M$) to ensure that each agent doesn't perform more than one evaluation with a high probability.
\begin{assumption}[Task-independent strategies]\label{asp:task-cont}
Assume that each agent $j$ picks a reporting strategy $\bq^j$ {\it before} the allocation of evaluation tasks, with the assumption that the same reporting strategy will be applied to every task that the agent performs.
\end{assumption}

Such assumptions are often justifiable in crowdsourcing applications and hence are commonly made in the multi-task peer-prediction literature as we discussed in Section~\ref{sec:related}. We can argue that under such assumptions, the fact that SRA is asymptotically strongly truthful across symmetric equilibria implies that it is, in fact, asymptotically strongly truthful. To state our result, we need the notion of the population average reporting strategy given a strategy profile, defined as $\bar{\bq}(y) =\frac{1}{M}\sum_{j\in \mathcal{M}}\bq^{j}(y)$ for all $y\in\cY$. We have the following result.
\begin{theorem}\label{thm:new}
Consider an $\alpha$-separated generating model $(P_X, \bp)$ that is commonly known to the agents. Suppose that Assumptions~\ref{asp:random} and \ref{asp:task-cont} are satisfied. Then for any $\omega>0$ and $\eta>0$, there is a positive integer $N_0$ that depends on $\alpha$, $\eta$, $\omega$, $m$, $n$, $K$, and $|\cY|$, such that for any $N>N_0$, under SRA, the following holds.
\begin{enumerate}
\item Any symmetric fully informative strategy profile is a strict Bayes-Nash equilibrium.
\item For any Bayes-Nash equilibrium strategy profile such that probability of reporting any answer $y\in \cY$ is either $0$ or at least $\eta$ under the population average reporting strategy, the expected payoff of any agent is at most $\omega$ higher than that under any symmetric fully informative strategy profile.
\item Consider any Bayes-Nash equilibrium strategy profile such that probability of reporting any answer $y\in \cY$ is either $0$ or at least $\eta$ under the population average reporting strategy, and that yields a weakly higher expected payoff to any agent than that under any symmetric fully informative strategy profile. Then the population average reporting strategy under this strategy profile is at most $\omega$-uninformative.
\end{enumerate} 
\end{theorem}
Note that for any $\omega>0$, we are not able to achieve $\omega$-domination of the fully informative equilibrium over {\it all} equilibria for some finite but large enough $N$, but only those equilibria in which the population average reporting probabilities are either $0$ or bounded away from $0$. This is because if the average reporting probability for an answer across the population becomes arbitrarily small in the number of tasks $N$ as $N$ grows, then it is not possible to obtain decaying concentration bounds on the agreement rewards; refer to the similar discussion in the context of Theorem~\ref{thm:sym2} above. However, for any fixed $\omega>0$, in the limit as $N\rightarrow\infty$, {\it all} equilibria are $\omega$-dominated by the truthful equilibrium. Moreover, for any fixed $\omega>0$, in the limit as $N\rightarrow\infty$, {\it any} equilibrium that yields a weakly higher payoff than the truthful equilibrium is at most $\omega$-uninformative. Since $\omega$ is arbitrary, this shows that the mechanism is {\it asymptotically} strongly truthful.\footnote{We remark that agents are expected to use simple strategies in practice, and thus strategies in which the reporting probabilities of the answers become arbitrarily close to zero are unlikely to arise in finite but large N settings under an $\alpha$-separated generating model.}

Informally, the idea of the proof of Theorem~\ref{thm:new} is the following. The random allocation of tasks, as well as non-contingency of reporting strategies on task identities, imply that from the perspective of each agent, no task is special, i.e., they are expected to be paired with a generic peer from the population of agents irrespective of the identity of the task. The assumption that the number of tasks performed by each agent is bounded on average implies that one agent's reports are expected to have a vanishing impact on the population indices. Because of these assumptions, from the perspective of {\it each} agent, the remainder of the population with an asymmetric strategy profile can be replaced by a population where each agent utilizes the {\it population average} reporting strategy $\bar{\bq}$, while (approximately) preserving the payoff structure of the game. Additionally, the fact that {\it each} of the different strategies utilized by the agents is near-optimal against this {\it average} reporting strategy of the population implies that the average strategy $\bar{\bq}$ is {\it also} near-optimal against $\bar{\bq}$, because the payoff of an agent is linear in her reporting strategy. 
%(this fact is evident in Equation~\ref{eqn:linear} in the proof of Theorem~\ref{thm:main} in the Appendix). 
Hence, the agents' payoff is approximately the same as one in the strategy profile where everyone follows $\bar{\bq}$, which constitutes a symmetric strategy profile (with the approximation error in each of the above arguments vanishing in the large tasks limit). The result then essentially follows from the fact that SRA is asymptotically strongly truthful across symmetric equilibria.

We argue that Assumptions~\ref{asp:random} and~\ref{asp:task-cont} are generally not appropriate in the context of feedback elicitation on online platforms. In this context, tasks are not randomly allocated but are chosen by the agents themselves. Also, entities being evaluated have a multitude of extraneous identifying features, e.g., the name of a plumber, the race or ethnicity of the Airbnb property owner, etc. Given this information, one cannot preclude the possibility of the agents choosing their reporting strategy based on such extraneous features and achieving coordination in their reports, potentially achieving higher payoffs compared to truthful reporting of their observations. 

However, such extraneous coordination could be more or less likely depending on the context. For instance, in the example above, the possibility that agents coordinate their behavior with respect to this one plumber, John, out of potentially hundreds seems unlikely. However, such coordination may not be unreasonable in reviewing a popular neighborhood restaurant amongst people residing in that neighborhood.

To ensure that practitioners concerned with our target applications are not misled by SRA's properties reported in our work, we choose not to make restricting assumptions that disallow such coordination and content ourselves with the weaker property of strong truthfulness {\it across symmetric equilibria}. As we argued in Section~\ref{sec:related}, in the absence of such assumptions, the dominance properties of other mechanisms are also restricted. 
\section{Numerical Evaluation}\label{sec:numerics}
In this section, we numerically evaluate the performance of SRA, both, in the case of objective evaluations using synthetic data (Section~\ref{sec:objeval}) and in the case of subjective evaluations using real data from online platforms (Section~\ref{sec:subeval}). In the latter case, we also compare SRA's performance to other related mechanisms.% we evaluate and compare the performance of SRA and related mechanisms using ratings data from a few major online platforms. 
\subsection{Performance on objective evaluations}\label{sec:objeval}
In this section, we test SRA's robustness in inducing truthful behavior in the finite $N$ regime, in settings where population homogeneity may not necessarily hold exactly for objective evaluations due to observation or reporting biases. To do so, we assume the perspective of a single agent operating in a platform environment where SRA is deployed, and examine her incentives for truthful reporting given her beliefs about the other agents' generating model of the observations and their reporting behavior. This exercise also serves as a practical alternative to arguing about whether the common knowledge assumptions that are required for our results hold in practice; we demonstrate numerically that each agent has strong practical incentives to be truthful under SRA if she believes that agents are largely unbiased and they report their observations faithfully. We begin by defining our experimental setup and the performance measures that we consider.

{\bf Setting.} We consider the setting of an online service platform such as Thumbtack,\footnote{Thumbtack is an online service platform that matches customers with local professionals; see \url{https://www.thumbtack.com/}} on which a large number of moving companies offer local moving services. The platform seeks to collect information about how punctual the different moving companies are in adhering to the committed time to start the move. Poor scheduling and information collection practices can result in large variability in the start and finish times of different moves in a day, potentially leading to disgruntled customers. 

In our simulation setup, we assume that a moving company's delay in showing up is an exponentially distributed random variable with a certain mean.  We assume that there are $|\cX| = 5$ types of moving companies, where 
$$\cX = \left\{1\textup{ (Mostly Punctual)}, \,2 \textup{ (Somewhat Punctual)}, \,3 \textup{ (Mostly Tardy)},\, 4 \textup{ (Tardy)},\, 5 \textup{ (Very Tardy)} \right\}.$$ Each of these types is associated with a mean for the distribution of delay. We denote these means as $\bm{\mu} = \left\{\mu_1,\mu_2, \mu_3, \mu_4, \mu_5\right\},$ where we assume that $\mu_1 <\mu_2<\mu_3<\mu_4<\mu_5$. The distribution of these five types across the population is denoted as $P_X$. The platform asks the customers the following question: ``How long (in minutes) after the scheduled time of appointment did the movers show up?'' We consider two settings that differ in the possible answers to this question.
\begin{enumerate}
\item The number of answers is $|\cY| = 3$\ where, $\cY = \left\{ 0 \textup{ to } 30, \,30 \textup{ to } 60,\, 60 \textup{ to }\infty\right\}$.
\item The number of answers is $|\cY| = 5$\ where, $\cY = \left\{ 0 \textup{ to } 15,\,15 \textup{ to } 30,\, 30 \textup{ to } 45,\, 45 \textup{ to } 60,\, 60 \textup{ to }\infty\right\}$.
\end{enumerate}
Given the type of a mover $x\in \cX$, the delay is exponentially distributed with mean $\mu_x$, and hence the conditional probability of making an observation $y = l \textup{ to }u$ is given by:
\begin{equation}
p_y(x) = \exp(-\frac{l}{\mu_x}) - \exp(-\frac{u}{\mu_x}).\label{eq:cond}
\end{equation}

{\bf Instances.} An instance is specified by the probability distribution of the types $P_X$, and the mean delays associated with the types $\bm{\mu}$. We generate $10000$ instances. In each instance, $P_X$ is determined by sampling $5$ numbers independently and uniformly in $[0,1]$ and dividing them by their total to obtain a probability distribution. $\bm{\mu}$ is obtained by sampling $5$ numbers independently and uniformly in $[0,60]$ (hence the mean delay of a moving company can be at most $60$ min) and sorting them in an increasing order to satisfy the requirement that that $\mu_1 <\mu_2<\mu_3<\mu_4<\mu_5$. Once $P_X$ and $\bm{\mu}$ are thus specified, the conditional distributions over observations, $\bp(x)$ for each $x\in\cX$, get specified as well according to Equation~\ref{eq:cond}, for both the settings of $|\cY| = 3$ and $|\cY|=5$.
%\footnote{The generation code for these instances can be found at \url{https://tinyurl.com/y42y83uo}.} 
\begin{remark}\label{notsat}We note that for $|\cY| =3$, $9787$ out of the $10000$ instances thus generated failed to satisfy the self-predicting responses condition (see item~\ref{sec:selfpred} in Section~\ref{sec:mech}) and $9995$ instances failed to satisfy the categorical responses condition (see item~\ref{sec:dg} in Section~\ref{sec:mech}). For $|\cY| =5$, none of the $10000$ instances satisfied either of the two conditions. Hence, the PTSC mechanism \citep{radanovic2016incentives} and the mechanism of \cite{dasgupta2013crowdsourced} are inapplicable with a high frequency in this setting. This finding continues to hold if $P_X$ is chosen to be more ``regular.'' We sampled $10000$ instances in two additional settings when $|\cY| =5$: (a) one where $P_X(x)$ is decreasing in $x$, i.e., types with higher mean delay are more rare, and (b) $P_X(x)$ is decreasing in $x$, i.e., types with higher mean delay are more frequent. In both these settings, none of the instances satisfied the self-predicting responses or categorical responses conditions.  
\end{remark}
\begin{remark}\label{rem:clust}
The CA-HR mechanism described in Section~D.1 in the Appendix is informed truthful across symmetric equilibria in general. However, if the instance is such that the joint distribution of observations of a pair of agents for a common evaluation task is not ``clustered,'' as defined in Definition~10 in the Appendix, then the mechanism is strongly truthful across symmetric equilibria for that instance. We find that for $|\cY| =3$, $9995$ out of the $10000$ instances thus generated had clustered observations. For $|\cY|=5$, $9940$ out of the $10000$ instances had clustered observations. Thus, in this setting, with a very high frequency, SRA is the only applicable mechanism requiring one task per agent that is also strongly truthful across symmetric equilibria.
\end{remark}

{\bf Agent beliefs.} 
We focus on a single agent (a customer on the platform), whom we refer to as agent $j$, and examine her incentives for being truthful under SRA. Each instance that we define above represents a belief that agent $j$ has about the generating model for the movers' true delays. Additionally, we allow the agent to account for potential biases in the observation-making process of other agents. In particular, agent $j$ believes that, while she can make perfect observations, other agents do not observe the delay perfectly, but rather observe a biased version. Owing to this, she believes that given a mover type $x$, the conditional distribution of observations made by a generic agent in the population is not $\bp(x)$, but $\bp'(x)$, which is (slightly) different. We assume that 
\begin{equation}\bp'(x) = (1-\epsilon)\bp(x) + \epsilon \bq'(x),\label{eq:bpp}\end{equation}
where $\bq'$ represents agent $j$'s belief of the bias in the population, and $\epsilon$ represents the magnitude of the bias. In each of the $10000$ instances, for both the settings of $|\cY| = 3$ and $|\cY|=5$ answers, we independently sample an associated bias $\bq'(x)$ for each $x\in\cX$, again by generating $3$ and $5$ values uniformly in $[0,1]$ and dividing each by their sum to obtain a distribution. In our analysis, we will independently consider different values of $\epsilon \in \{0, 0.1, 0.2\}$, where $\epsilon = 0$ is the case where agent $j$ believes that everyone else makes perfect observations. 

{\bf Performance measures.}
We define two performance measures that capture the relative attractiveness of non-truthful behavior compared to being truthful from the perspective of agent $j$, assuming every other agent truthfully reports her (potentially biased) observation. Before we define these measures, we first compute the expected reward for agreement on each answer $y\in\cY$ under the mechanism from the perspective of agent $j$, i.e., compute $\textup{E}(e_j(y))$. For each instance, assuming $K =1$ and denoting $N-|\cN_j|=N'$, we denote $r(y,N',\epsilon)\triangleq \textup{E}(e_j(y))$ to be the expected payment that agent $j$ receives if she and her peer both give a matching response $y$, for each $y\in\cY$. We consider values of $N'$ in the set $\{200,400, 600, 800,1000\}$ and $\epsilon \in \{0, 0.1, 0.2\}$. Note that $r(y,N',\epsilon)$ can be computed exactly given the generating model.\footnote{In SRA, $\bar{f}_j(y)$ is a discrete random variable taking $N'+1$ possible values in the set $\{1/N',2/N',\cdots,(N'+1)/N'\}$. In particular, $\bar{f}_j(y)= 1/N' + Z(y)/N'$, where $Z(y)$ is a binomial random variable arising from $N'$ trials, with probability of success equalling the probability of agreement on $y$ (between two (potentially) biased agents). Thus the expectation of $e_j(y)$, which is also a discrete random variable and a function of $\bar{f}_j(y)$, can be exactly computed.} Also note that under SRA, $e_j(y)$ is computed based on answers of agents other than $j$; hence, from the perspective of agent $j$, $e_j(y)$ incorporates the bias of the agents in making their observations. This bias is reflected in the computation of $\textup{E}(e_j(y))=r(y,N',\epsilon)$.

Based on our calculation of $r(y,N',\epsilon)$, we next define three quantities that will be utilized to define our performance measures. In defining all of these quantities, we assume that all agents other than $j$ truthfully report their (potentially biased) observations.
\begin{enumerate}
\item {\bf Truthful reward.} First, for each instance, we define the expected reward of agent $j$ under truthful behavior:
\begin{align}
\textbf{truthful-reward}(N',\epsilon) &\triangleq \sum_{y\in\cY}P(Y_j =Y_{j'} = y)r(y,N',\epsilon).
\end{align}
Note that in defining this reward, we account for the fact that agent $j$ believes that she makes perfect observations while her peer $j'$ is potentially biased. In particular, we have,
$$P(Y_{j'} = Y_j = y) = \sum_{x\in\cX}P_X(x)p_y(x)p'_{y}(x),$$
where $\bp(x)$ is defined in Equation~\ref{eq:cond} and $\bp'(x)$ is defined in Equation~\ref{eq:bpp} (capturing the fact that the peer agent is biased).

\item {\bf Optimal Reward.} Next, we define the expected reward of agent $j$ from the optimal reporting strategy that maximizes her expected reward (which could potentially entail lying). In doing so, we address an important consideration. Although the question simply asks for the interval in which the true delay of the mover lies, agent $j$ can actually observe the true delay. Thus the optimal report must be determined conditioned not on the true answer, but the true delay (from which the true answer can be determined). We assume that agent $j$ can accurately observe the delay to within a minute and she stops observing if the delay is larger than 180 min (i.e., 3 hours).\footnote{This assumption is for simplicity of computation of the optimal reporting strategy as a function of the observed delay. Our findings are not expected to change significantly if the optimal reports are computed conditioned on finer feedback information.} Formally, we assume that the observations of the delay lie in the finite set $\overline{\cY}= \{a \textrm{ to } a+1; \textrm{ for } a \in \{0,1,\cdots,179\}\}\cup \{180 \textrm{ to } \infty\}$. We denote the observed delay of agent $j$ by the random element $\overline{Y}_j\in \overline{\cY}$. As before, since the delay is exponentially distributed with mean $\mu_x$ given the type $x\in\cX$, the conditional probability of observing the delay $y = l \textup{ to }u \in\overline{\cY}$ is given by:
\begin{equation}
\bar{p}_y(x) = \exp(-\frac{l}{\mu_x}) - \exp(-\frac{u}{\mu_x}).\label{eq:cond2}
\end{equation}
Accounting for this consideration, we finally define the expected reward from the optimal reporting strategy as:
\begin{align}
\textbf{optimal-reward}(N',\epsilon) &\triangleq \sum_{y\in\overline{\cY}}P(\overline{Y}_j = y)\max_{y'\in \cY}P(Y_{j'} = y'|\overline{Y}_j = y)r(y',N',\epsilon)\\
&= \sum_{y\in\overline{\cY}} \max_{y'\in \cY}P(Y_{j'} = y',\,\overline{Y}_j = y)r(y',N',\epsilon).
\end{align}
Here we have,
$$P(Y_{j'} = y',\,\overline{Y}_j = y) = \sum_{x\in\cX}P_X(x)\bar{p}_y(x)p'_{y'}(x),$$
where $\bp'(x)$ is defined in Equation~\ref{eq:bpp} (capturing the fact that the peer agent is biased) and $\bar{\bp}(x)$ is defined in Equation~\ref{eq:cond2} (capturing the fact that the agent observes the true delay in the set $\overline{Y}$). 

\item {\bf Effortless reward.} Finally, as a baseline, we define the reward obtained by agent $j$ by choosing an answer $y\in\cY$ uniformly at random without making any observation:
\begin{align}
\textbf{effortless-reward}(N',\epsilon) &\triangleq \sum_{y\in\cY}\frac{1}{|\cY|}P(Y_{j'} = y)r(y,N',\epsilon),
\end{align}
where we have,
$$P(Y_{j'} = y) = \sum_{x\in\cX}P_X(x)p'_{y}(x),$$
where $\bp'(x)$ is defined in Equation~\ref{eq:bpp}.
\end{enumerate}
We finally define our two main performance measures.
\begin{enumerate}
\item {\bf \underline{Lying gain}.} The first measure we define captures the percentage gain in expected reward of agent $j$ by reporting optimally rather than simply being truthful. It is defined as,
\begin{align}
\textbf{lying-gain}(N',\epsilon) &\triangleq \frac{\textbf{optimal-reward}(N',\epsilon) - \textbf{truthful-reward}(N',\epsilon)}{\textbf{truthful-reward}(N',\epsilon)}\times 100\%.\label{lying-gain}
\end{align}
Ideally, we would like this gain to be small.
\item {\bf \underline{Truthful coverage}.} Although the measure that we define above is a natural one to consider, it provides at best a partial picture of the incentives generated by the mechanism. In particular, one way of ensuring a small lying-gain is to simply add a very large fixed reward to the reward under the mechanism so that the denominator in Equation~\ref{lying-gain} becomes large. By scaling this fixed reward, one can ensure that the lying-gain is as small as one desires without changing the incentive properties of the mechanism. By doing so, even mechanisms with poor incentives for truthful behavior can result in small lying-gain. In other words, although the lying-gain is invariant to multiplicative scaling of the rewards under the mechanism, it is not invariant to additive shifts of the rewards. To address this concern, we define the following relative performance measure, which is invariant to both additive shifts as well as multiplicative scaling of the rewards.
\begin{align}
\textbf{truthful-coverage}(N',\epsilon) &\triangleq \frac{\textbf{truthful-reward}(N',\epsilon) - \textbf{effortless-reward}(N',\epsilon)}{\textbf{optimal-reward}(N',\epsilon)- \textbf{effortless-reward}(N',\epsilon)}\times 100\%.\label{truthful-coverage}
\end{align}
This quantity measures the fraction of the incremental gain resulting from optimal reporting as compared to reporting randomly, that can be attained by truthful reporting. This quantity should ideally be large.
\end{enumerate}

We note that due to the bias in the population, and given that agent $j$ evaluates the incentives for lying conditioned on the observed delay as opposed to the true answer, being truthful is not guaranteed to be optimal for agent $j'$ under SRA even in the $N'\rightarrow\infty$ limit. Hence, it is expected that $\textbf{lying-gain}(N',\epsilon)>0\%$ and $\textbf{truthful-coverage}(N',\epsilon)<100\%$.

We define $\textbf{avg-lying-gain}(N',\epsilon)$ to be the average across $10000$ instances of the $\textbf{lying-gain}(N',\epsilon)$. Similarly, we define $\textbf{avg-truthful-coverage}(N',\epsilon)$ to be the average across $10000$ instances of the $\textbf{truthful-coverage}(N',\epsilon)$. When the context is clear, for notational convenience, we will refer to these aggregate quantities as avg-lying-gain and avg-truthful-coverage respectively. 

 \begin{figure}[h]
  \begin{subfigure}[b]{0.5\textwidth}
    \centering
    \includegraphics[width=0.9\textwidth]{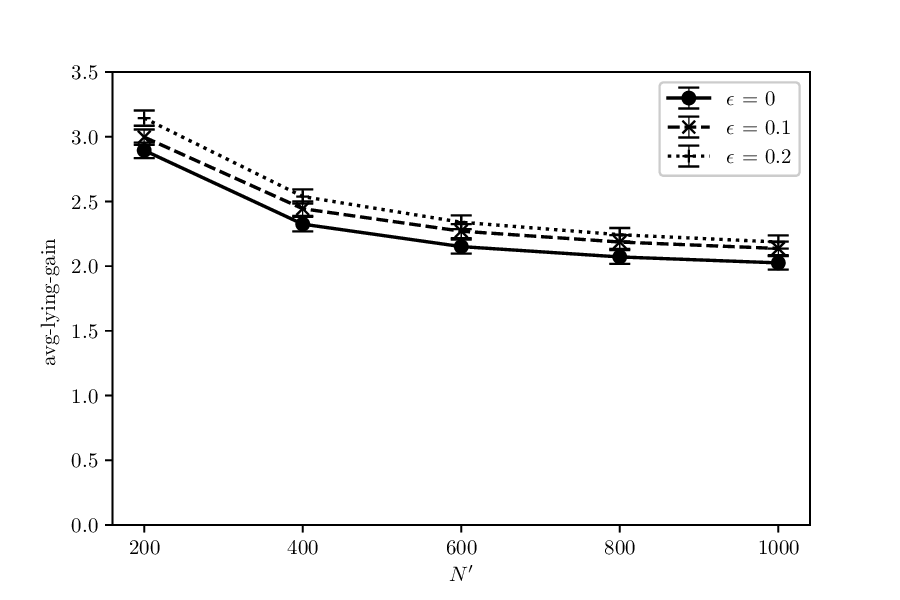}
    \caption{avg-lying-gain ($|\cY|=3$)}
    \label{perf:a3}
  \end{subfigure}%%
  \begin{subfigure}[b]{0.5\textwidth}
    \centering
    \includegraphics[width=0.9\textwidth]{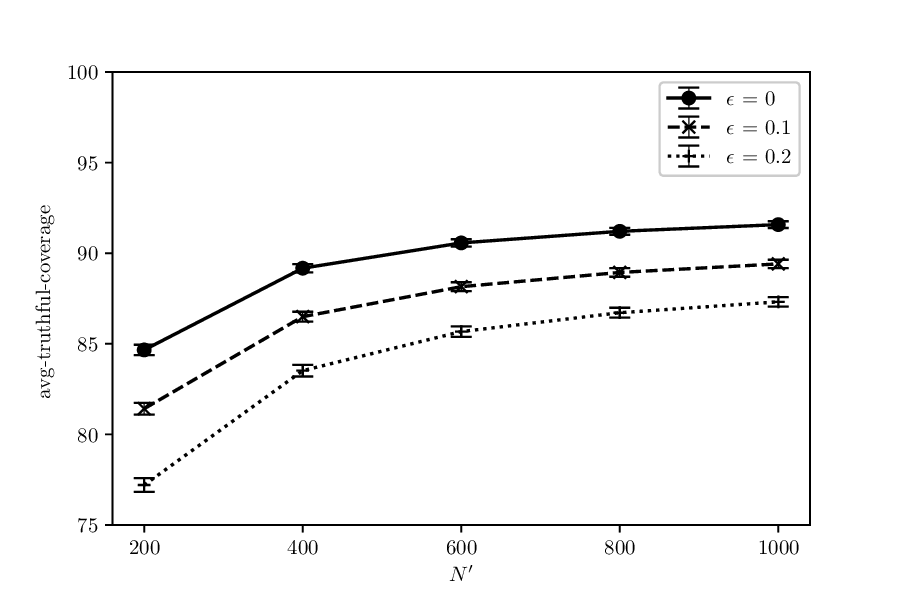}
    \caption{avg-truthful-coverage ($|\cY|=3$)}
    \label{perf:b3}
  \end{subfigure}
  \caption{The $\textup{avg-lying-gains}$ along with standard errors under SRA for $N'\in \{100,200,300,400,500\}$ (X-axis) for different values of $\epsilon$ and for the $|\cY|=3$ setting.}
  \label{fig:perf3}
\end{figure}
  
\begin{figure}[h]
  \begin{subfigure}[b]{0.5\textwidth}
    \centering
    \includegraphics[width=0.9\textwidth]{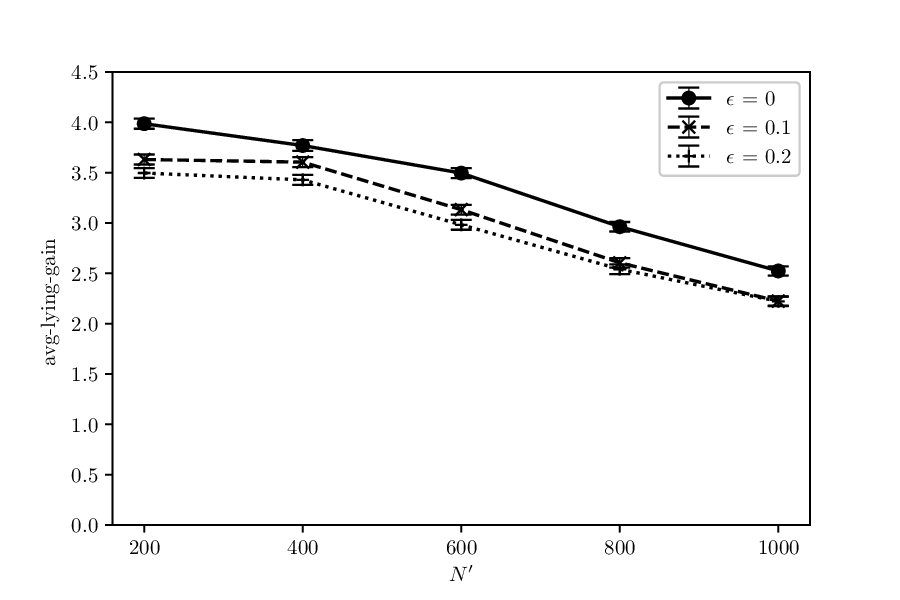}
    \caption{avg-lying-gain ($|\cY|=5$)}
    \label{perf:a5}
  \end{subfigure}%%
  \begin{subfigure}[b]{0.5\textwidth}
    \centering
    \includegraphics[width=0.9\textwidth]{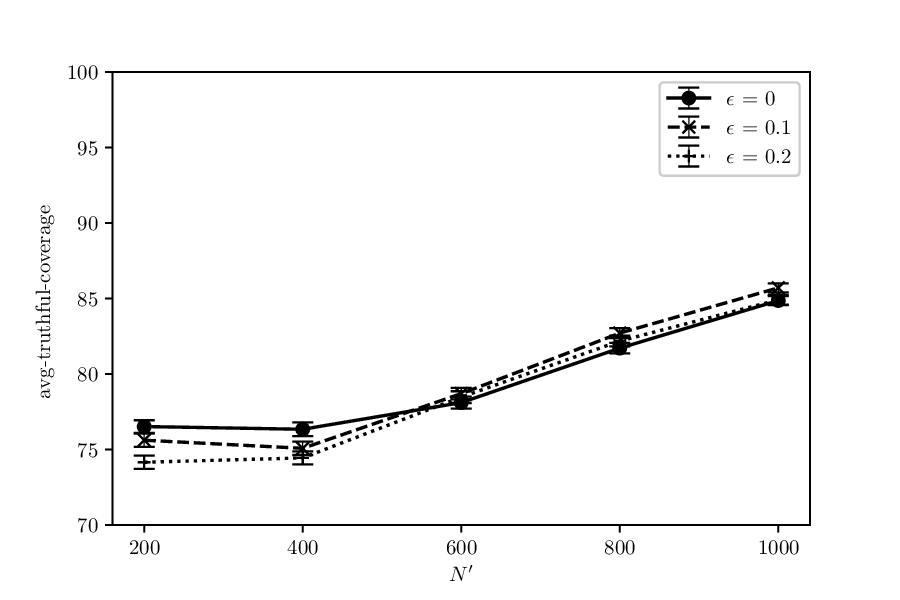}
   \caption{avg-truthful-coverage ($|\cY|=5$)}
    \label{perf:b5}
  \end{subfigure}
  \caption{The $\textup{avg-lying-gains}$ (Y-axis) along with standard errors under SRA for $N'\in \{100,200,300,400,500\}$ (X-axis) for different values of $\epsilon$, for the $|\cY|=5$ setting.}
  \label{fig:perf5}
\end{figure}

{\bf Results.} The aggregate performance measures and standard errors\footnote{The standard errors of the avg-lying-gain and avg-truthful-coverage are the standard errors of these mean quantities, defined as $1.96$ times the empirical standard deviation of the lying-gains or truthful-coverages divided by the square root of the sample size ($10000$).} for the different values of $N'$ and $\epsilon$ are presented for the $|\cY| = 3$ setting in Figure~\ref{fig:perf3} and for the $|\cY| = 5$ setting in Figure~\ref{fig:perf5}. We note two main observations. 

First, SRA displays reasonably good performance in incentivizing truthful behavior despite the limited number of tasks, even when agent $j$ believes that the rest of the population is biased. For instance, for $|\cY|=3$ and $N'=1000$, the avg-lying-gain is at most about $2.5\%$ across all settings of biases. Correspondingly, for $|\cY|=5$ and $N'=1000$, the avg-lying-gain is at most about $2.7\%$. Additionally, SRA displays reasonably good truthful-coverage, e.g., for $|\cY|=3$ and $N'=1000$, truthful behavior attains about $86\%$ of the maximal potential gain over random reporting on average. For $|\cY|=5$ and $N'=1000$, truthful behavior attains about $84\%$ of the maximal potential gain over random reporting on average.

Second, we note an interesting phenomenon in the $|\cY|=5$ setting: when agent $j$'s belief about the bias in the population increases, i.e., $\epsilon$ increases from $0$ to $0.2$, her incentive to lie seems to {\it decrease} under almost all performance measures, especially for $N'\in \{600,800,1000\}$. In particular, the lying-gain is smaller and truthful-coverage is larger in aggregate for $\epsilon = 0.1$ as compared to $\epsilon = 0$. This is quite unlike the $|\cY|=3$ setting, in which the incentive to lie {\it increases} with $\epsilon$, as expected, under all measures. 

\begin{figure}[h]
 \centering
 \includegraphics[width=0.45\textwidth]{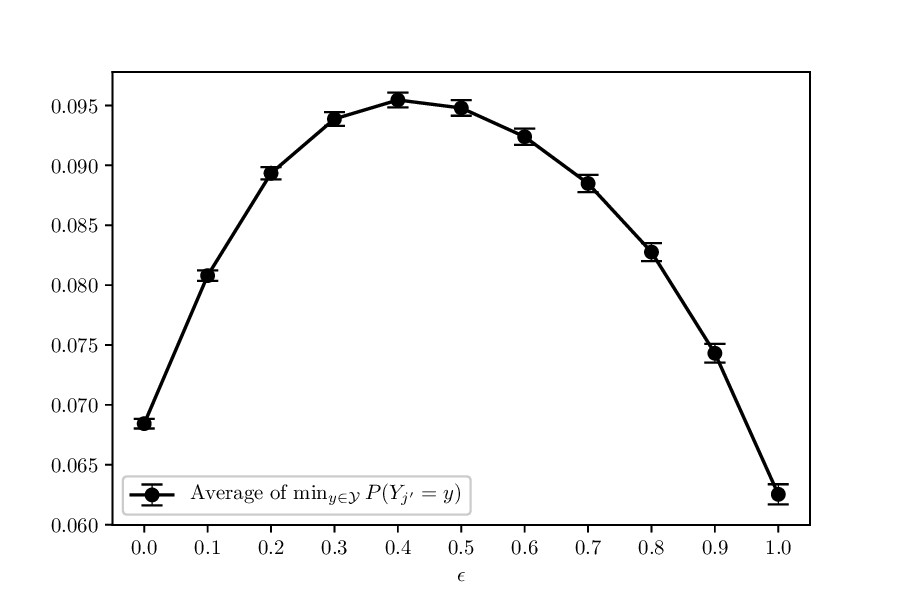}
 \caption{ For the $|\cY|=5$ setting, the average across $10000$ instances of $\min_{y\in\cY} P(Y_{j'} = y)$ (Y axis) for some generic agent $j'$ in the population as $\epsilon$ (X axis), i.e., the magnitude of the bias in the population, increases.}
  \label{fig:avg-y}
 \end{figure}

This observation is surprising, given that population homogeneity is crucial to the properties of SRA. It can be explained in light of the fact that adding a small random bias typically increases the probability of rare answers in the population. Informally, adding a small observation noise makes the distribution of answers ``better mixed,'' since the reporting probability of answers with a low probability of occurrence improves simply because of the noise. This results in a higher inequality gap in the Cauchy-Schwarz inequality, as shown in Proposition~\ref{prop:equi}. In other words, the biased generating model is better separated on average than the unbiased model. In turn, this fact results in faster convergence of the popularity indices of the answers to their stable values. In Figure~\ref{fig:avg-y}, we show the average across the $10000$ instances of the smallest probability of observation (in $\cY$) made by a generic biased agent in the population, where we observe that this average indeed increases with $\epsilon$ when $\epsilon$ is small, before decreasing (the mixing effect is highest around $\epsilon \approx 0.4$). Thus, in finite $N'$ settings, the incentive to lie can potentially be higher when the population is assumed to be unbiased since the population indices are expected to be farther from their asymptotic stable values, as compared to the case where the population is expected to be mildly biased. 
  
\begin{figure}[h]
  \begin{subfigure}[b]{0.5\textwidth}
    \centering
    \includegraphics[width=0.9\textwidth]{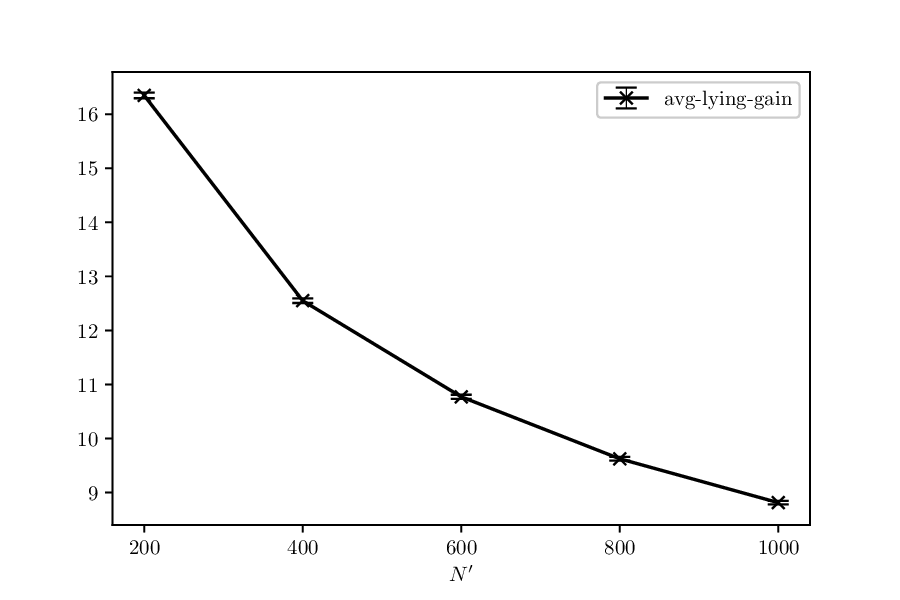}
    \caption{ $|\cY|=5$ and $\epsilon =1$ (aggregate lying-gains)}
    \label{perf:eps}
  \end{subfigure}%%
  \begin{subfigure}[b]{0.5\textwidth}
    \centering
    \includegraphics[width=0.9\textwidth]{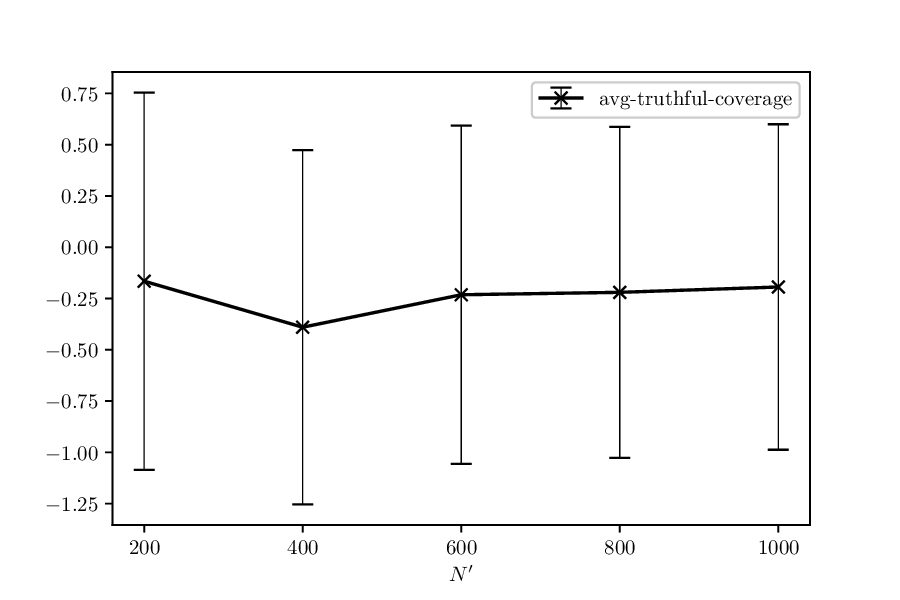}
    \caption{$|\cY|=5$ and $\epsilon =1$ (aggregate truthful-coverages)}
    \label{perf:eps1}
  \end{subfigure}
  \caption{For $|\cY|=5$, the different aggregate measures (Y-axis) along with standard errors under SRA for $N'\in \{100,200,300,400,500\}$ (X-axis) in the fully biased setting, i.e., $\epsilon = 1$ (the distributions of observations of agent $j$ and her randomly chosen peer are independent).}
  \label{fig:perf-eps}
\end{figure}

However, as the bias in the population becomes large, i.e., $\epsilon$ increases, this effect is overpowered by the increased incentive to lie, since the generating model of the peer agent's observations starts to look starkly different from agent $j$'s model, i.e., the response homogeneity assumption is violated to a higher degree. We indeed verify this to be the case. In Figure~\ref{fig:perf-eps}, we plot the various aggregate measures in the $|\cY|=5$ setting for $\epsilon = 1$, i.e., when the distributions of observations conditioned on the type are completely uncorrelated for agent $j$ and her randomly chosen peer agent. As expected, the incentive to lie is significantly higher across all values of $N'$ in this case compared to settings with smaller values of $\epsilon$. %The aggregate measures of truthful-coverage indicate that there are instances where truthful behavior, in fact, yields a lower reward than random reporting (see Figure~\ref{perf:eps1}).

Overall, these observations suggest that the belief that there exists a mild observation bias in the population may in fact improve incentives for truthful behavior in finite $N'$ settings, as long as this bias is small enough.

\subsection{Performance on subjective evaluations}\label{sec:subeval}
As we have discussed in Section~\ref{sec:related}, strongly truthful or informed truthful mechanisms for eliciting subjective evaluations from heterogeneous agents require multiple evaluations from each agent. Moreover, these mechanisms require that each agent uses the same strategy for each evaluation. These constraints may hinder the practical applicability of these mechanisms in many platform environments. In the face of these drawbacks, mechanisms tailored to homogeneous response settings that require a single task per agent could be a practical alternative. In this section, we hence evaluate the performance of SRA, PTSC, and CA-HR for incentivizing truthful responses to subjective evaluations in real settings. 

 {\bf Datasets.} We test these mechanisms using publicly available rating datasets from different online platforms. In particular, we consider the following three datasets.
\begin{enumerate}
\item {\bf Goodreads.} We consider book rating data from Goodreads, which is a popular book review platform.\footnote{The data is publicly available at \url{https://sites.google.com/eng.ucsd.edu/ucsdbookgraph/home}. The source requires us to cite \cite{wan2018item} and \cite{wan2019fine}.} We restrict our attention to books belonging to two largest and similarly-sized categories: (a) romance and (b) fantasy and paranormal. We assume books in each of these categories to be a priori statistically similar and we test the performance of different mechanisms for these two categories independently.
\item {\bf Amazon.} We next consider product rating data from the e-commerce platform, Amazon.\footnote{The data is publicly available at \url{https://nijianmo.github.io/amazon/index.html}. The source requires us to cite \cite{ni2019justifying}.} We restrict ourselves to the ``Clothing, Shoes, and Jewelery'' (CSJ) category, which is by far the largest product category other than books.\footnote{The Goodreads data is richer than Amazon's rating data for books.} 
%The definition of this category is nevertheless quite broad and significant diversity is expected across the products in this category; hence, the assumption of the products being a priori statistically similar may not be appropriate. 
\item {\bf Netflix.} We finally consider movie rating data from the streaming platform, Netflix, that was released as part of the Netflix Prize challenge.\footnote{The data is publicly available at \url{https://www.kaggle.com/netflix-inc/netflix-prize-data}. Information about the Netflix prize is available at \url{https://en.wikipedia.org/wiki/Netflix_prize}.} 
\end{enumerate}
In all of the above cases, the ratings are integers on a scale from 1 to 5. Moreover, the ratings are expected to have a strongly subjective influence, especially so in the case of books and movies. Table~\ref{tbl:stats} provides some basic information about these datasets. %The same goes for the Netflix ratings dataset where no meta-data about the movies is available.  
 
\begin{table}[ht]
\centering
\caption{Properties of datasets. The rating strength represents the highest lower bound on the number of ratings given by the top 1000 high-contributing users. \label{tbl:stats}}
\begin{tabular}[t]{lcccc}
\toprule
&No. of entities & No. of users & No. of ratings & Rating strength\\
\midrule
Goodreads: romance&334957 books&198141&3565378& 472\\
Goodreads: fantasy/paranormal &258212 books&256088& 3424641&278\\
Amazon: CSJ &2681297 products&12483678&32292099& 104\\
Netflix &17770 movies& 480189 &100480507&2087\\
\bottomrule
\end{tabular}
\end{table}%

%from the online marketplace, Amazon.com. To consider a setting where subjectivity in ratings is a central feature, we restrict our attention to books in the ``Literature and Fiction" (LF) category, sold to customers online via the Kindle tablet device. Restricting to online books eliminates any aspect of the reviewing emanating from sources exogenous to the content of the book, such as delivery speed, physical damage, problems arising during shipping, or post-sales customer service, which could lend objectivity to the evaluations. We moreover restrict attention to entries involving books that have been rated by at least 5 users, such that every user has reviewed at least 5 books: this pre-processed ``5-core" dataset is publicly available online. There were a total of 1055532 entries of this type in the LF category, involving 42952 books rated by 121679 users.

{\bf Testing procedure.} In each of the above cases, we focus on the top 1000 users who have rated the most number of entities (books, movies, or products). Assuming that all ratings in the dataset are truthful, we estimate the reporting behavior of these users and investigate their incentives for lying under the various mechanisms. Formally, let $\cH$ denote the set of high-contributing users and consider a user $i\in\cH$. Based on $i$'s ratings across the books they have rated and the ratings of randomly chosen peers for these books, we estimate the joint distribution of the rating of $i$ and that of a randomly chosen peer agent for a random book that they rate. Let us denote this estimate as $\left(Q_i(y,y')\right)_{y,y'\in \cY}$. Similarly, we estimate the joint distribution of the ratings of two randomly chosen agents for a random book by sampling two agents at random for each book in the data set and computing the empirical distribution of the resulting answers. Let us denote this estimate as $\left(\overline{Q}(y,y')\right)_{y,y'\in \cY}$. The estimates $Q_i$ and $\overline{Q}$ are expected to be different, in line with the expectation that agent $i$'s responses are statistically different from a randomly chosen agent in the population due to the subjectivity of responses.

Based on the estimates $Q_i$ and $\overline{Q}$, we can estimate the truthful-coverage of each mechanism for each agent $i\in \cH$, assuming that (a) $i$'s belief about the joint distribution of her rating and that of a random peer for a random book is identical to $Q_i$, and (b) her belief about the joint distribution of the ratings of two randomly chosen agents for a random book is identical to $\overline{Q}$. Note that we focus on truthful-coverage since, as we discussed earlier, the lying-gain is not invariant to additive shifts in the rewards. %In particular, each user knows precisely how statistically different they are as compared to an average user in the population. 
\\

\begin{figure}[ht]
  \begin{subfigure}[b]{0.5\textwidth}
    \centering
    \includegraphics[width=0.8\textwidth]{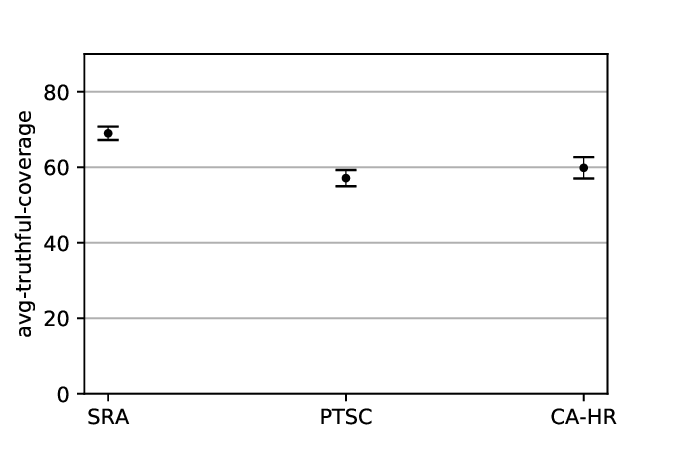}
    \caption{Goodreads: romance}
    \label{perf:goodreads1}
  \end{subfigure}%%
  \begin{subfigure}[b]{0.5\textwidth}
    \centering
    \includegraphics[width=0.8\textwidth]{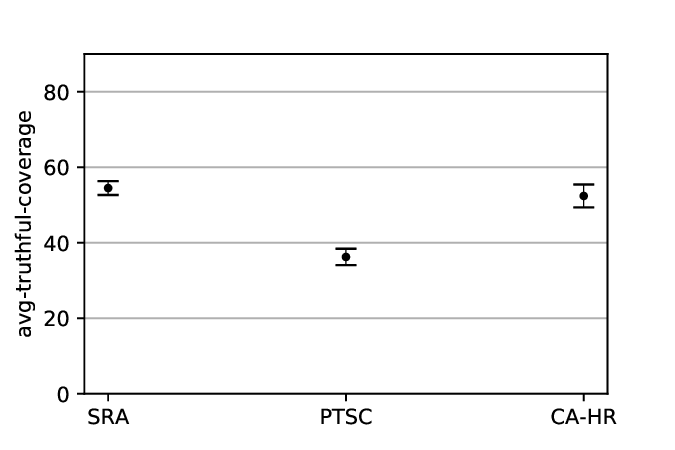}
    \caption{Goodreads: fantasy/paranormal}
    \label{perf:goodreads2}
  \end{subfigure}
   \begin{subfigure}[b]{0.5\textwidth}
    \centering
    \includegraphics[width=0.8\textwidth]{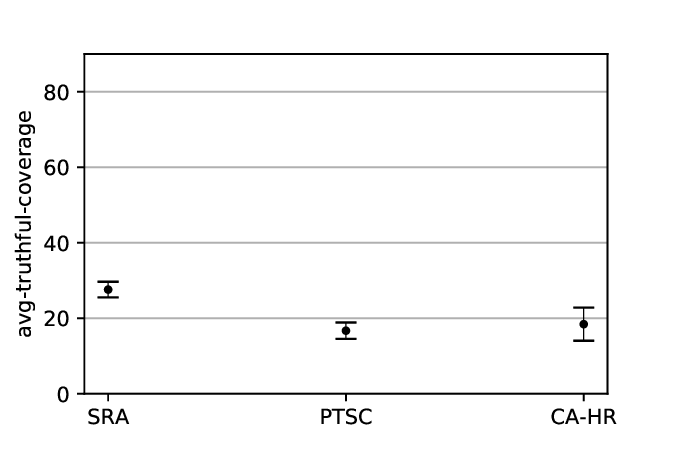}
    \caption{Amazon: CSJ}
    \label{perf:amazon}
  \end{subfigure}
   \begin{subfigure}[b]{0.5\textwidth}
    \centering
    \includegraphics[width=0.8\textwidth]{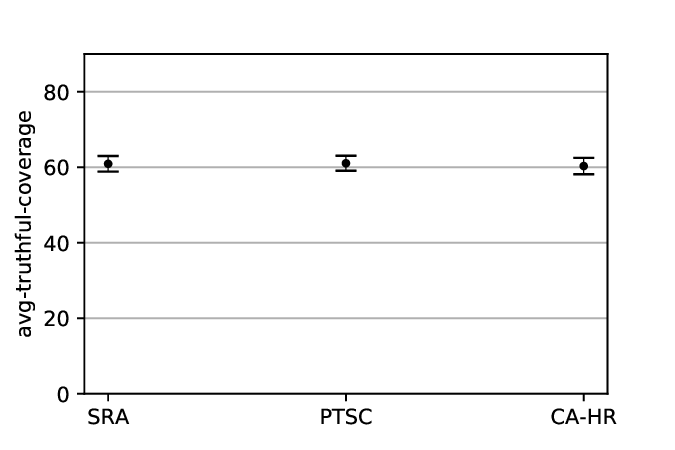}
    \caption{Netflix}
    \label{perf:netflix}
  \end{subfigure}
  \caption{The $\textup{avg-truthful-coverages}$ along with standard errors under the different mechanisms in different settings.}
  \label{fig:real-perf}
\end{figure}

{\bf Results.} The results are shown in Figure~\ref{fig:real-perf}. First, we observe that SRA achieves an average truthful-coverage of at least 50\% across all settings except for Amazon. In this latter case, the performance of all mechanisms is relatively poor. The reason for this may be that the definition of the CSJ category is quite broad, and significant diversity is expected across the products in this category; hence, the assumption of the products being a priori statistically similar likely doesn't hold. 

Next, we observe that SRA outperforms PTSC in all settings except for the case of Netflix, where their performance is statistically similar. SRA outperforms CA-HR in the Goodreads setting for the romance category and in the case of Amazon, while their performance is statistically similar in the other two cases. To investigate the difference in SRA and PTSC, we consider the estimate $\overline{Q}$ of the joint distribution of two agents' responses to a common evaluation. The incentive to lie for an agent $i$ stems from two sources: (a) $Q_i$ may be different from $\overline{Q}$, and (b) $\overline{Q}$ may not satisfy the conditions necessary for inducing truthful behavior even when all agents are identical. We find that in all settings, $\overline{Q}$ satisfies the Cauchy-Schwarz property required for SRA to be truthful (Equation~\ref{eq:csprop}), while the self-prediction property that is needed for PTSC to be truthful (see item~\ref{sec:selfpred} in Section~\ref{sec:mech}) is not satisfied in any of the settings; see Table~\ref{tbl:q-prop}. %This may explain its relatively poor performance in the other three settings. 

Moreover, we find that the response distribution $\overline{Q}$ is `clustered' in all settings, as defined in Definition~10 (from \cite{shnayder2016informed}) in the Appendix. This implies that under the CA-HR mechanism, the agents can achieve the same payoff by merging their responses. For example, considering the $\overline{Q}$ from the Netflix setting, we find that agents need not distinguish between the ratings 4 and 5 under the CA-HR mechanism (see Table~\ref{tbl:q-prop}). This points to the importance of the distinction between strong and informed truthfulness in these settings.

\begin{table}[ht]
\centering
\caption{Properties of population average joint distribution of responses $\overline{Q}$. \label{tbl:q-prop}}
\begin{tabular}[t]{lccc}
\toprule
&Satisfies CS property & Satisfies self-prediction & Clustered ratings\\
\midrule
Goodreads: romance&Yes&No&(1, 2)\\
Goodreads: fantasy/paranormal &Yes&No& (1, 2, 3)\\
Amazon: CSJ &Yes&No& (2, 3)\\
Netflix &Yes& No & (4, 5)\\
\bottomrule
\end{tabular}
\end{table}

\section{Discussion and Conclusion}\label{sec:conclusion}
In the paper, we focus on the practical setting of reputation systems in online platforms where objective evaluations must be strongly incentivized; ideally, without imposing any constraints on the number of evaluations performed by each agent. Our results show that SRA is the first mechanism that achieves this goal. 

While there are other mechanisms, such as those of \cite{kong2016framework}, \cite{kong2020dominantly}, or CA \citep{shnayder2016informed}, that incentivize truthful behavior despite response heterogeneity across agents, these mechanisms incur a high operational cost of requiring multiple evaluations from each agent, which could be prohibitive in many scenarios, including in online platforms.  On the other hand, our numerical evaluations show that the truthfulness property of SRA is robust to mild degrees of heterogeneity and subjectivity in the population. This observation overall suggests that SRA can be a simpler alternative to these more complex mechanisms in settings where response homogeneity is a reasonable approximation to the mild degree of heterogeneity and subjectivity expected in the evaluations. Eliciting objective evaluations in online platforms is one such setting that we focused on in the paper. Additionally, our tests on real data show that SRA generates strong incentives for truthful behavior even when evaluations are expected to be highly subjective.

At the same time, we acknowledge that there are settings where other mechanisms could be preferable over SRA. Moreover, metrics such as lying-gain or truthful-coverage may not adequately inform the practical utility of mechanisms in these settings and other operational considerations may take precedence. For example, in applications such as crowdsourcing and peer-grading,\footnote{Peer-grading is the idea of having students grade each others' assignments and examinations. This idea is key in obtaining a scalable solution to the problem of grading in massive open online courses (MOOC). Incentivizing students to grade truthfully is an important concern in such settings.} agents typically perform several evaluations in a short span of time. Moreover, subjectivity in evaluations could be a major concern in settings like peer-grading for courses in the arts and the humanities. In this case, Kong (2020)'s mechanism would provide strong truthfulness guarantees without requiring the homogeneity assumption and hence could be preferable over SRA, even if, hypothetically, it turns out to be the case that SRA achieves better truthful-coverage or lying-gain than Kong's mechanism in homogeneous settings with comparable data. As another example, if the responses can be validated to be self-predicting, then PTSC may be preferable owing to the simpler description of the agreement rewards. 

Effective feedback and reputation systems are fundamental to the efficient functioning of online
platforms. The impact of user
feedback and peer-reviews on customer decisions is evident in the success of independent reputation
systems like Yelp and TripAdvisor, which are used by millions of people across the world. But
as has been recently shown, these systems are currently fraught with several operational,
behavioral, and strategic concerns \citep{hu2017self,filippas2018reputation,nosko2015limits}. We believe that appropriate incentive mechanisms
that are simple and intuitive can go a long way in addressing some of these concerns, and hence our mechanism has strong practical significance. We emphasize here that rather than
thinking of our mechanism as a fully specified solution in any setting, it is more useful to think of it
as a framework that provides conceptual guidelines for platform designers as they undertake their
design decisions.

Our work presents many avenues for future exploration. For instance, in our model, we assume
that the task allocations are exogenously specified. But for a platform that is interested in learning
the underlying distributions of responses for each task, some of these distributions may be more
difficult to learn than others, and thus may need more evaluations. Moreover, the agents may be
willing to strategically respond to differences in potential rewards across tasks by choosing which
tasks to evaluate. It is important to understand the fundamental tradeoffs faced
by dynamic mechanisms that balance incentives with different statistical accuracy objectives in such situations. We
are optimistic that our framework and insights can be used as building blocks in
this pursuit.

\bibliographystyle{apalike}
\bibliography{crowdsourcing}
\newpage
%\setcounter{page}{1}
%\ECHead{\centering The Square Root Agreement Rule for Incentivizing Truthful Feedback on Online Platforms: Online Appendix}% by Kamble, Shah, Marn, Parekh, and Ramchandran}
\begin{APPENDIX}{}
\section{Proofs}

\begin{proof}{Proof of Proposition~\ref{prop:equi}}
Let $m = \min_{y\in\cY} \|\bv(y)\|$. To see that 2 implies 1, note that $\theta(\bv(y),\bv(y'))\geq \gamma$ for all $y,\,y'\in\cY$ such that $y\neq y'$, implies that $m>0$ and that
$$\frac{\bv(y).\bv(y')}{\|\bv(y)\|\|\bv(y')\|} \leq \cos\gamma,$$
Multiplying throughout by $\|\bv(y)\|\|\bv(y')\|$, we have:
$$\|\bv(y)\|\|\bv(y')\|-\bv(y).\bv(y')\geq (1- \cos\gamma)\|\bv(y)\|\|\bv(y')\|\geq (1- \cos\gamma)m^2>0.$$
%Here in the last inequality, we use that fact that 
%$$\|v(s)\|= \sqrt{\sum_{h\in \mathcal{H}}P_X(h)p(s|h)^2} \geq \sum_{h\in \mathcal{H}}P_X(h)p(s|h) >\gamma,$$
%which follows from the Jensen's inequality. 
To show that $1$ implies $2$ is less straightforward and this is where we need to use the fact that $\|\bv(y)\|\leq 1$ for all $y\in\cY$. First of all 
$$|\bv(y).\bv(y')|\leq \|\bv(y)\|\|\bv(y')\| - \alpha,$$
implies that both $\|\bv(y)\|$ and $\|\bv(y')\|$ are non-zero. Then dividing on both sides, we get:
\begin{align*}
\frac{|\bv(y).\bv(y')|}{\|\bv(y')\|\|\bv(y)\|}&\leq 1-\frac{\alpha}{\|\bv(y')\|\|\bv(y)\|}\\
&\leq 1-\alpha
\end{align*}
where the last inequality holds since $\|\bv(y)\|\leq 1$ for all $y\in\cY$. In other words:
$$\cos \theta(\bv(y),\bv(y')) \leq  1-\alpha,$$
%or 
%$$\|\bv(y)\|(1-\cos \theta(\bv(y),\bv(y')))\geq \alpha.$$
This implies that $\theta(\bv(y),\bv(y')))\geq \arccos (1-\alpha)$. %Finally we have $\sum_{x\in \cX}P_X(x)p_y(x) \geq \|\bv(y)\|^2 \geq \alpha^2$. 
Note that $\alpha > 0$ so that $\arccos (1-\alpha)>0$.\label{fn:1}  \hfill $\Box$

\end{proof}
\bigskip
\begin{proof}{Proof of Theorem~\ref{thm:main}}
First, note that the payments $e_j(y)$ for the different $y\in \cY$ are independent of the reports of agent $j$ for any reporting strategy. This is because $\{e_j(y):y\in\cY\}$ are computed only based on evaluation tasks that $j$ does not perform. Next, suppose that everyone but agent $j$ is truthful. Recalling the definition of $\bv(y)\triangleq \big(\sqrt{P_X(x)}p_y(x);\,x\in\cX\big)$, we have,
$$\textup{E}(\bar{f}_j(y)- \frac{1}{N-|\cN_j|})=\textup{E}\bigg[\frac{1}{N-|\cN_j|}\sum_{i\in\cN\setminus\cN_j}\mathbf{1}_{\{r^{i}_{j_1(i')}=y\}}\mathbf{1}_{\{r^{i}_{j_2(i')}=y\}} \bigg]= \sum_{x\in\cX}P_X(x)p_y(x)^2= \|\bv(y)\|^2 \triangleq g(y).$$

In the proof of Proposition~\ref{prop:equi}, we have seen that $\delta(P_X,\bp)>\alpha$ implies that $\|\bv(y)\|>\alpha$, and thus we have $g(y)> \alpha^2>0$ for all $y\in\cY$. Next, recall that

$$e_j(y)=\frac{K}{\sqrt{\bar{f}_j(y)}}.$$
Let $N' = N-|\cN_j|$. Then we have for any $\epsilon \in (0,1)$:
\begin{align*}
\textup{E}(e_j(y)) &\geq P(\bar{f}_j(y) -1/N'\in [g(y)(1-\epsilon), g(y)(1+\epsilon)])\frac{K}{\sqrt{g(y)(1+\epsilon)+1/N' }}\\
& \stackrel{(a)}{\geq} (1-2\exp(-\epsilon^2g(y)^2N'))\frac{K}{\sqrt{g(y)(1+\epsilon)+1/N' }}\\
&  \geq (1-2\exp(-\epsilon^2\alpha^4N'))\frac{K}{\sqrt{g(y)(1+\epsilon)+1/N' }}\\
& \geq \frac{K}{\sqrt{g(y)(1+\epsilon)+1/N' }}-2\exp(-\epsilon^2\alpha^4N')\frac{K}{\alpha\sqrt{(1+\epsilon)}}\\
& \geq \frac{K}{\sqrt{g(y)(1+\epsilon+1/(g(y)N') }}-2\exp(-\epsilon^2\alpha^4N')\frac{K}{\alpha\sqrt{(1+\epsilon)}}\\
& \geq \frac{K}{\sqrt{g(y)(1+\epsilon+1/(\alpha^2N') }}-2\exp(-\epsilon^2\alpha^4N')\frac{K}{\alpha\sqrt{(1+\epsilon)}}\\
& \stackrel{(b)}{\geq} \frac{K}{\sqrt{g(y)}}(1 - \epsilon-1/(\alpha^2N')) -2\exp(-\epsilon^2\alpha^4N')\frac{K}{\alpha}&\textrm{ (for large enough $N'$)}\\
& \geq \frac{K}{\sqrt{g(y)}}(1 - \epsilon-1/(\alpha^2(N-n))) -2\exp(-\epsilon^2\alpha^4(N-n))\frac{K}{\alpha}&\textrm{ (for large enough $N$)}\numberthis\label{eqn:upper}.
%& \geq \frac{K}{\sqrt{g(y)}}(1-\frac{\epsilon}{2}+h(\epsilon))-2\exp(-\epsilon^2\delta_0^4N)\frac{K}{\tau_0\sqrt{(1+\epsilon)}}.
\end{align*}
Here (a) follows from Hoeffding's inequality, and (b) is because $\frac{1}{\sqrt{1+a}} \geq 1 - a$ for every $a \in (0,1)$. The other inequalities result from the fact that $g(s)\geq \alpha^2$ and $|\cN_j|\leq n$. 
Taking $\epsilon = (N-n)^{-1/4}$, we obtain:
\begin{align*}
\textup{E}(e_j(y)) & \geq \frac{K}{\sqrt{g(s)}} - \textup{o}(N).
\end{align*}
Next, we also have,
\begin{align*}
\textup{E}(e_j(y))&\leq P(\bar{f}_j(y)-1/N' \in [g(y)(1-\epsilon), g(y)(1+\epsilon)])\frac{K}{\sqrt{g(y)(1-\epsilon)}} \\
&~~+ \textup{E}\left(\mathbf{1}_{\{\bar{f}_j(y)-1/N'\notin [g(y)(1-\epsilon), g(y)(1+\epsilon)] \}}\frac{K}{\sqrt{\bar{f}_j(y)}}\right)\\
& \stackrel{(a)}{\leq} \frac{K}{\sqrt{g(y)(1-\epsilon)}} + P\left(\mathbf{1}_{\bar{f}_j(y)-1/N'\notin [g(y)(1-\epsilon), g(y)(1+\epsilon)] }\right)K\sqrt{N'}\\
& \stackrel{(b)}{\leq} \frac{K}{\sqrt{g(y)(1-\epsilon)}} + 2K\sqrt{N'}\exp(-\epsilon^2g(y)^2N')\\
&\leq \frac{K}{\sqrt{g(y)(1-\epsilon)}} + 2K\sqrt{N'}\exp(-\epsilon^2\alpha^4N')\\
& \stackrel{(c)}{\leq}  \frac{K}{\sqrt{g(y)}}(1+\frac{\epsilon}{2}+ w(\epsilon))+ 2K\sqrt{N'}\exp(-\epsilon^2\alpha^4N')\\
&\leq  \frac{K}{\sqrt{g(y)}} + \frac{\epsilon K}{2\alpha}+ \frac{|w(\epsilon)|K}{\alpha}+ 2K\sqrt{N'}\exp(-\epsilon^2\alpha^4N')\\
&\leq  \frac{K}{\sqrt{g(y)}} + \frac{\epsilon K}{2\alpha}+ \frac{|w(\epsilon)|K}{\alpha}+ 2K\sqrt{N}\exp(-\epsilon^2\alpha^4(N-n))\numberthis\label{eqn:lower}.
\end{align*}
Here, (a) results from the fact that $\bar{f}_j(y)\geq 1/N'$ (because of the smoothing). (b) follows from Hoeffding's inequality, and (c) follows from the Taylor approximation of the function $1/\sqrt{1-\epsilon}$, where $w(\epsilon) =\textup{o}(\epsilon)$. Now choosing $\epsilon = (N-n)^{-1/4}$, we get:
\begin{align*}
\textup{E}(e_j(y)) & \leq \frac{K}{\sqrt{g(y)}} + \textup{o}(N).
\end{align*}

Thus, we finally have $|\textup{E}(e_j(y)) - \frac{K}{\sqrt{g(y)}}| \leq \sigma(N) = \textup{o}(N)$,
where $\sigma(N)\geq 0$ is a function of $N$ that depends only on $\alpha$, $n$ and $K$ and not on $y$ (note that our bounds explicitly define this function: we have $w(\epsilon)<\epsilon/2$ for $\epsilon<1/2$ and thus $w(\epsilon)$ can be replaced by $\epsilon/2$ for $N-n\geq 2^4 = 16$)). 
%\begin{equation}
%\textup{E}(r_j(s_k)) = \frac{K}{\sqrt{g(s_k)}} + m(N),
%\end{equation}
%where $m(N)= \textup{o}(1)$ and $m(N)$ depends only on $K$ and $\delta_0$, and not on the particular choice of a generating model in $\mathcal{C}_{hom}$.

Assuming everyone else is truthful, the expected reward of person $j$ for evaluating object $i$ if she chooses a reporting strategy $\bq^{ij}$ is,
\begin{align}
R(\bq^{ij})&\triangleq \sum_{y\in\cY}P(Y^i_{j'}=y,Y^i_{j}=y)\textup{E}(r_j(y)) = \sum_{y\in\cY}\textup{E}(r_j(y))\sum_{x\in \cX}P_X(x)p_{y}(x)\sum_{y'\in\cY}p_{y'}(x)q^{ij}_y(y').\label{eqn:linear}
\end{align}
Thus the agent solves $\max_{\bq^{ij}}R(\bq^{ij})$. The objective is linear in $\bq^{ij}$, and further, $\bq^{ij}(y)$ lies on a unit simplex for each $y\in\cY$. Thus the optimal reporting strategy chooses $\bq^{ij}(y)$ to be one of the extreme points of the simplex for each $y\in\cY$, i.e., the optimal reporting strategy is deterministic. Now let $\bt$ be the truthful strategy, i.e., $t_{y'}(y) = \mathbf{1}_{\{y=y'\}}$. Then for any deterministic reporting strategy $\bq^{ij}$, we have,
\begin{align}
R(\bq^{ij})&=\sum_{y\in\cY}\textup{E}(e_j(y))\sum_{y'\in\cY}q^{ij}_y(y')\sum_{x\in \cX}P_X(x)p_{y}(x)p_{y'}(x)\nonumber\\
&\stackrel{(a)}{\leq} \sum_{y\in\cY}\textup{E}(e_j(y))\sum_{y'\in\cY}q^{ij}_y(y')\bigg(\sqrt{\sum_{x\in \cX}P_X(x)p_{y}(x)^2}\sqrt{\sum_{x\in \cX}P_X(x)p_{y'}(x)^2}-\alpha\mathbf{1}_{\{y\neq y'\}}\bigg)\nonumber\\
&\leq \sum_{y\in\cY}(\frac{K}{\sqrt{g(y)}} + \sigma(N))\sum_{y'\in\cY}q^{ij}_y(y')\bigg(\sqrt{g(y)g(y')}-\alpha\mathbf{1}_{\{y\neq y'\}}\bigg)\nonumber\\
&\leq K\sum_{y'\in\cY}\sqrt{g(y')} - \alpha K \sum_{y'\in\cY}\sum_{y\in\cY}\mathbf{1}_{\{y\neq y'\}}q^{ij}_y(y') + \sigma(N) \sum_{y\in\cY}\sum_{y'\in\cY}q^{ij}_y(y')\sqrt{g(y)g(y')}\label{eq:strict}\\
&\stackrel{(b)}{\leq}  K\sum_{y'\in\cY}\sqrt{g(y')} - \alpha K \mathbf{1}_{\{\bq^{ij}\neq \bt\}} + |\cY|\sigma(N).
\end{align}
Here, $(a)$ follows from the Cauchy-Schwarz inequality, from the definition of $\delta(P_x,\bp)$, and the fact that $\delta(P_x,\bp)>\alpha$. (b) follows from the fact that $\bq^{ij}$ is deterministic and so is $\bt$. While we have,
\begin{align*}
R(\bt)&=\sum_{y\in\cY}\textup{E}(e_j(y))\sum_{x\in \cX}P_X(x)p_{y}(x)^2\\
&\geq \sum_{y\in\cY}(\frac{K}{\sqrt{g(y)}} - \sigma(N))g(y)\\
&\geq \sum_{y\in\cY}\sqrt{g(y)} - |\cY|\sigma(N).
\end{align*}
Thus we have,
$$R(\bq^{ij}) \leq  R(\bt)- \alpha K \mathbf{1}_{\{\bq^{ij}\neq \bt\}} + 2|\cY|\sigma(N)$$
% $r^i_j=y'$ when her true evaluation is $y$ is
%\begin{align*}
%R(y',y)&\triangleq P(Y^i_{j'}=y'|Y^i_{j}=y)\textup{E}(r_j(y')) =\frac{\sum_{x\in \cX}P_X(x)p_{y'}(x)p_{y}(x)}{\sum_{x\in\cX}P_X(x)p_y(x)}\textup{E}(r_j(y')).
%\end{align*}
%Similarly, we have $R(y,)=\frac{\sum_{h\in \mathcal{H}}P_X(h)p(s_k|h)^2}{\sum_{h\in \mathcal{H}}P_X(h)p(s_k|h)}\textup{E}(r_j(s_k))$. 
%Thus, lying is strictly worse for person $j$ if $R(y',y)<R(y,y)$, which is true if 
%\begin{align*}
%&\frac{\sum_{x\in \cX}P_X(x)p_{y'}(x)p_{y}(x)}{\sum_{x\in\cX}P_X(x)p_y(x)}\bigg(\frac{K}{\sqrt{\sum_{x\in\cX}P_X(x)p_{y'}(x)^2}} + \sigma(N)\bigg)\\5&< \frac{\sum_{x\in \cX}P_X(x)p_{y}(x)^2}{\sum_{x\in\cX}P_X(x)p_y(x)}\bigg(\frac{K}{\sqrt{\sum_{x\in \cX}P_X(x)p_{y}(x)^2}} - \sigma(N)\bigg).
%\end{align*}
%This is true if, 
%\begin{eqnarray*}
%\sum_{x\in\cX}P_X(x)p_y(x)p_{y'}(x) & < & \sqrt{(\sum_{x\in\cX}P_X(x)p_y(x)^2)(\sum_{x\in\cX}P_X(x)p_{y'}(x)^2)} -\frac{2\sigma(N)}{K}.
 %\end{eqnarray*}
%But we already have
%\begin{eqnarray*}
%	\sum_{x\in\cX}P_X(x)p_y(x)p_{y'}(x) & < & \sqrt{(\sum_{x\in\cX}P_X(x)p_y(x)^2)(\sum_{x\in\cX}P_X(x)p_{y'}(x)^2)} -\alpha.
%\end{eqnarray*} 
%since $\delta(P_X,\bp)>\alpha>0$. Thus being truthful gives a strictly better payoff if $\frac{2\sigma(N)}{K}<\alpha$. 
Since $\sigma(N)$ depends only on $\delta$ and $K$ and $\sigma(N)=\textup{o}(1)$, there is an $N_1$ that depends only on $\alpha$, $K$, $n$ and $|\cY|$ such that for all $N>N_1$, $2|\cY|\sigma(N)<K\alpha$, which means that truthful behavior is a strict Bayes-Nash equilibrium. To prove the second statement, suppose that $\bq^{ij}$ is a strategy in which reports are chosen independently of the true answers. Denote $q^{ij}_y \triangleq q^{ij}_y(y')$ since $q^{ij}_y(y')= q^{ij}_y(y'')$ for all $y,\,y',\,y''\in\cY$. Then in \eqref{eq:strict},
\begin{align*}
\alpha K \sum_{y'\in\cY}\sum_{y\in\cY}\mathbf{1}_{\{y\neq y'\}}q^{ij}_y(y') &=\alpha K \sum_{y'\in\cY}\sum_{y\in\cY}\mathbf{1}_{\{y\neq y'\}}q^{ij}_y\\
&=\alpha K(|\cY|-1).
\end{align*} 
And thus, 
$$R(\bq^{ij}) \leq  R(\bt)- \alpha K (|\cY|-1) + 2|\cY|\sigma(N).$$
Thus for any $\omega \in(0,\alpha K (|\cY|-1))$, there is a positive integer $N_2$ depending on $\omega$, $\alpha$, $K$, $n$ and $|\cY|$ such that for any $N>N_2$, $R(\bq^{ij}) \leq  R(\bt) -\omega$. Choosing $N_0 = \max(N_1, N_2)$ proves the result.
\end{proof}
\bigskip
\begin{proof}{Proof of Proposition~\ref{prop:monotone}}
We have,
\begin{align*}
\Gamma(Z_1,Z_2)&= \sum_{y\in \cY}\sqrt{\sum_{x\in\cX,y_1\in \cY, y_2\in\cY} P_X(x)p_{y_1}(x)p_{y_2}(x)q_y(y_1)q_y(y_2)}\\
&= \sum_{y\in \cY}\sqrt{\sum_{y_1\in \cY, y_2\in\cY}q_y(y_1)q_y(y_2)\sum_{x\in\cX} P_X(x)p_{y_1}(x)p_{y_2}(x)}\\
&\stackrel{(a)}{\leq} \sum_{y\in \cY}\sqrt{\sum_{y_1\in \cY, y_2\in\cY}q_y(y_1)q_y(y_2)\bigg(\sqrt{\sum_{x\in\cX} P_X(x)p_{y_1}(x)^2}\sqrt{\sum_{x\in\cX} P_X(x)p_{y_2}(x)^2} -\delta(P_X,\bp)\mathbf{1}_{\{y_1\neq y_2\}}\bigg)}\\
&=\sum_{y\in \cY}\sqrt{\bigg(\sum_{y_1\in \cY}q_y(y_1)\sqrt{\sum_{x\in\cX} P_X(x)p_{y_1}(x)^2}\bigg)^2 - \delta(P_X,\bp)\sum_{y_1\in \cY,y_2\in\cY}q_y(y_1)q_y(y_2)\mathbf{1}_{\{y_1\neq y_2\}}}\\
&\stackrel{(b)}{\leq} \sum_{y\in \cY,y_1\in \cY}q_y(y_1)\sqrt{\sum_{x\in\cX} P_X(x)p_{y_1}(x)^2} - \frac{\delta(P_X,\bp)}{2}\sum_{y\in \cY}\frac{\bigg(\sum_{y'\in \cY,y''\in\cY}q_y(y')q_y(y'')\mathbf{1}_{\{y'\neq y''\}}\bigg)}{\sum_{y_1\in \cY}q_y(y_1)\sqrt{\sum_{x\in\cX} P_X(x)p_{y_1}(x)^2}}\\
&\stackrel{(c)}{\leq} \Gamma(Y_1,Y_2) - \frac{\delta(P_X,\bp)}{2}\sum_{y\in \cY}\frac{\bigg(\sum_{y'\in \cY,y''\in\cY}q_y(y')q_y(y'')\mathbf{1}_{\{y'\neq y''\}}\bigg)}{\Gamma(Y_1,Y_2)}\\
&\stackrel{(d)}{\leq} \Gamma(Y_1,Y_2) - \frac{\delta(P_X,\bp)}{2\sqrt{|\cY|}}\sum_{y\in \cY}\sum_{y'\in \cY,y''\in\cY}q_y(y')q_y(y'')\mathbf{1}_{\{y'\neq y''\}}\\
&\stackrel{(e)}{\leq}\Gamma(Y_1,Y_2) - \frac{\delta(P_X,\bp)\Omega(\bq)^2(|\cY|-1)}{2\sqrt{|\cY|}}.
\end{align*}
Here, (a) follows from the Cauchy-Schwarz inequality and the definition of $\delta(P_X,\bp)$. (b) follows from the fact that for $a,\,b>0$ and $a>b$, $\sqrt{a-b}\leq \sqrt{a}-b/(2\sqrt{a})$. $(c)$ follows from the fact that $q_y(y_1)\leq 1$ and from the definition of $\Gamma(Y_1,Y_2)$. (d) follows from the fact that $\Gamma(Y_1,Y_2)\leq |\cY|$. (e) holds since, by Jensen's inequality,
\begin{align*}
\Omega(\bq)^2 &= \bigg(\frac{|\cY|}{|\cY|^2(|\cY|-1)}\sum_{y\in \cY}\sum_{y'\in \cY,y''\in\cY}\sqrt{q_y(y')q_y(y'')\mathbf{1}_{\{y'\neq y''\}}}\bigg)^2\\
&\leq \frac{1}{|\cY|-1}\sum_{y\in \cY}\sum_{y'\in \cY,y''\in\cY}q_y(y')q_y(y'')\mathbf{1}_{\{y'\neq y''\}}
\end{align*} \hfill $\Box$
\end{proof}
\bigskip
\begin{proof}{Proof of Theorem~\ref{thm:sym}}
The first statement follows from Theorem~\ref{thm:main}: there is an $N_1$ such that for all $N\geq N_1$, the truthful strategy profile is a Bayes-Nash equilibrium. We focus on the second claim. With some abuse of notation, we denote $e^t_j(y)$ to be the agreement scores for an agent $j$ under the truthful equilibrium, and $e^s_j(y)$ to be the scores under a fixed symmetric equilibrium strategy profile where each agent follows the reporting strategy $\bq$. 

We have shown in the proof of Theorem~\ref{thm:main} that if everyone is truthful, then $|\textup{E}(e^t_j(y)) - \frac{K}{\sqrt{g(y)}}| \leq \sigma(N) = \textup{o}(1)$, where $\sigma(N)\geq 0$ is some function of $N$ that depends only on $\alpha$, $n$ and $K$ and not on $y$. 

Let us denote $\sum_{x\in\cX}P_X(x)(\sum_{y'\in\cY}p_{y'}(x)q_{y}(y'))^2\triangleq s(y)$ and denote  $\sum_{x\in\cX}P_X(x)\sum_{y'\in\cY}p_{y'}(x)q_{y}(y')\triangleq b(y)$. %And denote $\min_{y;\, s(y)>0} s(y) = \eta$. 
By Jensen's inequality, we have $s(y)\geq b(y)^2$. Then using arguments similar to the ones leading up to~\eqref{eqn:lower} in the proof of Theorem~\ref{thm:main}, we can show that for all $y\in\cY$ such that $b(y)\geq\delta(P_X,\bp)^2/|\cY|$ (and hence, $s(y)\geq  \delta(P_X,\bp)^4/|\cY|^2$), and for any $\epsilon \in (0,1)$,
\begin{align*}
\bigg|\textup{E}(e^s_j(y)) - \frac{K}{\sqrt{s(y)}} \bigg| &\leq \sigma'(N),
\end{align*}
where $|\sigma'(N)|=\textup{o}(1)$, and it depends on $\alpha$, $K$, $n$ and $|\cY|$. 
Consider the strategy $\bq$ and consider a $y\in\cY$, such that $b(y)>0$ but $b(y)<\delta(P_X,\bp)^2/|\cY|$. Then one can construct another strategy $\bq'$ such that a) a fixed agent $j$ is indifferent between choosing $\bq$ and $\bq'$ assuming everyone else is playing $\bq$, and, 2) for all $y$ such that $b(y)< \delta(P_X,\bp)^2/|\cY|$, $q'_y(y') = 0$ for all $y'\in\cY$. To show this, observe that for each $y'$, $\bq(y')$ cannot have support only on those $y$ for which $b(y)< \delta(P_X,\bp)^2/|\cY|$. This is because if that is the case then $P(Y^i_j= y') = P(Y^i_j= y')\sum_{y\in\cY; b(y)< \delta(P_X,\bp)^2/|\cY|}q_y(y') \leq \sum_{y\in\cY; b(y)< \delta(P_X,\bp)^2/|\cY|}b(y) <\delta(P_X,\bp)^2$, which contradicts the fact that $P(Y^i_j= y')\geq\delta(P_X,\bp)^2$ as we have seen in the proof of Proposition~\ref{prop:equi}. So then define $\bq'(y')$ to have support only on the $y\in\cY$ for which $b(y)\geq \delta(P_X,\bp)^2/|\cY|$ by transferring the probability masses. If we define $G(\bq)$ to be the expected payment to a fixed agent $j$ for a fixed task $i$ under the symmetric equilibrium under strategy $\bq$, and define $G(\bq', \bq^{-j})$ to be the expected payment to $j$ if she plays $\bq'$ while others play $\bq$, then we have $G(\bq) =G(\bq',\bq^{-j})$. Let us define $\sum_{x\in\cX}P_X(x)(\sum_{y'\in\cY}p_{y'}(x)q'_{y}(y'))^2\triangleq s'(y)$.
Then we have,
\begin{align*}
G(\bq)&= G(\bq',\bq^{-j}) \\
&\leq \sum_{y\in\cY;\,b(y)\geq \delta(P_X,\bp)^2/|\cY|} \textup{E}(e^s_j(y))\sum_{x\in\cX}P_X(x)[\sum_{y_1\in\cY}p_{y_1}(x)q'_y(y_1)][\sum_{y_2\in\cY}p_{y_2}(x)q_y(y_2)]\\
&\stackrel{(a)}{\leq} \sum_{y\in\cY;\,b(y)\geq \delta(P_X,\bp)^2/|\cY|} \textup{E}(e^s_j(y))\sqrt{s(y)s'(y)}\\
&\leq \sum_{y\in\cY;\,b(y)\geq \delta(P_X,\bp)^2/|\cY|} (\frac{K}{\sqrt{s(y)}} +\sigma'(N))\sqrt{s(y)s'(y)}\\
&\leq \sum_{y\in\cY;\,b(y)\geq \delta(P_X,\bp)^2/|\cY|} K\sqrt{s'(y)} +|\cY|\sigma'(N)\\
&\stackrel{(b)}{=} K\sum_{y\in\cY}\sqrt{s'(y)} +|\cY|\sigma'(N)\numberthis\label{eqn:final2}.
\end{align*}
Here (a) follows from the Cauchy-Schwarz inequality and (b) follows from the fact that $s'(y)=0$ for all $y$ such that $b(y)< \delta(P_X,\bp)^2/|\cY|$ by construction of the strategy $\bq'$.
Let $G(\bt)$ be the expected payment to agent $j$ for task $i$ under the truthful equilibrium.  Let $j'$ be $j$'s peer for task $i$. Then we have,
\begin{align*}
G(\bt)&= \sum_{y\in\cY} \textup{E}(e^t_j(y))g(y) \\
&\geq \sum_{y\in\cY} K\sqrt{g(y)} - \sum_{y\in\cY}\sigma(N)g(y)\\
&\geq K\Gamma(Y^i_j,Y^i_{j'}) -|\cY|\sigma(N)\\
&\stackrel{(a)}{\geq} K\sum_{y\in\cY}\sqrt{s'(y)} -|\cY|\sigma(N)\numberthis\label{eqn:final1}.
\end{align*}
Here, (a) follows from Proposition~\ref{prop:monotone}. 
Finally,~\eqref{eqn:final1} and~\eqref{eqn:final2} together imply that, for a large enough $N$,
\begin{align*}
G(\bt)&\geq G(\bq)- |\cY|(\sigma(N)+\sigma'(N)).
\end{align*}
Thus for any $\omega>0$, there exists some $N_2$ such that for any $N\geq N_2$, the payoff under the truthful equilibrium is less than that under any other symmetric strategy profile by at most $\omega$. Taking $N_0 = \max(N_1,N_2)$ proves our claim. \hfill $\Box$
 %$|\textup{E}(e^s_j(y)) - \frac{K}{\sqrt{s(y)}}| \leq \gamma(N) = \textup{o}(N)$, where $\gamma(N)\geq 0$ is some function of $N$ that depends only on $\eta$, $n$ and $K$ and not on $y$.
\end{proof}
\bigskip
\begin{proof}{Proof of Theorem~\ref{thm:sym2}}
 As before, we denote $e^t_j(y)$ to be the agreement scores for an agent $j$ under a fully informative  equilibrium, and $e^s_j(y)$ to be the scores under a fixed symmetric strategy profile where each agent follows the reporting strategy $\bq$. We denote $\sum_{x\in\cX}P_X(x)(\sum_{y'\in\cY}p_{y'}(x)q_{y}(y'))^2\triangleq s(y)$ and denote  $\sum_{x\in\cX}P_X(x)\sum_{y'\in\cY}p_{y'}(x)q_{y}(y')\triangleq b(y)$. By our assumption, $b(y)\geq \eta$ if $b(y)\neq 0$, and since $s(y)\geq b(y)^2$, we have $s(y)\geq \eta^2$ if $b(y) \neq 0$. Then using arguments similar to the ones leading up to~\eqref{eqn:lower} in the proof of Theorem~\ref{thm:main}, we can show that for all $y\in\cY$,
$|\textup{E}(e^t_j(y)) - \frac{K}{\sqrt{g(y)}}| \leq \sigma(N) = \textup{o}(1)$, and for all $y\in \cY$ such that $b(y)\neq 0$,  $|\textup{E}(e^s_j(y)) - \frac{K}{\sqrt{s(y)}}| \leq \sigma'(N) = \textup{o}(1)$, where $\sigma(N)\geq 0$ is some function of $N$ that depends only on $\alpha$, $n$ and $K$, and $\sigma'(N)\geq 0$ is some function of $N$ that depends only on $\alpha$, $\eta$, $n$ and $K$. Neither of these functions depend on $y$.
Let $G(\bt)$ and $G(\bq)$ be the expected payments to agent $j$ for task $i$ under the truthful strategy profile and the symmetric profile $\bq$, respectively.  Let $j'$ be $j$'s peer for task $i$. Let $Z^i_j$ and $Z^i_{j'}$ be the reported answers of $j$ and $j'$ for task $i$ under $\bq$. Then we have,
\begin{align*}
G(\bq)&= \sum_{y\in\cY} \textup{E}(e^s_j(y))s(y) \\
&\overset{(a)}{\leq} \sum_{y\in\cY} K\sqrt{s(y)} + \sum_{y\in \cY}s(y)\sigma'(N))\\
&\leq K\Gamma(Z^i_j,Z^i_{j'})+ |\cY|\sigma'(N).\numberthis\label{eqn:fin}
\end{align*}
Here, (a) follows from the fact that if $b(y) = 0$, then $s(y) = 0$ and moreover, for any $y$ such that $b(y) \neq 0$, we have $|\textup{E}(e^s_j(y)) - \frac{K}{\sqrt{s(y)}}| \leq \sigma'(N)$ from above. Similarly, we can show that 
\begin{align*}
G(\bt)&= \sum_{y\in\cY} \textup{E}(e^s_t(y))g(y) \\
&\geq \sum_{y\in\cY} K\sqrt{g(y)} - \sum_{y\in \cY}g(y)\sigma(N)\\
&\geq K\Gamma(Y^i_j,Y^i_{j'})- |\cY|\sigma(N)\\
&\geq K\Gamma(Z^i_j,Z^i_{j'}) + \frac{K\delta(P_X,\bp)\Omega(\bq)^2(|\cY|-1)}{2\sqrt{|\cY|}}- |\cY|\sigma(N).\numberthis\label{eqn:fin2}
\end{align*}
Thus if $G(\bq)\geq G(\bt)$ for any strategy $\bq$, then this implies that,
$$K\Gamma(Z^i_j,Z^i_{j'}) + \frac{K\delta(P_X,\bp)\Omega(\bq)^2(|\cY|-1)}{2\sqrt{|\cY|}} -|\cY|\sigma(N)\leq  K\Gamma(Z^i_j,Z^i_{j'})+ |\cY|\sigma'(N),$$
which implies that
$$\frac{K\delta(P_X,\bp)\Omega(\bq)^2(|\cY|-1)}{2\sqrt{|\cY|}}\leq |\cY|(\sigma(N)+\sigma'(N)),$$
or that,
\begin{equation}
\Omega(\bq) \leq \sqrt{\frac{2|\cY|^{3/2}(\sigma(N)+\sigma'(N))}{K\delta(P_X,\bp)(|\cY|-1)}}<\sqrt{\frac{2|\cY|^{3/2}(\sigma(N)+\sigma'(N))}{K\alpha(|\cY|-1)}}.
\end{equation}
Now the quantity on the right is $\textup{o}(1)$ (depending only on $\alpha$, $\eta$, $n$, $|\cY|$, and $K$). Thus for any $\omega>0$ and $\eta>0$, there exists some $N_0$ such that for any $N\geq N_0$, any symmetric strategy profile in which  the probability of reporting any answer $y\in \cY$ is either $0$ or at least $\eta$, and that gives a higher expected payoff to each agent than the truthful strategy profile, is at most $\omega-$uninformative. Since truthful reporting is a Bayes-Nash equilibrium for a large enough $N$, this implies the result.  \hfill $\Box$\\

\end{proof}

\begin{proof}{Proof of Theorem~\ref{thm:new}.}
We will use the following notion for the proof.
\begin{definition}
For any strategy profile $(\bq^j)_{j\in\cM}$ across agents, the average reporting strategy excluding the set of agents $\mathcal{J}$ is defined as $$\bar{\bq}^{-\mathcal{J}}(y) =\frac{1}{M-|\mathcal{J}|}\sum_{j'\in \mathcal{M}\setminus \mathcal{J}}\bq^{j'}(y).$$
\end{definition}
%Observe that 
%\begin{align}
%\bar{\bq}(y) = \frac{M-|\mathcal{J}|}{M}\bar{\bq}^{-\mathcal{J}}(y)+\frac{1}{M}\sum_{j'\in\mathcal{J}}\bq^j(y)
%\end{align}
Manipulating this definition, we have,
\begin{align}
\bar{\bq}^{-\mathcal{J}}(y) = \frac{M}{M-|\mathcal{J}|}\left(\bar{\bq}(y) - \frac{1}{M}\sum_{j'\in\mathcal{J}}\bq^j(y) \right).
\end{align}
We then directly have that
$$\bar{\bq}^{-\mathcal{J}}(y)\geq \bar{\bq}(y) - \frac{|\mathcal{J}|}{M}.$$
Next, since $1/(1-\frac{|\mathcal{J}|}{M})$ is $1+ \frac{|\mathcal{J}|}{M} + \textup{o}(\frac{|\mathcal{J}|}{M})$ as $M\rightarrow \infty$ (from the Taylor series expansion), we can conclude that there exists some $\kappa>1$ such that for any $M$ large enough, we have 
\begin{align}
 \bar{\bq}^{-\mathcal{J}}(y) \leq \bar{\bq}(y) + \frac{\kappa|\mathcal{J}|}{M}.
 \end{align}
To summarize, for some $\kappa>1$ and any $M$ large enough, we thus have 
\begin{align}
 \bar{\bq}(y) - \frac{\kappa|\mathcal{J}|}{M}\leq \bar{\bq}^{-\mathcal{J}}(y) \leq \bar{\bq}(y) + \frac{\kappa|\mathcal{J}|}{M}.\label{eqn:mainbd}
\end{align}
We now present the proof of Theorem~\ref{thm:new}. 
Let $j$ be a fixed agent evaluating a fixed task $i$. Let $J$ be her (random) peer on task $i$. Now upon observing $y$, her expected reward on reporting any $y'$ such that the probability of reporting $y'$ is $0$ under the population average strategy is $0$ (since there is no hope of matching $y'$ with any peer).  We thus focus on only those $y'\in\cY$ such that their reporting probability is at least $\eta$. For any such $y'$, $j$'s expected reward on reporting $y'$ when she observed $y$ is given by, 
\begin{align}
G^j_i(y,y') &= \textup{E}\left[\mathbbm{1}_{r^i_{J}=y'}\frac{K\sqrt{N -|\cW_j|}}{\sqrt{1+\sum_{i'\in \mathcal{N}\setminus \cW_j}  \mathbbm{1}_{r^{i'}_{J_1(i')}=y'}\mathbbm{1}_{r^{i'}_{J_2(i')}=y'}}}\mid Y^i_j = y\right]\\
 &\overset{(a)}{=} \underbrace{\textup{E}\left(\frac{K\sqrt{N -|\cW_j|}}{\sqrt{1+\sum_{i'\in \mathcal{N}\setminus \cW_j}  \mathbbm{1}_{r^{i'}_{J_1(i')}=y'}\mathbbm{1}_{r^{i'}_{J_2(i')}=y'}}}\right)}_{A \textrm{ (expected reward from matching $y'$)}} \underbrace{\textup{E}\left[\mathbbm{1}_{r^i_{J}=y'}\mid Y^i_j = y\right]}_{B \textrm{ (probability of matching $y'$)}}.\label{eqn:decomp}
\end{align}
Here, $J_1(i')$ and $J_2(i')$ are the (random) agents who have evaluated task $i'$, chosen to compute the agreement rewards for $j$. $(a)$ results from the fact that the agreement rewards are independent of $Y^i_j$ and $Y^i_{J}$: the former because of the fact that the agreement rewards only depend on the tasks that $j$ does not perform, and the latter because of the random task allocation policy ($Y^i_{J}$ may contain information about $J$, but that doesn't give any information about agents who will be utilized in computing the agreement rewards since the agent allocation to each task is i.i.d.). %We next focus on approximating terms A and B independently.

We first focus on term A in Equation~\ref{eqn:decomp}, which is the expected reward for matching on the answer $y'$.  Note that the random variables $\mathbbm{1}_{r^{i'}_{J_1(i')}=y'}\mathbbm{1}_{r^{i'}_{J_2(i')}=y'}$ across $i'$ are i.i.d. owing to our random task allocation policy with 
$\textup{E}(\mathbbm{1}_{r^{i'}_{j_1(i')}=y'}\mathbbm{1}_{r^{i'}_{j_2(i')}=y'})$ defined as follows (for notational simplicity we drop the dependence on $i'$).
\begin{align}
\textup{E}(\mathbbm{1}_{r_{J_1}=y'}\mathbbm{1}_{r_{J_2}=y'})&=\textup{E}\left[\sum_{x\in\cX, y_1, y_2\in \cY }P_X(x) p_{y_1}(x)p_{y_2}(x)q^{J_1}_{y'}(y_1)q^{J_2}_{y'}(y_2)\right].
\end{align}
Here, the latter expectation is over the random choice of $J_1$ and $J_2$. 
Now we have for a large enough $M$, 
\begin{align}
&\textup{E}\left[\sum_{x\in\cX, y_1, y_2\in \cY }P_X(x) p_{y_1}(x)p_{y_2}(x)q^{J_1}_{y'}(y_1)q^{J_2}_{y'}(y_2)\right]\notag\\
&~~ \overset{(a)}{=} \textup{E}\left[\sum_{x\in\cX, y_1, y_2\in \cY }P_X(x) p_{y_1}(x)p_{y_2}(x)q^{J_1}_{y'}(y_1)\bar{q}^{-\{J_1,j\}}_{y'}(y_2)\right]\notag\\
&~~\overset{(b)}{\leq} \textup{E}\left[\sum_{x\in\cX, y_1, y_2\in \cY }P_X(x) p_{y_1}(x)p_{y_2}(x)q^{J_1}_{y'}(y_1)\left(\bar{q}_{y'}(y_2) + 2\kappa/M\right)\right]\notag\\
&~~\overset{(c)}{\leq} \textup{E}\left[\sum_{x\in\cX, y_1, y_2\in \cY }P_X(x) p_{y_1}(x)p_{y_2}(x)q^{J_1}_{y'}(y_1)\bar{q}_{y'}(y_2)\right]+ 2\kappa/M\notag\\
&~~\overset{(d)}{\leq} \sum_{x\in\cX, y_1, y_2\in \cY }P_X(x) p_{y_1}(x)p_{y_2}(x)\bar{q}^{-j}_{y'}(y_1)\bar{q}_{y'}(y_2)+ 2\kappa/M\notag\\
&~~\overset{(e)}{\leq} \sum_{x\in\cX, y_1, y_2\in \cY }P_X(x) p_{y_1}(x)p_{y_2}(x)(\bar{q}_{y'}(y_1)+\kappa/M)\bar{q}_{y'}(y_2)+ 2\kappa/M\notag\\
&~~\leq \sum_{x\in\cX, y_1, y_2\in \cY }P_X(x) p_{y_1}(x)p_{y_2}(x)\bar{q}_{y'}(y_1)\bar{q}_{y'}(y_2)+ 3\kappa/M.
\end{align}
Here, (a) follows from the fact that, $J_2$ is equally likely to be any of the remaining agents other than $J_1$ and $j$, again by the random task allocation policy. (b) follows from Equation \ref{eqn:mainbd}. (c) follows from the fact that the coefficient of $2\kappa/M$ after the expansion is at most $1$. (d) follows from taking expectation over $J_1$, who is equally likely to be any agent other than $j$. Finally, (e) again follows from Equation \ref{eqn:mainbd}. Similarly, we have 
\begin{align}
\textup{E}\left[\sum_{x\in\cX, y_1, y_2\in \cY }P_X(x) p_{y_1}(x)p_{y_2}(x)q^{J_1}_{y'}(y_1)q^{J_2}_{y'}(y_2)\right] & \geq \sum_{x\in\cX, y_1, y_2\in \cY }P_X(x) p_{y_1}(x)p_{y_2}(x)\bar{q}_{y'}(y_1)\bar{q}_{y'}(y_2)- 3\kappa/M
\end{align}
for any large enough $M$. 
Let $\textup{E}(\mathbbm{1}_{r^{i'}_{J_1(i')}=y}\mathbbm{1}_{r^{i'}_{J_2(i')}=y})$ be denoted as $h(y)$, and define 
$$s(y) = \sum_{x\in\cX, y_1, y_2\in \cY }P_X(x) p_{y_1}(x)p_{y_2}(x)\bar{q}_{y}(y_1)\bar{q}_{y}(y_2).$$ 
We have then concluded that $|h(y') -s(y')|\leq 3\kappa/M$ for each $y'\in\cY$. We also have that $s(y)\overset{(a)}{\geq} \left(\sum_{x\in\cX, y'\in\cY }P_X(x) p_{y'}(x)\bar{q}_{y}\right)^2\geq \eta^2$, where (a) follows from Jensen's inequality applied to the function $f(x) = x^2$.
Now by the multiplicative Hoeffding's inequality, for any $\epsilon>0$, we have,
\begin{align}
P\left(\sum_{i'\in \mathcal{N}\setminus \cW_j}  \mathbbm{1}_{r^{i'}_{J_1(i')}=y'}\mathbbm{1}_{r^{i'}_{J_2(i')}=y'}\geq (N - |\cW_j|)h(y')(1+\epsilon)\right)&\leq \exp(-\epsilon^2h(y')/3), \textrm{ and,}\label{eqn:hoeffding1}\\
P\left(\sum_{i'\in \mathcal{N}\setminus \cW_j}  \mathbbm{1}_{r^{i'}_{J_1(i')}=y'}\mathbbm{1}_{r^{i'}_{J_2(i')}=y'}\leq (N - |\cW_j|)h(y')(1-\epsilon)\right)&\leq \exp(-\epsilon^2h(y')/3).\label{eqn:hoeffding2} 
\end{align}
Thus, for any $\epsilon>0$, and $N$ large enough, we have,
\begin{align}
&\textup{E}\left(\frac{\sqrt{N -|\cW_j|}}{\sqrt{1+\sum_{i'\in \mathcal{N}\setminus \cW_j}  \mathbbm{1}_{r^{i'}_{J_1(i')}=y'}\mathbbm{1}_{r^{i'}_{J_2(i')}=y'}}}\right)\nonumber\\
&~~\overset{(a)}{\leq} \textup{E}\left(\frac{\sqrt{N - |\cW_j|}}{\sqrt{1+(N - |\cW_j|) h(y') (1-\epsilon)}} + \exp(-\epsilon^2h(y')(N -|\cW_j|)/3)\sqrt{N -|\cW_j|}\right)\nonumber\\
&~~\overset{(b)}{\leq}  \textup{E}\left(\frac{\sqrt{N - |\cW_j|}}{\sqrt{1+(N - |\cW_j|) (s(y')-\frac{3\kappa}{M}) (1-\epsilon)}} + \exp(-\frac{\epsilon^2(s(y')-3\kappa/M)(N -|\cW_j|)}{3})\sqrt{N -|\cW_j|}\right)\nonumber\\
&~~\overset{(c)}{\leq} \textup{E}\left(\frac{\sqrt{N - |\cW_j|}}{\sqrt{1+(N - |\cW_j|) (s(y')-\frac{3\kappa n}{mN}) (1-\epsilon)}}\right) + \textup{E}\left(\exp(-\frac{\epsilon^2(\eta^2-\frac{3\kappa n}{mN})(N -|\cW_j|)}{3})\sqrt{N -|\cW_j|}\right)\nonumber\\
&~~\overset{(d)}{\leq} \frac{1}{\sqrt{(s(y')-\frac{3\kappa n}{mN}) (1-\epsilon)}} + \textup{E}\left(\exp(-\frac{\epsilon^2(\eta^2-\frac{3\kappa n}{mN})(N -|\cW_j|)}{3})\right)\sqrt{N}\nonumber\\
&~~= \frac{1}{\sqrt{(s(y')-\frac{3\kappa n}{mN}) (1-\epsilon)}} +\sqrt{N}\exp(-\frac{\epsilon^2(\eta^2-\frac{3\kappa n}{mN})N}{3}) \textup{E}\left(\exp(\frac{\epsilon^2(\eta^2-\frac{3\kappa n}{mN})|\cW_j|}{3})\right).\label{eqn:inter}
\end{align}
Here, (a) results from Equation~\ref{eqn:hoeffding2} and the fact that in the worst case, the left hand side is at most $\sqrt{N-|\cW_j|}$. (b) results from the fact that $|h(y') -s(y')|\leq 3\kappa/M$, and (c) results from the fact that $s(y)\geq \eta^2$ and $M>mN/n$. All the expectations are with respect to the randomness in $|\cW_j|$.  (d) follows from (i) noting that $s(y')\geq \eta^2> \frac{3\kappa n}{mN}$ for $N$ large enough, (ii) ignoring the constant 1 in the denominator of the first term, and (iii) ignoring $|\cW_j|\geq 0$ in the second term.

Now, due to the randomized task allocation policy, $|\cW_j|$ is distributed as Binomial$(N, m/N)$. Since the moment generating function of a Binomially distributed random variable $X$ with parameters $(n,p)$ is $\textup{E}(\exp(Xt)) = (1-p +pe^t)^n$, we have that 
\begin{align}
\textup{E}\left(\exp(\frac{\epsilon^2(\eta^2-\frac{3\kappa n}{mN})|\cW_j|}{3})\right)= \left(1-\frac{m}{N} + \frac{m}{N}\exp(\frac{\epsilon^2(\eta^2-\frac{3\kappa n}{mN})}{3})\right)^N.\label{eqn:willuse}
\end{align}
Choosing $\epsilon = N^{-1/4}$, we have that 
\begin{align} 
\lim_{N\rightarrow\infty} \left(1-\frac{m}{N} + \frac{m}{N}\exp(\frac{(\eta^2-\frac{3\kappa n}{mN})}{3\sqrt{N}})\right)^N= \lim_{N\rightarrow\infty} \left(1 + \frac{m}{N}(\exp(\frac{(\eta^2-\frac{3\kappa n}{mN})}{3\sqrt{N}})-1)\right)^N\leq  \lim_{N\rightarrow\infty} \left(1 + \frac{m}{N}\right)^N = \exp(m).\label{eqn:inter2}
\end{align}
Thus, choosing $\epsilon = N^{-1/4}$ in Equation~\ref{eqn:inter}, and combining Equation~\ref{eqn:inter2} with the fact that 
$$\lim_{N\rightarrow\infty} \sqrt{N}\exp(-\frac{(\eta^2-\frac{3\kappa n}{mN})\sqrt{N}}{3})= 0,$$
 we have that 
%\begin{align}
%C_N\overset{a.s.}{\to}\frac{1}{\sqrt{s(y')}} \quad\textrm{and}\quad  D_N\overset{a.s.}{\to} 0.
%\end{align}
%Hence, by Lebesgue's dominated convergence theorem,
\begin{align}
\lim_{N\rightarrow \infty} \textup{E}\left(\frac{\sqrt{N -|\cW_j|}}{\sqrt{1+\sum_{i'\in \mathcal{N}\setminus \cW_j}  \mathbbm{1}_{r^{i'}_{J_1(i')}=y'}\mathbbm{1}_{r^{i'}_{J_2(i')}=y'}}}\right) \leq \frac{1}{\sqrt{s(y')}}. \label{eqn:agreeub}
\end{align}
Next, we also have that, for a large enough $N$,
\begin{align}
&\textup{E}\left(\frac{\sqrt{N -|\cW_j|}}{\sqrt{1+\sum_{i'\in \mathcal{N}\setminus \cW_j}  \mathbbm{1}_{r^{i'}_{J_1(i')}=y'}\mathbbm{1}_{r^{i'}_{J_2(i')}=y'}}}\right)\nonumber\\
&~~\overset{(a)}{\geq} \textup{E}\left( \frac{\sqrt{N -|\cW_j|}}{\sqrt{1+ (N -|\cW_j|)h(y') (1+\epsilon)}}(1- \exp(-\epsilon^2h(y')(N -|\cW_j|)/3))\right)\nonumber\\
&~~\overset{(b)}{\geq}  \textup{E}\left(\frac{\sqrt{N -|\cW_j|}}{\sqrt{1+ N (s(y')+\frac{3\kappa}{M}) (1+\epsilon)}} - \exp(-\frac{\epsilon^2(s(y')-3\kappa/M)(N -|\cW_j|)}{3})\frac{\sqrt{N -|\cW_j|}}{\sqrt{1+ N (s(y') -\frac{3\kappa}{M}) (1+\epsilon)}}\right)\nonumber\\
&~~\overset{(c)}{\geq}  \textup{E}\left(\frac{\sqrt{N -|\cW_j|}}{\sqrt{1+ N (s(y')+\frac{3\kappa n}{mN}) (1+\epsilon)}}\right) - \textup{E}\left(\exp(-\frac{\epsilon^2(\eta^2-\frac{3\kappa n}{mN})(N -|\cW_j|)}{3})\frac{\sqrt{N -|\cW_j|}}{\sqrt{1+ N (\eta^2 -\frac{3\kappa n}{mN}) (1+\epsilon)}}\right)\nonumber\\
%&~~\overset{}{\geq}  \frac{\textup{E}\left(\sqrt{1 -\frac{|\cW_j|}{N}}\right)}{\sqrt{1/N+  (s(y')+\frac{3\kappa n}{mN}) (1+\epsilon)}} - \textup{E}\left(\exp(-\frac{\epsilon^2(\eta^2-\frac{3\kappa n}{mN})(N -|\cW_j|)}{3})\right)\frac{\sqrt{N}}{\sqrt{1+ N (\eta^2 -\frac{3\kappa n}{mN}) (1+\epsilon)}}\\
&~~\overset{}{\geq}  \frac{\textup{E}\left(\sqrt{1 -\frac{|\cW_j|}{N}}\right)}{\sqrt{1/N+  (s(y')+\frac{3\kappa n}{mN}) (1+\epsilon)}} - \textup{E}\left(\exp(\frac{\epsilon^2(\eta^2-\frac{3\kappa n}{mN})(|\cW_j|)}{3})\right)\exp(-\frac{\epsilon^2(\eta^2-\frac{3\kappa n}{mN})N}{3})\frac{\sqrt{N}}{\sqrt{1+ N (\eta^2 -\frac{3\kappa n}{mN}) (1+\epsilon)}}\nonumber\\
&~~\overset{(d)}{\geq}  \frac{\textup{E}\left(\sqrt{1 -\frac{|\cW_j|}{N}}\right)}{\sqrt{1/N+  (s(y')+\frac{3\kappa n}{mN}) (1+\epsilon)}} -  \left(1-\frac{m}{N} + \frac{m}{N}\exp(\frac{\epsilon^2(\eta^2-\frac{3\kappa n}{mN})}{3})\right)^N\exp(-\frac{\epsilon^2(\eta^2-\frac{3\kappa n}{mN})N}{3})\frac{\sqrt{N}}{\sqrt{1+ N (\eta^2 -\frac{3\kappa n}{mN}) (1+\epsilon)}}\label{eqn:tbd}.
\end{align}
Here, (a) follows from Equation~\ref{eqn:hoeffding1}, and the fact that the agreement rewards are always positive. (b) results from the fact that $|h(y') -s(y')|\leq 3\kappa/M$ and by ignoring the $|\cW_j|$ term in the denominator, and (c) results from the fact that $s(y)\geq \eta^2$ and $M>mN/n$. (d) follows from Equation~\ref{eqn:willuse}. We once again choose $\epsilon = N^{-1/4}$. Then, by Equation~\ref{eqn:inter2}, the second term in Equation~\ref{eqn:tbd} converges to $0$ as $N\rightarrow \infty$. We now focus on the first term. The denominator of this term clearly converges to $\sqrt{s(y')}$. It is now easy to show that the numerator converges to $1$. This is because $|\cW_j|$ is distributed as $\textup{Binomial}(N, m/N)$, and thus $|\cW_j|/N$ converges in distribution to  the constant $0$. Since $f(x) = \sqrt{1-x}$ is a bounded, continuous function on the domain $[0,1]$, it follows (by the Portmanteau's theorem on the equivalence of definitions of convergence in distribution) that $\textup{E}\left(\sqrt{1 -|\cW_j|/N}\right)$ converges to $1$ as $N\rightarrow\infty$.
%Then both the random variables $E_N$ and $F_N$ are bounded by a constant independent of $N$. Because $\frac{|W_j|}{N}\overset{a.s.}{\to}0$, we have that 
%\begin{align}
%E_N\overset{a.s.}{\to}\frac{1}{\sqrt{s(y')}} \quad\textrm{and}\quad  F_N\overset{a.s.}{\to} 0.
%\end{align}
Thus, we finally have,
\begin{align}
\lim_{N\rightarrow \infty} \textup{E}\left(\frac{\sqrt{N -|\cW_j|}}{\sqrt{1+\sum_{i'\in \mathcal{N}\setminus \cW_j}  \mathbbm{1}_{r^{i'}_{J_1(i')}=y'}\mathbbm{1}_{r^{i'}_{J_2(i')}=y'}}}\right) \geq \frac{1}{\sqrt{s(y')}}.\label{eqn:agreelb}
\end{align} 
Thus, from Equations~\ref{eqn:agreeub} and \ref{eqn:agreelb}, we finally have,
\begin{align}
\left| \textup{E}\left(\frac{K\sqrt{N -|\cW_j|}}{\sqrt{1+\sum_{i'\in \mathcal{N}\setminus \cW_j}  \mathbbm{1}_{r^{i'}_{J_1(i')}=y'}\mathbbm{1}_{r^{i'}_{J_2(i')}=y'}}}\right)-\frac{K}{\sqrt{s(y')}}\right| \leq \sigma(N). \label{eqn:agree}
\end{align}
where $\sigma(N) = \textup{o}(1)$. Now before we proceed, note that the convergence of the expected matching reward for answer $y$ to $K/\sqrt{s(y)}$ for each $y\in\cY$ is all that is required for strict truthfulness to follow for a large enough $N$, as we show in the proof of Theorem~\ref{thm:main}. For the truthful strategy profile, by the $\alpha$-separation assumption, we have that $s(y)\geq \alpha^2$ for all $y\in\cY$. Thus, by replacing $\eta$ with $\alpha$ in the arguments leading up to Equation~\ref{eqn:agree}, we can conclude the convergence of the matching rewards to $K/\sqrt{s(y)}$ for {\it each} $y\in\cY$. Thus, there exists $N_1$ such that for $N\geq N_1$, the truthful strategy profile is a Bayes-Nash equilibrium. We will not repeat the proof here for conciseness. The first statement of the theorem thus follows and we hence focus on proving the second and third statement.

To that effect, we now proceed to focus on term B in Equation~\ref{eqn:decomp}.  We have
\begin{align}
P(r^i_{J}=y',\, Y^i_j = y) &\overset{(a)}{=} \sum_{x\in\cX, y_2\in \cY }P_X(x) p_{y}(x)p_{y_2}(x)\bar{q}^{-\{j\}}_{y'}(y_2)\notag\\
&\overset{(b)}{\leq} \sum_{x\in\cX, y_2\in \cY }P_X(x) p_{y}(x)p_{y_2}(x)(\bar{q}_{y'}(y_2) + \kappa/M)\notag\\
&\leq \sum_{x\in\cX, y_2\in \cY }P_X(x) p_{y}(x)p_{y_2}(x)\bar{q}_{y'}(y_2) + \kappa/M.
\end{align}
Here, (a) follows from the fact that $J$ is equally likely to be any agent other than $j$, by the random task allocation policy. (b) again follows from Equation \ref{eqn:mainbd}. Similarly, we have
\begin{align}
P(r^i_{J}=y',\, Y^i_j = y) &\geq \sum_{x\in\cX, y_2\in \cY }P_X(x) p_{y}(x)p_{y_2}(x)\bar{q}_{y'}(y_2) - \kappa/M.
\end{align}
Thus, we have, for all $y'$ such that the population average probability of reporting is at least $\eta$, we have  
\begin{align}
P(Y^i_j = y)G^j_i(y,y') &= (\frac{K}{\sqrt{s(y')}}+ \textup{o}(1)) (\sum_{x\in\cX, y_2\in \cY }P_X(x) p_{y}(x)p_{y_2}(x)\bar{q}_{y'}(y_2) + \textup{o}(1))\notag\\
&= \frac{K(\sum_{x\in\cX, y_2\in \cY }P_X(x) p_{y}(x)p_{y_2}(x)\bar{q}_{y'}(y_2))}{\sqrt{s(y')}}+\textup{o}(1).\label{eqn:fb}
\end{align}
Here, we use the fact that $s(y)\geq \eta^2$. For every other $y'$ such that the population average probability of reporting is $0$, we have 
\begin{align}
P(Y^i_j = y)G^j_i(y,y') &= 0.
\end{align}
%Similarly, we have
%\begin{align}
%P(Y^i_j = y)G^j_i(y,y') &\geq  \frac{K(\sum_{x\in\cX, y_2\in \cY }P_X(x) p_{y}(x)p_{y_2}(x)\bar{q}_{y'}(y_2))}{\sqrt{s(y')}}+\textup{o}(1).
%\end{align}
Let $\cY'$ denote the set of responses such that the population average probability of reporting the response is at least $\eta$. Then the expected payoff of agent $j$ on task $i$ under policy  $\bq^j$ (fixing everyone else's policy) is:
\begin{align}
\mathbf{G}^j_i(\bq^j) &= \sum_{y,y'\in\cY} P(Y^i_j = y)G^j_i(y,y')q^j_{y'}(y)\\
&\overset{(a)}{=} \sum_{y\in\cY,\, y'\in\cY'} \frac{K(\sum_{x\in\cX, y_2\in \cY }P_X(x) p_{y}(x)p_{y_2}(x)\bar{q}_{y'}(y_2))}{\sqrt{s(y')}}q^j_{y'}(y) + \textup{o}(1)\label{eqn:tosum}\\
&\overset{(b)}{=}\max_{\bq} \sum_{y,y'\in\cY} P(Y^i_j = y)G^j_i(y,y')q^j_{y'}(y)\\
& \overset{(c)}{=}  \max_{\bq} \sum_{y\in\cY,\, y'\in\cY'} \frac{K(\sum_{x\in\cX, y_2\in \cY }P_X(x) p_{y}(x)p_{y_2}(x)\bar{q}_{y'}(y_2))}{\sqrt{s(y')}}q_{y'}(y)+\textup{o}(1).\label{eqn:out}
%&=  \frac{K(\sum_{x\in\cX, y_2, y, y'\in \cY }P_X(x) p_{y}(x)p_{y_2}(x)\bar{q}_{y'}(y_2))q_{y'}(y)}{\sqrt{s(y')}}+\textup{o}(1)
\end{align}
Here, (a) follows from Equation~\ref{eqn:fb}, (b) follows from the fact that $\bq^j$ is a best-response strategy, and (c) again follows from Equation~\ref{eqn:fb}.  Note that the final right hand side neither depends on the identity of agent $j$ nor does it depend on the identity of task $i$. It only depends on the population average strategy $\bar{\bq}$. Since, each policy $\bq^{j'}$ optimizes $\mathbf{G}^{j'}_i(\bq)$, we have that 
\begin{align}
\mathbf{G}^j_i(\bar{\bq}) &= \sum_{y,y'\in\cY} P(Y^i_j = y)G^j_i(y,y')\bar{q}_{y'}(y)\\
&\overset{(a)}{=} \sum_{y\in\cY,\, y'\in\cY'} \frac{K(\sum_{x\in\cX, y_2\in \cY }P_X(x) p_{y}(x)p_{y_2}(x)\bar{q}_{y'}(y_2))}{\sqrt{s(y')}}\bar{q}_{y'}(y) + \textup{o}(1)\\
& \overset{(b)}{=}  \max_{\bq} \sum_{y\in\cY,\, y'\in\cY'} \frac{K(\sum_{x\in\cX, y_2\in \cY}P_X(x) p_{y}(x)p_{y_2}(x)\bar{q}_{y'}(y_2))}{\sqrt{s(y')}}q_{y'}(y)+\textup{o}(1)\\
& = \mathbf{G}^j_i(\bq^j) +\textup{o}(1).
%&=  \frac{K(\sum_{x\in\cX, y_2, y, y'\in \cY }P_X(x) p_{y}(x)p_{y_2}(x)\bar{q}_{y'}(y_2))q_{y'}(y)}{\sqrt{s(y')}}+\textup{o}(1)
\end{align}
Here, (a) results from Equation \ref{eqn:fb}. (b) results from averaging the expression in Equation~\ref{eqn:tosum} and the expression in Equation~\ref{eqn:out} across all agents and realizing that the expression in Equation~\ref{eqn:out} is identical across the agents. Hence, we have that $|\mathbf{G}^j_i(\bq^j) - \mathbf{G}^j_i(\bar{\bq})| = \textup{o}(1)$. Hence, we finally have,
\begin{align}
\mathbf{G}^j_i(\bq^j) &=  \mathbf{G}^j_i(\bar{\bq})+ \textup{o}(1)\\
& = \sum_{y\in\cY,\, y'\in\cY'} \frac{K(\sum_{x\in\cX, y_2\in \cY }P_X(x) p_{y}(x)p_{y_2}(x)\bar{q}_{y'}(y_2))}{\sqrt{s(y')}}\bar{q}_{y'}(y) +\textup{o}(1)\\
& = \sum_{y'\in\cY'} \frac{K\sum_{x\in\cX, y_2\in \cY, y\in\cY }P_X(x) p_{y}(x)p_{y_2}(x)\bar{q}_{y'}(y_2)\bar{q}_{y'}(y) }{\sqrt{s(y')}}+\textup{o}(1)\\
& = \sum_{y'\in\cY'} \frac{K s(y')}{\sqrt{s(y')}}+\textup{o}(1)\\
& \overset{(a)}{=} \sum_{y'\in\cY} K\sqrt{s(y')}+\textup{o}(1)\label{eqn:fino}\\\
&\overset{(b)}{\leq} K\Gamma(Y^i_j,Y^i_{j'}) - \frac{K\delta(P_X,\bp)\Omega(\bar{\bq})^2(|\cY|-1)}{2\sqrt{|\cY|}} +\textup{o}(1)\\
& \overset{(c)}{=} \mathbf{G}(\bt) - \frac{K\delta(P_X,\bp)\Omega(\bar{\bq})^2(|\cY|-1)}{2\sqrt{|\cY|}} +\textup{o}(1)\label{eqn:abbas}
\end{align}
where $\mathbf{G}(\bt)$ is the expected reward to each agent for an evaluation task under the truthful strategy profile. Here, (a) follows from the fact that $s(y) = 0$ for all $y\in \cY\setminus\cY'$. (b) follows from Proposition~\ref{prop:monotone}, and (c) follows from the fact that $\mathbf{G}(\bt) = K \Gamma(Y^i_j,Y^i_{J})  + \textup{o}(1)$. This latter conclusion results from the fact that under the truthful strategy profile, because of $\alpha$-separation of the generating model, the probability of reporting any answer $y\in\cY$ is at least $\alpha^2$ (see proof of Proposition~\ref{prop:equi}). We can thus use the same arguments as that used for deriving the expression in Equation~\ref{eqn:fino} as the expected payoff of each agent, while replacing $\eta$ with $\alpha$. 

Now the second statement of the theorem immediately follows, since for any $\omega>0$, there is an $N_2$ such that for any $N>N_2$ we have that $\mathbf{G}^j_i(\bq^j)\leq \mathbf{G}(\bt) +\omega$. Moreover, we have that if $\mathbf{G}^j_i(\bq^j)> \mathbf{G}(\bt)$, then, Equation~\ref{eqn:abbas} allows us to conclude that 
\begin{align}
\frac{K\delta(P_X,\bp)\Omega(\bar{\bq})^2(|\cY|-1)}{2\sqrt{|\cY|}} &\leq \textup{o}(1), \textrm{ or,}
\end{align}
\begin{align}
\Omega(\bar{\bq}) \leq\sqrt{\frac{2\sqrt{|\cY|}\textup{o}(1)}{K\delta(P_X,\bp)(|\cY|-1)}}\leq \sqrt{\frac{2\sqrt{|\cY|}\textup{o}(1)}{K\alpha(|\cY|-1)}}.
\end{align}
Thus for any $\omega>0$, there is an $N_3$ such that for any $N>N_3$, we have that $\Omega(\bar{\bq}) \leq \omega$ for any population strategy profile where (a) the average probability of reporting any answer $y$ is either $0$ or at least $\eta$, and (b) there exists an agent whose expected payoff is larger than the expected payoff under the truthful strategy profile. Thus, all the statements of the theorem hold for any $N$ larger than $N_0 = \max(N_1, N_2, N_3)$. \hfill $\Box$\\
\end{proof}

\section{Properties of the square-root agreement measure}\label{apx:agreeprop}
The SRAM has the following properties.
\begin{enumerate}
\item $\Gamma(Y_1,Y_2)\geq 1$. To see this, note that Jensen's inequality implies that 
$$\sum_{y\in \cY} \sqrt{ \sum_{x\in \cX}P_X(x)p_y(x)^2}\geq \sum_{y\in \cY} \sum_{x\in \cX}P_X(x)p_y(x) = 1.$$
In fact $\Gamma(Y_1,Y_2)= 1$ only when $Y_1$ and $Y_2$ are independent.
\item $\Gamma(Y_1,Y_2)\leq \sqrt{|\cY|}$. To see this, note that Jensen's inequality implies that 
\begin{align*}
\sum_{y\in \cY} \sqrt{\sum_{x\in \cX}P_X(x)p_y(x)^2} &\leq |\cY| \sqrt{\frac{1}{|\cY|} \sum_{y\in \cY} \sum_{x\in \cX}P_X(x)p_y(x)^2}\\
&\leq |\cY| \sqrt{\frac{1}{|\cY|} \sum_{y\in \cY} \sum_{x\in \cX}P_X(x)p_y(x)} =  \sqrt{|\cY|}.
\end{align*}
In fact $\Gamma(Y_1,Y_2) = \sqrt{|\cY|}$ only when $Y_1$ and $Y_2$ are identical and they are distributed uniformly, i.e., $Y_1=Y_2$ and $P(Y_1=y)=1/|\cY|$ for all $y\in\cY$.
\end{enumerate}

We also prove the following inequality satisfied by the SRAM, which generalizes Proposition~\ref{prop:monotone} without the characterizing the inequality gap.

\begin{prop}[{\bf A general monotonicity property}]\label{prop:monotone2}
Consider a generating model $(P_X,\bp)$ defined over $\cX$ and $\cY$, and consider two random responses $Y_1$ and $Y_2$ drawn from this model. Also, consider two random responses $Z_1$ and $Z_2$ obtained by applying a reporting strategies $\bq$ and $\bq'$ independently to $Y_1$ and $Y_2$ respectively. Then, 
\begin{equation}
\sum_{y\in\cY}\sqrt{P(Z_1 = Z_2 = y)}\leq \Gamma(Y_1,Y_2). \label{eq:monotone2}
\end{equation}
Moreover, if $\delta(P_X,\bp)>0$, then the above inequality is an equality if and only if $\bq = \bq'$ and $\Omega(\bq)=0$, i.e., if and only if the two reporting strategies are identical and fully informative.
\end{prop}
\begin{proof}{Proof of Proposition~\ref{prop:monotone2}}
We have,
\begin{align*}
\sum_{y\in\cY}\sqrt{P(Z_1 = Z_2 = y)}&= \sum_{y\in \cY}\sqrt{\sum_{x\in\cX,y_1\in \cY, y_2\in\cY} P_X(x)p_{y_1}(x)p_{y_2}(x)q_y(y_1)q'_y(y_2)}\\
&= \sum_{y\in \cY}\sqrt{\sum_{y_1\in \cY, y_2\in\cY}q_y(y_1)q'_y(y_2)\sum_{x\in\cX} P_X(x)p_{y_1}(x)p_{y_2}(x)}\\
&\stackrel{(a)}{\leq} \sum_{y\in \cY}\sqrt{\sum_{y_1\in \cY, y_2\in\cY}q_y(y_1)q'_y(y_2)\bigg(\sqrt{\sum_{x\in\cX} P_X(x)p_{y_1}(x)^2}\sqrt{\sum_{x\in\cX} P_X(x)p_{y_2}(x)^2}}\bigg)\\
&=\sum_{y\in \cY}\sqrt{\bigg(\sum_{y_1\in \cY}q_y(y_1)\sqrt{\sum_{x\in\cX} P_X(x)p_{y_1}(x)^2}\bigg)\bigg(\sum_{y_2\in \cY}q'_y(y_2)\sqrt{\sum_{x\in\cX} P_X(x)p_{y_2}(x)^2}\bigg)}\\
&\stackrel{(b)}{\leq} \frac{1}{2}\sum_{y\in \cY,y_1\in \cY}q_y(y_1)\sqrt{\sum_{x\in\cX} P_X(x)p_{y_1}(x)^2} +\frac{1}{2}\sum_{y\in \cY,y_2\in \cY}q'_y(y_2)\sqrt{\sum_{x\in\cX} P_X(x)p_{y_2}(x)^2}\\
&=\frac{1}{2}\sum_{y_1\in \cY}\sqrt{\sum_{x\in\cX} P_X(x)p_{y_1}(x)^2} +\frac{1}{2}\sum_{y_2\in \cY}\sqrt{\sum_{x\in\cX} P_X(x)p_{y_2}(x)^2}\\
&= \frac{\Gamma(Y_1,Y_2)}{2} + \frac{\Gamma(Y_1,Y_2)}{2}\\
&= \Gamma(Y_1,Y_2)
\end{align*}
Here, (a) follows from the Cauchy-Schwarz inequality, and (b) results from the fact that the arithmetic mean of two numbers is no less than the geometric mean. 

Now suppose that $\delta(P_X,\bp)>0$. Then (a) is an equality if and only if $q_y(y_1)q'_y(y_2)=0$ for every $y$ and every $y_1\neq y_2$. Further, (b) is an equality, i.e., arithmetic mean equals geometric mean, if and only if all the terms are equal. This means for all $y\in\cY$, 
\begin{align}
\sum_{y_1\in \cY}q_y(y_1)\sqrt{\sum_{x\in\cX} P_X(x)p_{y_1}(x)^2} = \sum_{y_2\in \cY}q'_y(y_2)\sqrt{\sum_{x\in\cX} P_X(x)p_{y_2}(x)^2},\nonumber
\end{align}
i.e., if 
\begin{align}
\sum_{y' \in \cY}(q_y(y')-q'_y(y'))\sqrt{\sum_{x\in\cX} P_X(x)p_{y'}(x)^2} = 0.
\end{align}
Squaring both sides, we obtain, for all $y\in\cY$,
\begin{align}
&\sum_{y' \in \cY}(q_y(y')-q'_y(y'))^2(\sum_{x\in\cX} P_X(x)p_{y'}(x)^2)\nonumber\\
&~~+\sum_{y'\neq y''} (q_y(y')-q'_y(y'))(q_y(y'')-q'_y(y''))\sqrt{\sum_{x\in\cX} P_X(x)p_{y'}(x)^2}\sqrt{\sum_{x\in\cX} P_X(x)p_{y''}(x)^2}= 0.
\end{align}
Substituting $q_y(y')q'_y(y'') = 0$ for all $y'\neq y''$, we obtain,
\begin{align}
&\sum_{y' \in \cY}(q_y(y')-q'_y(y'))^2(\sum_{x\in\cX} P_X(x)p_{y'}(x)^2)\nonumber\\
&~~+\sum_{y'\neq y''} (q_y(y')q_y(y'')+q'_y(y')q'_y(y''))\sqrt{\sum_{x\in\cX} P_X(x)p_{y'}(x)^2}\sqrt{\sum_{x\in\cX} P_X(x)p_{y''}(x)^2}= 0.\label{bad}
\end{align}
But if $\delta(P_X,\bp)>0$, then we know from Proposition~\ref{prop:equi} that $\sqrt{\sum_{x\in\cX} P_X(x)p_{y'}(x)^2}\geq \sum_{x\in\cX} P_X(x)p_{y'}(x) >0$ for all $y'\in\cY$. Hence, we conclude that if $\delta(P_X,\bp)>0$, then $\sum_{y\in\cY}\sqrt{P(Z_1 = Z_2 = y)} = \Gamma(Y_1,Y_2)$ holds, if and only if all the terms in Equation~\ref{bad} are $0$, i.e., if and only if
\begin{enumerate}
\item $q_y(y')=q'_y(y')$ for all $y,y'\in\cY$, i.e., $q$ and $q'$ are identical, and,
\item $q_y(y')q_y(y'') = 0$ for all $y\in\cY$ and $y'\neq y''$, i.e., $\Omega(q)=0$.
\end{enumerate}

This finishes the proof.  \hfill $\Box$
\end{proof}
\subsection{Utility of the square-root agreement measure beyond our work}\label{apx:agree}
Definition~\ref{def:agree} essentially defines an agreement measure between any two random variables that are independent and identically distributed conditioned on some latent random variable. But we could just as well define an agreement measure between any two random variables that take values in some common finite set.
\begin{definition}
Consider two random variables $X$ and $X'$, which take values in a finite set $\cS$. Then the square-root agreement measure between $X$ and $X'$ is defined as  
\begin{equation*}
\Gamma(X,X') = \sum_{s\in \cS}\sqrt{P(X=X'=s)}.
\end{equation*}
\end{definition}

Proposition~\ref{prop:monotone2} implies that if $X\rightarrow X'\rightarrow Y$ form a Markov chain, i.e., $X$ is conditionally independent of $Y$ given $X'$, and if, conditioned on some latent random variable $U$, $X$ and $X'$ are independent and identically distributed random variables, then, 
$$\Gamma(X,Y) \leq \Gamma(X,X').$$
Inequalities of this form are called {\it data processing} inequalities and they have several applications in information theory, statistics, causal inference, and related fields. For example, such inequalities provide testable hypotheses to determine the validity of conditional independence assumptions across variables from data. Several {\it mutual information} measures between two random variables are known to satisfy such inequality. These measures are typically constructed from two classes of divergences or distance notions between probability distributions, called f-divergences and Bregman divergences; see \cite{kong2016framework} and references therein. It is interesting to note that our SRAM does not result from such a construction, and to the best of our knowledge, the resulting data-processing inequality was not known in the literature. Moreover, typical mutual information measures depend on the entire joint distribution of two variables, i.e., to estimate these measures from data, one typically needs to learn $|\cS|^2$ probability values where $\cS$ is the support set of each variable. On the other hand, the SRAM only depends on the diagonal values of the joint probability distribution, i.e., only the probabilities of agreement matter. Hence, to estimate the SRAM from data, one only needs to learn $|\cS|$ probability values. It is important to note that for the data processing inequality to hold for the SRAM, $X$ and $X'$ need to be conditionally independent and identically distributed (conditioned on some latent random variable). There is typically no such requirement for other measures. To show that this condition is necessary, consider the following counterexample. Suppose that $X$ is uniformly distributed on the discrete set $\{-1,+1\}$, and $X'=-X$. Thus $\Gamma(X,X') = 0$. Whereas if $Y=-X'$, then it is true that $X\rightarrow X'\rightarrow Y$ forms a Markov chain, and $\Gamma(X,Y)= \Gamma(X,X) = \sqrt{2}$. Hence, $\Gamma(X,X') <\Gamma(X,Y)$. \\

\section{Properties of the uninformativeness measure}\label{apx:uninformprop}
The uninformativeness measure has the following properties.
\begin{enumerate}
\item Clearly, $\Omega(\bq)=0$ if and only if $(\bq(y);\, y\in\cY)$ have disjoint supports across all $y\in\cY$, i.e., if and only if $\bq$ is fully informative. 
\item $\Omega(\bq)$ attains its highest value of $1$, if and only if $\bq(y) = \bq(y')$ for any $y\neq y'$, i.e., if the report is chosen independently of the true answer. To see this, observe that, 
%\begin{align*}
%&\bigg(\frac{1}{|\cY|}\sum_{y\in \cY}\sqrt{\sum_{y'\in \cY,y''\in\cY}q_y(y')q_y(y'')\mathbf{1}_{\{y'\neq y''\}} +\sum_{y_1}q_y(y_1)^2}\bigg)^2\\%
%&=1\\
%&\leq \frac{1}{|\cY|}\sum_{y\in \cY}\sum_{y'\in \cY,y''\in\cY}q_y(y')q_y(y'')\mathbf{1}_{\{y'\neq y''\}} + \frac{1}{|\cY|}\sum_{y\in \cY}\sum_{y_1}q_y(y_1)^2\\
%&= \sqrt{\frac{1}{|\cY|}\sum_{y'\in \cY,y''\in\cY}\mathbf{1}_{\{y'\neq y''\}}\sum_{y\in \cY}q_y(y')q_y(y'')}\\
%&\leq \sqrt{\frac{1}{|\cY|}\sum_{y'\in \cY,y''\in\cY}\mathbf{1}_{\{y'\neq y''\}}\sqrt{\sum_{y\in \cY}q_y(y')^2}\sqrt{\sum_{y\in \cY}q_y(y'')^2}}\\
%&\leq \frac{1}{|\cY|}\sum_{y\in \cY}\sqrt{\sqrt{\sum_{y'\in \cY,y''\in\cY}q_y(y')^2\mathbf{1}_{\{y'\neq y''\}}}\sqrt{\sum_{y'\in \cY,y''\in\cY}q_y(y'')^2\mathbf{1}_{\{y'\neq y''\}}}}\\
%&=\frac{1}{|\cY|}\sum_{y\in \cY}\sqrt{(|\cY|-1)\sum_{y'\in \cY}q_y(y')^2}\\
%&=|\cY|-1
%\end{align*}
\begin{align*}
&\frac{1}{|\cY|(|\cY|-1)}\sum_{y\in \cY}\sum_{y'\in \cY,y''\in\cY}\sqrt{q_y(y')q_y(y'')\mathbf{1}_{\{y'\neq y''\}}}\\
&\stackrel{(a)}{\leq} \frac{1}{|\cY|(|\cY|-1)}\sum_{y\in \cY}\sqrt{\bigg(\sum_{y'\in \cY,y''\in\cY}q_y(y')\mathbf{1}_{\{y'\neq y''\}}\bigg)\bigg(\sum_{y'\in \cY,y''\in\cY}q_y(y'')\mathbf{1}_{\{y'\neq y''\}}\bigg)}\\
&=\frac{1}{|\cY|(|\cY|-1)}\sum_{y\in \cY}\sqrt{(|\cY|-1)^2\bigg(\sum_{y'\in \cY}q_y(y')\bigg)^2}\\
&=\frac{1}{|\cY|}\sum_{y\in \cY}\sum_{y'\in \cY}q_y(y')\\
&=1. \numberthis
\end{align*}
Here, (a) follows from the Cauchy-Schwarz inequality. \\
\end{enumerate}

\section{Miscellaneous remarks on existing mechanisms}\label{apx:misc}
%\subsection{The vanilla output agreement mechanism is not truthful for homogeneous responses}\label{apx:oa}
%An output agreement mechanism acts on a pair of agents, where the agents get a fixed positive payment only if their answers on a common evaluation task match, else they get a payment of 0. We provide an example to show that this mechanism does not yield a truthful equilibrium in general for homogeneous responses. We note that this observation is not new; we include it for completeness. We also note that this fact also follows from Proposition~\ref{prop:KS} below.
%\begin{example}\label{ex:oa}
%{\it Consider the setting in Example~\ref{ex1}. The question is ``Did the plumber show up within 5 minutes of his/her appointed time?'' with the possible observations/answers being $\cY=\{\textrm{Yes, No}\}$. Suppose an agent observes that the plumber did not arrive within 5 minutes of her appointed time, i.e., her observation was ``No.'' Then the conditional probability of the other agent, assumed to be truthful, replying ``Yes'' can be computed to be $0.54\overline{09}$, which is higher than the conditional probability of her replying ``No,'' which can be computed to be $0.459\overline{09}$. Hence, replying ``Yes,'' i.e., lying, results in a higher expected payoff than being truthful and replying ``No.'' }
%\end{example}
\subsection{An adaptation of the Correlated Agreement (CA) mechanism to the homogeneous responses setting}\label{apx:ca}
In this section, we present an adaptation of CA to the homogeneous responses setting, which induces truthful behavior while only requiring one evaluation per agent. %We will, however, show that this mechanism is no longer informed truthful; it is at best (asymptotically) informed truthful across symmetric equilibria. 
In order to define this mechanism, we first present the original CA mechanism.

{\bf Original CA mechanism.} CA operates on a pair of agents. Both agents perform a common ``bonus'' evaluation task, say A, and individually perform one independent ``penalty'' evaluation task that the other agent doesn't perform; say agent 1 performs B, and agent 2 performs C. In keeping with our notation, let $r^i_j$ for $j=1,2$, and $i \in\{A,B,C\}$, be the response of agent $j$ in task $i$, where the responses are taken to be the null $\phi$ if an agent doesn't perform the corresponding task (hence $r^C_1 = r^B_2 =\phi$). CA defines an intermediate scoring function that maps two responses to a real number, which, informally, is monotonically increasing in the expected correlation between the responses, i.e., $S$ is higher if the two responses are expected to frequently occur together. Formally, denoting $ \Delta_{ab}=P(Y^i_j = a, Y^i_{j'} = b)-P(Y^i_j = a)P(Y^i_{j'} = b)$, for any two responses $a,\, b$ (where it is assumed that agents $j$ and $j'$ have both performed task $i$), the intermediate score for these responses is defined to be:
$$S(a,b) = \sgn(\Delta_{ab}),$$ 
where $\sgn$ is the sign function. In the multi-task, homogeneous responses setting, this scoring function can be estimated from the response data obtained from the large number of other participants (i.e., excluding the two agents under consideration) operating on the platform, and relying on the ``self-fulfilling prophecy of truthfulness'' to assume truthful behavior.\footnote{In the general setting with non-homogeneous responses, this function can be estimated by having the two agents perform a large number of overlapping and disjoint tasks.} The final score/payment to an agent $i$ is then defined to be 
$$S(r^A_1, r^A_2) - S(r^B_1, r^C_2),$$
i.e., the final score/payment is the difference between the bonus score and the penalty score. The intuition is that the payment scheme rewards {\it incremental} correlation in the responses to the bonus task over what is expected anyway from the responses to two independent evaluation tasks.

{\bf An adaptation of CA for homogeneous responses requiring one evaluation per agent.} Consider the following adaptation of CA. Consider an agent, say $1$, who has performed evaluation task $A$, and whose payment needs to be determined. Let $2$ be another agent who has performed task $A$. Let $3$ be a third agent who has performed some task $B$ that $1$ hasn't performed. Then the payment to agent $1$ is defined to be:

$$S(r^A_1, r^A_2) - S(r^A_1, r^B_3).$$
Here, it is assumed that these scoring functions are calculated as in the original CA mechanism based on the responses data from all tasks other than $A$ and $B$. In a natural practical implementation of this mechanism, to calculate the payment of an agent $j$, the platform would randomly pick a peer agent who has performed the same evaluation to calculate the bonus score and similarly, randomly pick another agent who has performed some task that the agent $j$ hasn't performed to calculate the penalty score. These scoring functions are calculated from the response data of all tasks that $j$ hasn't performed. We call this mechanism CA for homogeneous responses (CA-HR).

It is clear that this mechanism only requires each agent to perform one evaluation as long as (a) each task is performed by at least two agents, and (b) there is a large number of tasks while each agent performs only a small number (so that the scores can be estimated accurately, independently of the agent's reports). These assumptions are almost the same as that required by SRA for its properties, and they are easily satisfied on most platforms. 

 It is easy to argue that truthful behavior is an equilibrium under CA-HR (in the large tasks regime where the scoring function estimates are reasonably accurate) in the homogeneous responses setting. This is because, due to the statistical indistinguishability of agent $2$ and $3$'s responses to an arbitrary task assuming that they are truthful, replacing agent $2$ with agent $3$ in the calculation of the penalty score is inconsequential from the perspective of agent $1$. All that matters from the perspective of agent $1$ (in terms of aligning with the incentives generated by the original CA mechanism) is that this penalty score is computed on the basis of some agent's response to a task that $1$ hasn't performed. Thus, the fact that truthful behavior is a best response under the original CA mechanism implies that it is a best response under this modification as well.

\subsection{Remarks on the properties of CA/CA-HR in our setting.} \label{apx:ca2}
%This example shows that the reduction in the number of responses required per agent comes at the cost of the loss of an important incentive property of CA. 
Although CA is informed truthful in the setting in which it is originally defined, neither CA nor CA-HR are informed truthful in our setting. This is because in our setting, task allocations are exogenously specified and agents can choose task-contingent reporting strategies based on task identities. We present an example below that shows this for CA-HR.
\begin{example}
{\it Consider the setting in Example~\ref{ex1} again. For the sake of the present discussion, suppose that the plumbers are numbered $i=1,\cdots, N$ (just as the tasks are numbered in our formal model). If everyone is truthful, the accurate scoring function is $S(a,b) = \mathbf{1}_{\{a=b\}} -\mathbf{1}_{\{a\neq b\}}$. Suppose that $j$ has evaluated plumber $A$. $j'$ is her randomly chosen peer who also has also evaluated $A$. Let $j''$ be another randomly chosen peer who has evaluated plumber $B$, whom $j$ hasn't evaluated. Then the (random) payment of agent $j$ under CA-HR is 
$$ \mathbf{1}_{\{Y^A_j=Y^A_{j'}\}} -\mathbf{1}_{\{Y^A_j\neq Y^A_{j'}\}} - \mathbf{1}_{\{Y^A_j = Y^B_{j''}\}} + \mathbf{1}_{\{Y^A_j \neq Y^B_{j''}\}}.$$
Thus, the expected payment of agent $j$ can be determined to be 
$$2\left(P(Y^A_j=Y^A_{j'}) - P(Y^A_j=Y^B_{j''})\right)$$
This can be computed to be $0.2025$, given the generating model. On the other hand, consider the following strategy profile. For all even tasks, agents report `Yes,' and for all odd tasks, agents report `No.' %Note that this is a valid strategy profile in our setting described in Section~\ref{sec:model} since agents are allowed to choose task-contingent reporting strategies.  
We first argue that this strategy profile is an equilibrium in the many tasks regime. Note that under this strategy profile, the scoring function that will be estimated by the platform is $\bar{S}(a,b) = \mathbf{1}_{\{a=b\}} -\mathbf{1}_{\{a\neq b\}}$, same as that under truthful behavior. Thus, the expected payment of an agent for reporting `No' on an even task (or reporting `Yes' on an odd task) is $-1$, whereas the expected payment from following the prescribed strategy is $1$. This argument shows both, that (a) this strategy profile constitutes an equilibrium and (b) the expected payoff to any agent under this strategy profile (which is $1$) is strictly higher than the expected payment under the truthful equilibrium (which is $0.2025$). Thus CA-HR is not informed truthful.}
\end{example}
The same construction of a non-truthful strategy profile also shows that CA is also not informed truthful in our setting. 

Moreover, unlike the mechanism of \cite{kong2016framework} (see Section~\ref{sec:kong}), in our setting, neither CA nor CA-HR are informed truthful across all equilibria where agents choose the same reporting strategy for each task they perform. This is because task allocations are exogenously specified: in the example above, it could very well be the case that every agent performs exactly one task under CA-HR. In this case, the non-truthful equilibrium strategy profile constructed above respects the constraint that each agent chooses the same reporting strategy for each task they perform, simply because each agent performs only one task. A similar argument shows this for CA by considering a situation in which each agent performs either even tasks only or odd tasks only. 
%; see Section~\ref{apx:imposs} below for a discussion and an example. 

Although CA and CA-HR are not informed truthful in our setting, Lemma~5.12 in \cite{shnayder2016informed} implies that these mechanisms are informed truthful across symmetric equilibria, i.e., they are informed truthful when restricted to symmetric equilibria, in the many tasks limit. 

Next, we discuss why CA and CA-HR are {\it not} (asymptotically) strongly truthful across symmetric equilibria in general for homogeneous responses, i.e., there could be symmetric strategy profiles that are {\it not} fully informative, that asymptotically yield the same payoff as the truthful equilibrium. %Hence CA-HR has a strictly weaker equilibrium dominance property as compared to SRA. 
The existence of such strategy profiles is related to the following notion of ``clustered observations'' as defined in \cite{shnayder2016informed}.
\begin{definition}\citep{shnayder2016informed}\label{apx:clust}
A distribution of two agents' observations for a common evaluation is said to be clustered if there exist at least two identical rows in the matrix $[\sgn(\Delta_{yy'})]_{y\in\cY,\,y'\in\cY}$. (Note that $[\sgn(\Delta_{yy'})]_{y\in\cY,\,y'\in\cY}$ is a symmetric matrix under homogeneous responses)
\end{definition}

In the presence of clustered observations, there are symmetric equilibrium strategy profiles that are not fully informative, that yield the same payoff asymptotically as the truthful equilibrium under CA/CA-HR.
To see this for CA-HR, suppose that $y$ and $\bar{y}$ are two observations for which the corresponding rows $(\sgn(\Delta_{yy'}); y'\in\cY)$ and $(\sgn(\Delta_{\bar{y}y'}); y'\in\cY)$ are identical. Then, if all agents report a fixed observation, e.g., $y$, irrespective of whether they observe $y$ or $\bar{y}$, the scoring function estimated by the platform under CA-HR is the same as that under truthful behavior, except with the answer $\bar{y}$ eliminated as a possible report. However, if everyone else was truthful, the bonus and penalty scores obtained by an agent would have anyway been identical irrespective of whether any of the three agents involved in computing a payment report $y$ or $\bar{y}$. Thus the payments to all agents remain the same if everyone reports $y$ irrespective of whether they observe $y$ or $\bar{y}$. It thus follows that this strategy profile is an equilibrium under CA-HR, which yields the same expected payoff to any agent as the truthful equilibrium. This strategy profile is not a fully informative strategy profile, and hence, CA-HR is not asymptotically strongly truthful across symmetric equilibria. The mechanism is essentially incapable of identifying the difference between $y$ and $y'$ since it depends only on the sign structure of the $\Delta$ matrix and not the values themselves.

If an instance does not possess clustered observations, CA and CA-HR are strongly truthful across symmetric equilibria. In our practically motivated experimental setup, however, we find that clustered observations are encountered with a high frequency; see Remark~\ref{rem:clust}. 

%\subsection{Comments on the \cite{kong2016framework} (KS) framework}
\subsection{Insufficiency of a single evaluation per agent with homogenous responses in the \cite{kong2016framework}  (KS) mechanism design framework}\label{apx:KS}
In this section, we show that it is impossible to design a mechanism within the KS mechanism design framework in the homogeneous responses setting, that incentivizes truthfulness with one evaluation per agent. The KS framework operates on a pair of agents and the payment of each agent is defined to be some scaling of an unbiased estimate of some mutual information measure constructed from their responses to a common set of tasks. The sufficiently of a single response per agent within this framework implies that the payment must be decided based only on the pair of agents' responses to a single task. We argue that such a payment scheme cannot strictly incentivize truthful behavior even in the homogenous, binary response setting. This result is not new; it has been shown in the general homogeneous responses setting in \cite{jurca2011incentives} (Theorem 1). We present a proof of the simpler binary responses case below for completeness. This result implies that there cannot be any mutual information measure satisfying information monotonicity, whose unbiased estimate can be constructed based on two agents' responses to a single evaluation task. 
\begin{prop}\label{prop:KS}\citep{jurca2011incentives}
In any truthful mechanism in the homogenous, binary responses setting that calculates the payment of an agent only as a function of the responses of the agent and her peer to a single evaluation task, the payment to the agent does not depend on her own responses.
\end{prop}
\begin{proof}{Proof.}
Consider an evaluation task with only two responses: $\cY=\{\textup{Yes},\,\textup{No}\}$. The payment scheme that depends on the responses of an agent and her peer to a common task is a specification of payment to the agent for every possible pair of responses. One of these payments can be $0$ without loss of generality since additive shifts of payments across all possibilities do not change the incentive structure of the game. Let us suppose that the payments are as shown in Table~\ref{tab:my-table}, where it is assumed that the agent is the row player.
\begin{table}[H]
\centering
\begin{tabular}{ccl}
                         & Yes                    & No                     \\ \cline{2-3} 
\multicolumn{1}{c|}{Yes} & \multicolumn{1}{c|}{a} & \multicolumn{1}{l|}{b} \\ \cline{2-3} 
\multicolumn{1}{c|}{No}  & \multicolumn{1}{c|}{c} & \multicolumn{1}{l|}{0} \\ \cline{2-3} 
\end{tabular}
\vspace{0.1in}
\caption{The payments to the row agent corresponding to the pair of responses for the common evaluation task.}
\label{tab:my-table}
\end{table}
Let the generating model have two possible types $\cX = \{A,B\}$, with $P_X = (1/2, 1/2)$, $\bp(A) = (p,1-p)$, and $\bp(B) = (q,1-q)$. %We show that for any fixed values of $a$, $b$, and $c$, there are $p$ and $q$ such that truthful reporting yields a strictly lower payoff for at least one of the two observations `Yes' and `No,' unless $a=c$ and $b=0$.
The expected payment of the agent if she reports `Yes' on observing `Yes' can be determined to be:
\begin{align}
\frac{a(p^2/2 + q^2/2) + b(p(1-p)/2 + q(1-q)/2)}{p/2 +q/2}.
\end{align}
The expected payment of the agent if she reports `No' on observing `Yes' can be determined to be:
\begin{align}
\frac{c(p^2/2 + q^2/2)}{p/2 +q/2}.
\end{align}
Thus reporting `Yes' on observing `Yes' yields a higher expected payment if 
%\begin{align}
%c(pp^2 + (1-p)r^2) <a(pp^2 + (1-p)r^2) + b(pp(1-p) + (1-p)r(1-r))
%\end{align}
%Or if
\begin{align}
(c-a)(p^2 + q^2) \leq b(p(1-p) + q(1-q)).\label{eq-1}
\end{align}
The expected payment of the agent if she reports `No' on observing `No' can be determined to be:
\begin{align}
\frac{c(p(1-p)/2 + q(1-q)/2)}{(1-p)/2 +(1-q)/2}.
\end{align}
The expected payment of the agent if she reports `Yes' on observing `No' can be determined to be:
\begin{align}
\frac{a(p(1-p)/2 + q(1-q)/2)+ b((1-p)^2/2 + (1-q)^2/2)}{(1-p)/2 +(1-q)/2}.
\end{align}
Thus reporting `No' on observing `No' yields a higher expected payment if 
\begin{align}
b((1-p)^2 + (1-q)^2)\leq (c-a)(p(1-p) + q(1-q)).\label{eq-2}
\end{align}
If we set $p$ and $q$ such that $p^2 + q^2 = p(1-p) +q(1-q)$ (e.g., $p=q=0.5$), then from Equation~\ref{eq-1} we obtain $b\geq c-a$ on the other hand, if we set $p$ and $q$ such that $(1-p)^2 + (1-q)^2 = p(1-p) +q(1-q)$ (e.g., $p=q=0.5$), then from Equation~\ref{eq-2} we obtain $b\leq c-a$. Thus, we have $b = c-a$. 

Next, if $b=c-a>0$, then Equations \ref{eq-1} and \ref{eq-2}, reduce to:
\begin{align}
p^2 + q^2 &\leq p(1-p) + q(1-q),\\
(1-p)^2 + (1-q)^2&\leq p(1-p) + q(1-q).\label{eq-2}
\end{align}
In this case, setting $p=q=0.25$ violates the second inequality.
If $b=c-a<0$, then Equations \ref{eq-1} and \ref{eq-2}, reduce to:
\begin{align}
p^2 + q^2 &\geq p(1-p) + q(1-q),\\
(1-p)^2 + (1-q)^2&\geq p(1-p) + q(1-q).\label{eq-2}
\end{align}
In this case, setting $p=q=0.25$ violates the first inequality. Hence, we have that $b = c-a = 0$, i.e., the mechanism's payments are independent of the reports of the agent.\hfill $\Box$
\end{proof}

\subsection{Infeasibility of a generic adaptation of the KS framework to multi-task, homogeneous responses settings and the special role of the square-root agreement measure (SRAM)}\label{apx:KS-shannon}
The design of SRA suggests that perhaps a generic adaptation of the KS mechanism to homogeneous responses setting that incentivizes single evaluations is possible under any mutual information measure. We argue that this is not true via the example of Shannon mutual information \citep{cover2006elements}. For two random variables $Y_1$ and $Y_2$ taking values in finite sets $\cY_1$ and $\cY_2$ respectively, the Shannon mutual information is defined to be,
\begin{align}
I(Y_1;Y_2) = \sum_{y\in \cY_1,\, y' \in \cY_2} P(Y_1=y,\,Y_2 = y')\log\frac{P(Y_1=y,\,Y_2 = y')}{P(Y_1=y)P(Y_2 = y')}.
\end{align}
Suppose that the distribution of two agents' responses to a common evaluation task is available to the platform (estimated from a large number of evaluation tasks). Then, along the lines of SRA, the mutual information measure above suggests the following mechanism.
\begin{enumerate}
\item Each agent $j$ is paired with another randomly chosen agent $j'$, and their responses are compared. 
\item If the response of agent $j$ is $y$ and that of agent $j'$ is $y'$, then $j$ gets a reward $K\log\left(\frac{P(Y_j=y,\,Y_{j'} = y')}{P(Y_j=y)P(Y_{j'} = y')}\right)$, where $K$ is some positive constant.  
\end{enumerate}
Under this mechanism, if $j$'s true response is $y$ and $j'$ is truthful, her expected reward for a truthful report is,
\begin{equation}
K\sum_{y'\in\cY}P(Y_{j'} = y'\mid Y_j = y)\log\frac{P(Y_j=y,\,Y_{j'} = y')}{P(Y_j=y)P(Y_{j'} = y')} = K\sum_{y'\in\cY}\frac{P(Y_{j'} = y',\,Y_j = y)}{P(Y_{j}=y)}\log\frac{P(Y_j=y,\,Y_{j'} = y')}{P(Y_j=y)P(Y_{j'} = y')}.\label{eq:truthful-shan}
\end{equation}
Similarly, her reward for any other report $\bar{y}$ is, 
\begin{equation}
K\sum_{y'\in\cY}\frac{P(Y_{j'} = y',\,Y_j = y)}{P(Y_{j}=y)}\log\frac{P(Y_j=\bar{y},\,Y_{j'} = y')}{P(Y_j=\bar{y})P(Y_{j'} = y')}.\label{eq:notruthful-shan}
\end{equation}
Thus being truthful yields a higher reward if for any $\bar{y}\neq y$, expression in Equation~\ref{eq:truthful-shan} is higher than the one in Equation~\ref{eq:notruthful-shan}, which simplifies to the condition,
%\begin{equation}
%K\sum_{y'\in\cY}P(Y_j = y,\,Y_{j'} = y')\log\frac{P(Y_j=y,\,Y_{j'} = y')P(Y_j = \bar{y})}{P(Y_j=\bar{y},\,Y_{j'} = y')P(Y_j=y)}\geq 0.\label{eq:istrue}
%\end{equation}
\begin{equation}
K\sum_{y'\in\cY}P(Y_j = y,\,Y_{j'} = y')\log\frac{P(Y_j=y,\,Y_{j'} = y')}{P(Y_j=\bar{y},\,Y_{j'} = y')} - K P(Y_j = y)\log\frac{P(Y_j=y)}{P(Y_j=\bar{y})}\geq 0.\label{eq:istrue}
\end{equation}
This inequality is not satisfied in general for homogeneous responses. We tested this condition in our experimental setup of Section~\ref{sec:numerics}. Assuming that there are $|\cY|=5$ responses as defined in that section, and two types of moving companies with delays exponentially distributed and mean delays drawn uniformly in $[0,60]$ (in minutes), we found that $629$ of $10000$ instances we generated violated the inequality in Equation~\ref{eq:istrue}.\\

%This underscores the importance of the new square-root agreement measure that we have proposed in our work, which is crucial to the unique properties of SRA.
%\input{endgame}
\section{Auxillary results}
%\begin{prop}\label{prop:aux}
%Homogenenous responses with $|\cY|=2$ satisfy the self-alignment condition assuming that every answer in $\cY$ has a positive probability of being observed.
%\end{prop}
%\begin{proof}{Proof.} Let the two responses be $\cY = \{a,b\}$. The self-alignment condition requires that 
%\begin{align}
%P(Y_{j'}=a\mid Y_j = a)&\geq P(Y_{j'}=a\mid Y_j = b),
%\end{align}
%which is the same as requiring that
%\begin{align}
%\frac{P(Y_{j'}=a,Y_j = a)}{P(Y_j= a)} &\geq \frac{P(Y_{j'}=a, Y_j = b)}{1-P(Y_j = a)},
%\end{align}
%which is the same as 
%\begin{align}
%P(Y_{j'}=a,Y_j = a) &\geq P(Y_j = a)\left(P(Y_{j'}=a, Y_j = b)+ P(Y_{j'}=a,Y_j = a)\right),
%\end{align}
%or, $P(Y_{j'}=a,Y_j = a) \geq P(Y_j = a)^2$. This inequality is always true for homogeneous responses by applying Jensen's inequality to the convex function $f(x) = x^2$. The same conclusion holds if we switch the role of $a$ and $b$, thus proving the result.
%\end{proof}

\begin{prop}\label{prop:aux}
If responses are categorical then they are self-predicting.
\end{prop}
\begin{proof}{Proof.} For any two responses $y$ and $y'$, the categorical responses condition says that,
\begin{align}
P(Y_{j'}=y'\mid Y_j = y)&\leq P(Y_{j'}=y').
\end{align}
However, this implies that $P(Y_{j'}=y') \leq P(Y_{j'}=y'\mid Y_j = y')$. This means that for any two responses $y$ and $y'$, 
\begin{align}
P(Y_{j'}=y'\mid Y_j = y)&\leq P(Y_{j'}=y'\mid Y_j = y').
\end{align}
But this is exactly the self-prediction condition. \hfill $\Box$\\
\end{proof}

\begin{prop}\label{prop:aux2}
Consider two exchangeable random variables, $Y_1$ and $Y_2$, taking values in a finite set $\cY$. If their distribution satisfies the {\it strict} Cauchy-Schwarz property:
\begin{align}\sqrt{P(Y_{1} =Y_2 = y)}\sqrt{P(Y_1 = Y_2 =y')} > P(Y_1 =y,\,Y_2 = y'),\end{align}
for each $y,\, y'\in\cY$, then $Y_1$ and $Y_2$ are stochastically relevant random variables.
\end{prop}
\begin{proof}{Proof.}
We will show that stochastic irrelevance for two values $y$ and $y'$ implies that the CS property is satisfied for these values with an equality. Stochastic irrelevance for $y$ and $y'$ implies that the conditional distributions of $Y_2$ given $Y_1=y$ and $Y_1 = y'$ are identical. This implies that there is some constant $C>0$ such that $\left(P(Y_1 = y', Y_2 = a); a\in\cY\right)= C\times \left(P(Y_1 = y, Y_2 = a); a\in\cY\right)$. In particular we have that:
\begin{align}
&P(Y_1 = Y_2 = y') = C\times P(Y_1 = y,\, Y_2 = y')\textrm{ and}\\
&P(Y_1 = y',\, Y_2 = y) = C\times P(Y_1 = Y_2 = y).
\end{align}
We thus have,
\begin{align}
\sqrt{ P(Y_1 = Y_2 = y)}\sqrt{ P(Y_1 = Y_2 = y')} &= \sqrt{ P(Y_1 = Y_2 = y)}\sqrt{ C\times P(Y_1 = y,\, Y_2 = y')}\\
& = \sqrt{ P(Y_1 = Y_2 = y)}\sqrt{ \frac{P(Y_1 = y',\, Y_2 = y)}{P(Y_1 = Y_2 = y)}\times P(Y_1 = y,\, Y_2 = y')}\\
& = \sqrt{P(Y_1 = y',\, Y_2 = y)P(Y_1 = y,\, Y_2 = y')}\\
&\overset{(a)}{=} P(Y_1 = y,\, Y_2 = y').
\end{align}
Here $(a)$ follows from exchangeability of $Y_1$ and $Y_2$. Thus the CS property is satisfied with an equality for $y$ and $y'$.  \hfill $\Box$\\
\end{proof}

\end{APPENDIX}

\end{document}